\documentclass[11pt]{article}
\usepackage{jcappub}
\usepackage{graphicx}
\usepackage{dcolumn}
\usepackage{bm}
\usepackage{latexsym}
\usepackage{amsfonts}
\usepackage{cancel}
\usepackage{umoline}
\usepackage{xspace} 
\usepackage{amssymb}
\usepackage{amsmath}
\usepackage{mathrsfs}
\usepackage{afterpage}
\usepackage{placeins}
\usepackage{color}
\usepackage{hyperref}
\usepackage{natbib} 
\usepackage{siunitx}
\usepackage{verbatim}
\usepackage{ dsfont }
\usepackage{float}
\usepackage{subcaption}
\usepackage[most]{tcolorbox}
\newcommand{\xv}{\mathbf{x}}

\newcommand{\kv}{\mathbf{k}}
\newcommand{\nv}{\mathbf{n}}

\newcommand{\qv}{\mathbf{q}}

\newcommand{\del}{\delta}

\usepackage[OT2,T1]{fontenc}
\DeclareSymbolFont{cyrletters}{OT2}{wncyr}{m}{n}
\DeclareMathSymbol{\Sha}{\mathalpha}{cyrletters}{"58}
\DeclareMathSizes{12}{20}{4}{10}

\newcommand{\be}{\begin{equation}}
\newcommand{\ee}{\end{equation}}
\newcommand{\bea}{\begin{eqnarray}}
\newcommand{\eea}{\end{eqnarray}}
\newcommand{\bdm}{\begin{displaymath}}
\newcommand{\edm}{\end{displaymath}}
\newcommand{\nn}{\nonumber}

\newcommand{\eos}{\textsc{Eos}\xspace}
\newcommand{\quijote}{\textsc{Quijote}\xspace}
\newcommand{\eone}{\textsf{E}$1$\xspace}
\newcommand{\qone}{\textsf{Q}$1$\xspace}

\newcommand{\qtwo}{\textsf{Q}$2$\xspace}

\definecolor{ForestGreen}{rgb}{0.13, 0.55, 0.13}
\definecolor{airforceblue}{rgb}{0.36, 0.54, 0.66}
\definecolor{orange}{rgb}{1.0, 0.5, 0.0}

\usepackage[normalem]{ulem}

\newcommand{\Mpc}{\, h^{-1} \, {\rm Mpc}}

\newcommand{\cGpc}{\, h^{-3} \, {\rm Gpc}^3}
\newcommand{\kMpc}{\, h \, {\rm Mpc}^{-1}}
\newcommand{\fnl}{f_{\rm NL}^{\rm loc}}

\title{The Covariance of Squeezed Bispectrum Configurations}

\author[a,b,c,d]{Matteo Biagetti,}
\author[e]{Lina Castiblanco,}
\author[e]{Jorge Nore\~na,}
\author[c,d,a]{and Emiliano Sefusatti}

\affiliation[a]{Institute for Fundamental Physics of the Universe, Via Beirut 2, 34151 Trieste, Italy}
\affiliation[b]{SISSA - International School for Advanced Studies, Via Bonomea 265, 34136 Trieste,  Italy}
\affiliation[c]{Istituto Nazionale di Astrofisica, Osservatorio Astronomico di Trieste, via Tiepolo 11, 34143 Trieste, Italy}
\affiliation[d]{Istituto Nazionale di Fisica Nucleare, Sezione di Trieste,  via  Valerio  2,  34127 Trieste,  Italy}
\affiliation[e]{Instituto de F\'isica, Pontificia Universidad Cat\'olica de Valpara\'iso, Casilla 4950, Valpara\'iso, Chile}

\abstract{	We measure the halo bispectrum covariance in a large set of N-body simulations and compare it with theoretical expectations. We find a large correlation among (even mildly) squeezed halo bispectrum configurations. A similarly large correlation can be found between squeezed triangles and the long-wavelength halo power spectrum.  This shows that the diagonal Gaussian contribution fails to describe, even approximately, the full covariance in these cases.  We compare our numerical estimate with a model that includes, in addition to the Gaussian one, only the non-Gaussian terms that are large for squeezed configurations. We find that accounting for these large terms in the modeling greatly improves the agreement of the full covariance with simulations. We apply these results to a simple Fisher matrix forecast, and find that constraints on primordial non-Gaussianity are degraded by a factor of $\sim 2$ when a non-Gaussian covariance is assumed instead of the diagonal, Gaussian approximation.}
	
\begin{document}

\maketitle

\clearpage

\section{Introduction}

The next generation of large scale structure surveys such as Euclid~\cite{EUCLID:2011zbd}, the Dark Energy Spectroscopic Instrument (DESI)~\cite{DESI:2016fyo}, the Legacy Survey of Space and Time (LSST)~\cite{Abate:2012za} and the Spectro-Photometer for the History of the Universe, Epoch of Reionization, and Ices Explorer (SPHEREx)~\cite{Dore:2014cca} are going to map an unprecedented number of galaxies at high redshifts. These maps will describe the distribution of the large-scale structure with great accuracy. 

Most of cosmological information is routinely extracted from the 2-point correlation function of the galaxy distribution, or its Fourier-space counterpart, the power spectrum. However, low-redshift galaxy density perturbations are a highly non-Gaussian random field. As such, their statistical properties are described, in addition to 2-point statistics, by higher-order correlation functions, starting with the 3-point correlation function. In this work, we consider in particular the galaxy bispectrum, i.e. the Fourier Transform of the 3-point correlation function. Significant effort has been made to provide an accurate theoretical description of this statistic \cite{PhysRevLett.73.215,Matarrese:1997sk,Scoccimarro:2000sp,Scoccimarro:2000ee,Takada:2002qq,Szapudi:2004gg,Gaztanaga:2005an,Nichol:2006mg,Sefusatti:2006pa,Sefusatti:2007ih,Smith:2007sb,Liguori:2010hx,Pollack:2011xp,Assassi:2014fva,Baldauf:2014qfa,Angulo:2014tfa,McCullagh:2015oga,Lazanu:2015rta,deBelsunce:2018xtd,Sugiyama:2018yzo,Eggemeier:2018qae,Floerchinger:2019eoj,Steele:2020tak,Alkhanishvili:2021pvy}, while several works attempted to quantify the additional information potentially provided by the bispectrum \cite{Schmittfull:2014tca,Karagiannis:2018jdt,Yankelevich:2018uaz,Gualdi:2021yvq}. On the other hand, its measurement and analysis has been performed from early data-sets \cite{Scoccimarro:2000sp, Verde:2001sf} up to the latest BOSS survey  \cite{Gil-Marin:2014sta, Gil-Marin:2014baa, Slepian:2015hca,Gil-Marin:2016wya,Slepian:2016kfz}.

Despite such efforts, the analysis of the galaxy bispectrum has not yet reached the same maturity as the treatment of the galaxy power spectrum. One of the main challenges is the accurate estimation of the bispectrum covariance properties. The straightforward method consists in a direct measurement from a large set of mocks, typically obtained with approximate large-scale structure distributions based on Lagrangian Perturbation Theory \cite{Monaco:2016pys, Colavincenzo:2018cgf}. This has been used to obtain a full covariance matrix for bispectrum measurements in tests of theoretical modeling or forecasts \cite{Scoccimarro:2000sn, Sefusatti:2004xz, Sefusatti:2006pa, Pollack:2011xp, Chan:2016ehg, Byun:2017fkz, Chan:2017fiv, Hahn:2019zob, Oddo:2019run, Gualdi:2020ymf, Sugiyama:2019ike,  Byun:2020rgl, Hahn:2020lou, Gualdi:2021yvq, Oddo:2021iwq},  as well as in actual data analysis \cite{Scoccimarro:2000sp, Gil-Marin:2016wya}. The clear advantage is the natural inclusion of systematic and window effects in the case of redshift surveys. Moreover, it accounts for all non-Gaussian contributions to the bispectrum and power spectrum-bispectrum cross-covariance matrices. \footnote{However, approximate schemes based on second-order Lagrangian perturbation theory can reproduce the large-scale (tree-level) bispectrum, but provide an incomplete description of the trispectrum and higher-order correlators.} Yet, current sets of mock catalogs need a large number of realizations (on the order of a few thousands, see e.g. \cite{Kitaura:2015uqa}) for the precise determination of the covariance of the power spectrum alone. Including thousands of bispectrum configurations require roughly an order of magnitude more realizations.

When a large set of mocks is not available it is common to limit the bispectrum covariance matrix to its diagonal.
This can be estimated from a limited set of realizations (see e.g.~\cite{Eggemeier:2021cam} for a recent implementation) or approximated by its Gaussian expression. This Gaussian approximation is written in terms of the power spectrum alone, in turn, obtained from simulations or in linear theory (see e.g.~\cite{MoradinezhadDizgah:2020whw, Ivanov:2021kcd}). These are good approximations, at least for the analysis of large-scale measurements from simulations in boxes with periodic boundary conditions aiming at the determination of bias parameters \cite{Oddo:2019run}.\footnote{An alternative approach to the high dimensionality of the covariance matrix relative to the number of mocks available has been to compress the bispectrum to make the covariance in this reduced space tractable purely with mocks; compression has been explored in \cite{Gualdi:2017iey,Child:2018kec,Child:2018klv,Gualdi:2018pyw,Gualdi:2019sfc,Gualdi:2019ybt,Gualdi:2019sfc,Gualdi:2021yvq}.}

It is reasonable to expect, however, that non-Gaussian contributions could become relevant as we consider smaller scales \cite{Chan:2016ehg, Barreira:2019icq, Gualdi:2020ymf, Shirasaki:2020vkk} or in combination with window (i.e. finite-volume) effects such as super-sample covariance and local-average (see \cite{Chan:2017fiv, Barreira:2021ueb} or \cite{Wadekar:2019rdu} and references therein for the power spectrum case). In addition, the leading contribution to the cross-covariance between power spectrum and bispectrum is non-Gaussian and can be quite relevant at large scales as well \cite{Sefusatti:2006eu, Byun:2017fkz, Oddo:2021iwq}.

These arguments motivated a few works in recent years to explore in more detail an analytic description of the full joint power spectrum and bispectrum ($P+B$) covariance. A first theoretical prediction of all non-Gaussian contributions to the $P+B$ covariance matrix, including finite-volume effects, has been studied by \cite{Kayo:2012nm, Kayo:2013aha}. They studied weak lensing statistics in the context of the halo model, therefore focusing on matter correlators at relatively small scales.  A comparison of a prediction of the non-Gaussian covariance for the 3D  bispectrum of matter and halos is presented instead in \cite{Chan:2016ehg} but limited to equilateral configurations, showing only a qualitative, overall agreement. It is shown that non-Gaussian contributions, for these triangles, are subdominant w.r.t. the Gaussian one with some exception in the case of sparse halo distributions, essentially due to shot-noise. In \cite{Chan:2017fiv},  the  super-sample covariance is estimated for the bispectrum in the response function formalism \cite{Takada:2013wfa} finding good agreement with simulations. They show it to be a small contribution to the bispectrum covariance, but it is particularly relevant for the power spectrum-bispectrum cross-covariance.  In \cite{Barreira:2019icq}, the focus is on the covariance of squeezed triangular configurations of the matter bispectrum, following the results of \cite{Barreira:2017sqa}. For these triangles, the non-Gaussian contributions are shown to be significantly larger than the Gaussian one, already in the quasi-linear regime, but without direct comparison to simulation results. Ref. \cite{Sugiyama:2019ike} computes a full perturbation theory prediction for the covariance of the redshift-space multipoles of the power spectrum and bispectrum, the latter measured with the estimator proposed in \cite{Sugiyama:2018yzo}. This prediction includes all correlators up to the 6-point correlation but in the approximation of linear bias and accounts for survey volume by a simple rescaling of the volume, without including super-sample effects. The comparison with the galaxy mocks produced for the BOSS survey by \cite{Kitaura:2015uqa} shows a qualitatively good agreement, with some significant discrepancies particularly for the $P\!\!-\!\!B$ cross-covariance. Finally, \cite{Gualdi:2020ymf} also provides a comparison of an analytical covariance for the redshift-space $P$ multipoles and the bispectrum monopole with the numerical estimate from the same light-cone BOSS mocks. In this case, predictions are calibrated against the mocks in terms of three free parameters accounting for shot-noise contributions and the overall amplitudes of the power spectrum and bispectrum submatrices. This set-up provides a qualitative prediction of the relative size of off-diagonal terms, except for extra contributions due to window function effects not included in the model.       

This work presents an accurate model for the theoretical covariance of the galaxy bispectrum and its detailed comparison with estimates from numerical simulations. We explore the regime where non-Gaussian contributions to the covariance are most important, that is the squeezed triangular configurations where the longest-wavelength mode $k_L$ is at least three times smaller than the other two.
For such triangles, it is possible to express all relevant non-Gaussian contributions in terms of power spectrum and bispectrum configurations that can be directly measured in the simulations themselves (or accurately modeled in perturbation theory). This allows a rather accurate prediction of the bispectrum covariance and power spectrum-bispectrum cross-covariance matrices.
We highlight three main features: first, the Gaussian approximation fails to order one \cite{dePutter:2018jqk, Barreira:2019icq}; second, off-diagonal elements corresponding to triangles sharing a long mode are large (as we can expect); third, there are large cross power spectrum-bispectrum covariance terms where the power spectrum shares the same long mode as the squeezed bispectrum.

We test our prescription against the \quijote suite of simulations \cite{Villaescusa-Navarro:2019bje} and find that the agreement is at the $20\%$ level for (even mildly) squeezed configurations and within $40\%$ for all configurations. This is much better than the Gaussian approximation that has an error $\sim 100\%$ for squeezed configurations, and completely misses off-diagonal correlations.

This has an important impact on observables that are particularly sensitive to squeezed configurations, such as primordial non-Gaussianity,\footnote{A breaking of adiabaticity, such as that induced by the presence of multiple light fields during inflation, is expected to leave a characteristic signal in the squeezed limit \cite{Salopek:1990jq,Bartolo:2001cw,Bernardeau:2002jf,Bernardeau:2002jy,Creminelli:2004yq}. This has the same shape in that limit as the local template with amplitude $f_{\mathrm{NL}}$ \cite{Komatsu:2001rj}, which is often used to constrain this effect (see \cite{Planck:2019kim,Castorina:2019wmr} for recent results).} and in general for  methods exploiting consistency relations in large scale structures \cite{Esposito:2019jkb,Marinucci:2019wdb,Marinucci:2020weg}. Given that there are a large number of triangles in the mildly squeezed regime we consider, our findings are potentially relevant for constraints on any cosmological parameter.

We organize this work as follows: in Section \ref{sec:pbdef} we define the power spectrum and bispectrum covariances and find an improved formula to compute non-Gaussian terms of the bispectrum covariance. In Section \ref{sect:theory}, we estimate the relative importance of non-Gaussian terms over Gaussian ones, determining that the former cannot be neglected for squeezed configurations. We then use response functions to propose a prescription to take these terms into account and find a formula to analytically invert the joint power spectrum-bispectrum covariance. In Section \ref{sec:sim_meas} we verify our findings against a large suite of N-body simulations. We show that indeed non-Gaussian terms are large in the regime we predicted and find good agreement of our prescription with the data. We then verify through a $\chi^2$ test that the inverse covariance of our theoretical prediction is reliable. In Section \ref{sect:fisher} we determine the impact of non-Gaussian terms using a Fisher matrix to determine the constraining power of primordial non-Gaussianity using joint power spectrum-bispectrum measurements. We confirm that non-Gaussian terms cannot be neglected, leading to a degradation in constraints by a factor of $\sim 2$. We finally conclude in Section \ref{sec:conclusions}.

\section{The power spectrum and bispectrum covariances}\label{sec:pbdef}

In this section we briefly review the theoretical description of the covariance matrix of the power spectrum and the bispectrum (along with their cross-covariance) of a generic, non-Gaussian random field $\delta$. This can represent the matter, galaxy or halo distributions.
In section \ref{section_sums_approx}, we show how some approximations used in the evaluation of the mode-counting factors in the theoretical covariance can lead to large errors, and propose an efficient method to fix such problems.

\subsection{Estimators}

In our comparison with simulation results we deal with finite-volume effects. Therefore, we start by introducing the Discrete Fourier Transform (DFT) of the density contrast $\delta(\xv)$ as 
\begin{equation}
\delta(\kv) \equiv \int_V \frac{d^3 x}{(2\pi)^3} e^{-i\kv\cdot\xv}\delta(\xv)\,,
 \end{equation}
with the inverse given by the series
\begin{equation}
    \delta(\xv) \equiv k_f^3\sum_\kv e^{i\kv\cdot\xv} \delta(\kv)\,.
\end{equation}
The $N$-point correlator $P_N(\kv_1, ..., \kv_N)$ in Fourier space is then generically defined as
\begin{equation}
    \langle \delta(\kv_1)...\delta(\kv_N)\rangle \equiv \frac{\delta_K\left( \kv_{1\dots N}\right)}{k^3_f} P_N(\kv_1, ..., \kv_N),
\end{equation}
where we adopt the notation $\kv_{1...N} \equiv \kv_1 + ... + \kv_N$, $k_f = 2\pi/L$ is the fundamental frequency of a cubic box of volume $L^3$ while $\delta_K(\kv)$ stands for the Kronecker symbol equal to one when the argument vanish, zero otherwise. The cases of $N=2$ and 3 correspond to the power spectrum $P(k)$ and the bispectrum $B(k_1,k_2,k_3)$, but the full expression for the bispectrum covariance includes contributions from correlation functions of up to $N=6$. 

An unbiased estimator for the power spectrum of a catalog of particles in a box with periodic boundary conditions can be written as
\be
\label{eq:estP}
\hat{P}(k)  \equiv  \frac{k_f^3}{N_k}\sum_{\qv \in k} \,\del(\qv)\,\del(-\qv)
\ee
where the sum runs over all wavenumbers $\qv$ 
in the shell of radius $k$, that is such that $k-\Delta k/2\le q < k+\Delta k/2$, $\Delta k$ being the radial size of the shell. 
The normalization factor $N_k$ gives the number of modes in the shell
\be\label{eq:Nk}
N_k  \equiv  \sum_{\qv \in k}
\ee
and is often approximated in the ``thin shell'' limit  $k\gg\Delta k$ by the integral 
\be
N_k \simeq \frac1{k_f^3} \int_{k-\Delta k/2}^{k+\Delta k/2} dq\,q^2 d\Omega
=4\pi \, \frac{k^2\, \Delta k}{k_f^3} + {\mathcal O}(\Delta k^3)\,. \label{eq:NkInt}
\ee

Similarly, an unbiased estimator for the bispectrum can be written as \cite{Scoccimarro:1997st} 
\be\label{eq:estB}
\hat{B}(k_1,k_2,k_3)  \equiv  \frac{k_f^3}{N_{tr}(k_1,k_2,k_3)}\sum_{\qv_1 \in k_1}\sum_{\qv_2 \in k_2}\sum_{\qv_3 \in k_3}\,\delta_K(\qv_{123})\, \,\del(\qv_1)\,\del(\qv_2)\,\del(\qv_3)
\ee
where the normalization factor $N_{tr}$ gives the number of ``fundamental triangles'' formed by the vectors $\qv_i$ satisfying the condition $\qv_{123}=0$ that fall in the ``triangle bin'' defined by the triplet of bin centers $(k_1,k_2,k_3)$ and width $\Delta k$. This is given by
\be\label{eq:Nt}
N_{tr}(k_1,k_2,k_3)  \equiv  \sum_{\qv_1 \in k_1}\sum_{\qv_2 \in k_2}\sum_{\qv_3 \in k_3}\,\delta_K(\qv_{123})\,.
\ee
We discuss how to approximate this sum in section \ref{section_sums_approx}.
Note that, according to this definition, we allow for ``open bins'', whose centers do not satisfy the triangle condition themselves (e.g. $k_3 > k_1 + k_2$), but contain fundamental triangles that do ($q_3 < q_1 + q_2$ and permutations)~\cite{Oddo:2019run}.

\subsection{Power spectrum and bispectrum covariance}

The power spectrum covariance is defined in terms of the estimator $\hat{P}(k_i)\equiv \hat{P}_i$ of eq.~(\ref{eq:estP}) as
\be
\label{eq:defcovP}
    C^P_{ij} \equiv \langle \delta \hat{P}_i \delta \hat{P}_j \rangle,
\ee
with $\delta \hat{P}_i = \hat{P}_i - \langle \hat{P}_i \rangle$ and the indices $i$ and $j$ denoting the wavenumbers bins. For realizations of the density field in a box with periodic boundary conditions, it is well known \cite{Scoccimarro:1999kp} that the covariance matrix is the sum of a Gaussian and a non-Gaussian contributions,
\be
\label{eq:cpp}
C^{P}_{ij} = C_{ij}^{P, (PP)} + C_{ij}^{P, (T)}\,.
\ee
The Gaussian term, depending only on the power spectrum of the distribution, is given by
\be
\label{eq:cppG}
C^{P,(PP)}_{ij} 
 = 2 \frac{\delta_{ij}^K}{N_{k_i}^2}\sum_{\qv\in k_i} P(q)^2  \simeq 2 \frac{\delta_{ij}^K}{N_{k_i}}P(k_i)^2 \,,
\ee
where $\delta_{ij}^K$ is a Kronecker symbol vanishing when the bins $k_i$ and $k_j$ do not coincide and where in the second step we assume that $P(q)\simeq P(k_i)$ for $\qv\in k_i$, as expected in the thin-shell approximation. The non-Gaussian term depends instead on the trispectrum of the distribution $T(\qv_1,\qv_2,\qv_3,\qv_4)$ and is given by 
\be
\label{eq:cppT}
C^{P,(T)}_{ij} =  k_f^3\,\bar{T}(k_i,k_j) \equiv
 \frac{k_f^3}{N_{k_i} N_{k_j}} \sum_{\qv^i \in k_i} \sum_{\qv^j \in k_j} T(\qv^i, -\qv^i, \qv^j, -\qv^j)\,. 
\ee
It is easy to see that the two sums, along with the normalization factors $N_{k}$, provide, in practice, an average of the trispectrum over the angle $\theta$ between the two vectors $\qv^i$ and $\qv^j$. In fact, for thin shells ($\Delta k \ll k_i$) we can approximate in the expression above $T(\qv^i, -\qv^i, \qv^j, -\qv^j)\equiv T(q^i, q^j, \theta)\simeq T(k_i, k_j, \theta)$. In addition, the sum is often replaced by an integral, as in eq.~(\ref{eq:NkInt}), so that
\be
\label{eq:Tij}
\bar{T}(k_i, k_j) \simeq \frac{1}{2}\int d\cos\theta\, T(\kv_i, -\kv_i, \kv_j, -\kv_j)\,.
\ee

Similarly, the bispectrum covariance is defined in terms of the estimator $\hat B$ in eq.~(\ref{eq:estB}) as
\be
C^{B}_{ij}  \equiv \langle \delta \hat B_i \delta \hat B_j \rangle\,,
\ee
where the indices $i$ and $j$ now denote triplets of wavenumbers, so that $\hat{B}_i\equiv\hat{B}(k_1^i, k_2^i, k_3^i)$. Again, the full expression can be written in terms of a Gaussian and a non-Gaussian contribution, the latter depending on the density field bispectrum $B(k_1,k_2,k_3)$, trispectrum $T(\kv_1,\dots,\kv_4)$ and pentaspectrum $P_6(\kv_1,\dots,\kv_6)$  \cite{Sefusatti:2006pa}, that is 
\begin{align}
C^{B}_{ij} &= C^{B, (PPP)}_{ij} + C^{B, (BB)}_{ij} + C^{B, (PT)}_{ij} + C^{B, (P_6)}_{ij}\,.\label{appeqCBB}
 \end{align}
The Gaussian contribution is given, in the thin-shell approximation, by 
\be
\label{eq:CBB-PPP}
C^{B,(PPP)}_{ij} 
 \simeq \frac{\delta_{ij}\,s_B}{k_f^3 N_{tr}^i}\, P(k_1^i)\,P(k_2^i)\,P(k_3^i)\,, 
\ee
where $s_B = 6, 2, 1$ for equilateral, isosceles and scalene triangles, respectively, $N_{tr}^i$ is the number of fundamental triangles in the triangle bin $\left\{k_1^i, k_2^i, k_3^i\right\}$, eq.~(\ref{eq:Nt}). Assuming that the correlators are slowly varying in the wavenumber shells, the non-Gaussian terms can be written as
\begin{align}
C^{B,(BB)}_{ij}  & \simeq B_i\,B_j\,\left(\Sigma^{11}_{ij} + 8~{\rm perm.} \right)\label{appeqCBB-BB} \,,\\
C^{B,(PT)}_{ij}  & \simeq  P(k_1^i)\,\tilde{T}(k_2^i, k_3^i, k_2^j, k_3^j, k_1^i)\Sigma^{11}_{ij} +8~{\rm perm.} \label{appeqCBB-PT}\,,\\
C^{B,(P_6)}_{ij} &\simeq  k_f^3\, \tilde{P}_6(k_1^i, k_2^i, k_3^i, k_1^j, k_2^j, k_3^j)\,,\label{appeqCBB-E}
 \end{align}
where we introduce the  mode-counting factor
\be
\label{sigma_definition}
 \Sigma_{ij}^{ab} \equiv \frac{1}{N_{tr}^{i} N_{tr}^{j}} \sum_{\qv_1^i \in k_1^i\dots \qv_3^i \in k_3^i} \sum_{\qv_1^j \in k_1^j\dots \qv_3^j \in k_3^j} \delta_K(\qv_{123}^i)\, \delta_K(\qv_{123}^j)\, \delta_K(\qv_a^i + \qv_b^j) \,.
\ee
This can be approximated as $\Sigma_{ij}^{ab}  \simeq \frac{\delta_{k_a^i k_b^j}}{N_{k_a^i}}$ (e.g. \cite{Gualdi:2020ymf}), although the approximation is very inaccurate for some configurations, as we see in section \ref{section_sums_approx}. Similarly to the case of the non-Gaussian contribution to the power spectrum covariance, eq.~(\ref{eq:cppT}), $\tilde{T}$ is an angle-averaged trispectrum defined in this case as
\be
\tilde{T}(k_2^i, k_3^i, k_2^j, k_3^j, q) \equiv \frac{1}{\Delta p} \int dp\ T(k_2^i, k_3^i, k_2^j, k_3^j, q, p)\,,
\ee
where we have written the trispectrum $T(\kv_2^i, -\kv_3^i, \kv_2^j, -\kv_3^j)$ as a function of the sides of the quadrilateral formed by the momenta, and the two diagonals $q = k_{12}$ and $p=k_{14}$. The range of integration, of size $\Delta p$, is over all allowed values of $p$. Finally, $\tilde{P}_6$ is an angle average of the pentaspectrum (see e.g. \cite{Gualdi:2020ymf}) that we ignore in our implementation, under the assumption that it is negligible w.r.t. the other contributions (we verify this in Section~\ref{sect:theory}).   
The full covariance matrix for a data vector including both power spectrum and bispectrum measurements can be written as the block matrix 
\begin{equation}
\label{eq:fullmat}
    \mathbf{C} = \begin{pmatrix}
\mathbf{C}^P & \mathbf{C}^{PB}\\
\mathbf{C}^{BP} & \mathbf{C}^{B}
\end{pmatrix},
\end{equation}
that, in addition to the power spectrum and bispectrum covariance matrices $\mathbf{C}^P$ and $\mathbf{C}^B$ described above, includes the cross-covariance between the two statistics $\mathbf{C}^{PB}$ (and its transpose $\mathbf{C}^{BP}$). These are defined as
\begin{equation}
\label{eq:defcov}
    C^{PB}_{ij} \equiv C^{BP}_{ji}\equiv \langle \delta \hat{P}_i \delta \hat{B}_j \rangle\,.
\end{equation}

The explicit expression is given by \cite{Sefusatti:2006pa}
\begin{align}
C^{PB}_{ij} 
& =  C^{PB, (PB)}_{ij} + C^{PB, (P_5)}_{ij}\nonumber\\
& \simeq \frac{2}{N_{k_i}}P(k_i)\,B_j\left(\delta^K_{k_i, k_1^j} + \delta^K_{k_i, k_2^j}  + \delta^K_{k_i, k_3^j} \right) + k_f^3 \,\tilde{P_5}(k_{1}^j, k_2^j, k_3^j, k_i)  \,,\label{eq:PB_cov}
\end{align}
where $\tilde{P}_5$ is an angle-average of the tetraspectrum that we again assume negligible w.r.t. the first contribution, for the reasons that we discuss in Section~\ref{sect:theory}.

\subsection{Approximation of the  mode-counting factors}
\label{section_sums_approx}

The theoretical predictions presented above should be evaluated over the Fourier-space grid defined by the discrete modes $\qv=\nv\, k_f$ for a proper comparison to N-body simulations. This would provide an exact determination of all factors accounting for the number of modes in each $k$-shell, or in the intersections of such shells when dealing with one or more triangular bins. In addition, the correlators themselves should, in principle, be calculated on the Fourier grid and the result summed over the shells. This procedure (see e.g. \cite{Oddo:2019run, Oddo:2021iwq}), however, is numerically rather expensive and sums over modes are often replaced by integrals over a continuum of modes. 

In our approach, we assume the thin-shell approximation for all the relevant expressions as described in eq.s~(\ref{eq:cppG}), (\ref{eq:CBB-PPP})-(\ref{appeqCBB-PT}) and (\ref{eq:PB_cov}). Moreover, we employ integrals instead of sums in the evaluation of the mode-counting factors. We are particularly careful in the definition of such integrals and their integration limits, since a naive approach could lead to significant errors for some subsets of triangular configurations. 

As an illustration of the problem, we present here the explicit calculation for the integral that approximates the number  of fundamental triangles $N_{tr}$. Consider a fundamental triangle $\left\{\qv_1,\qv_2,\qv_3\right\}$ in a given triangular bin $\left\{k_1,k_2,k_3\right\}$ for a single measurement of the bispectrum estimator \eqref{eq:estB}. From eq.~(\ref{eq:Nt}) we have
\begin{align}
N_{tr} & \equiv \sum_{\qv_1\in k_1}\sum_{\qv_2\in k_2}\sum_{\qv_3\in k_3} \delta_K(\qv_{123}) 
\simeq \frac{1}{k_f^6}\int_{\qv_1 \in k_1}\!\!d^3q_1 \int_{\qv_2 \in k_2}\!\! d^3q_2 \int_{\qv_3 \in k_3}\!\! d^3q_3\ \delta_D(\qv_{123})\nn\\
&= \frac{8\pi^2}{k_f^6} \int_{k_1 - \Delta k/2}^{k_1 + \Delta k/2}\!\!dq_1 \int_{k_2 - \Delta k/2}^{k_2 + \Delta k/2}\!\! dq_2 \int_{k_3 - \Delta k/2}^{k_3 + \Delta k/2}\!\! dq_3 \int_{-1}^1 \!\! d\mu \ q_1\, q_2\, q_3 \delta_D\left(\mu - \frac{q_3^2 - q_1^2 - q_2^2}{2q_1q_2}\right)\,.
\end{align}
After taking advantage of the Dirac delta function to get rid of the integral on $\mu$, this leads to the usual result \cite{Scoccimarro:1997st, Scoccimarro:2003wn}
\be
N_{tr}
\simeq \frac{8\pi^2}{k_f^6} \int_{k_1 - \Delta k/2}^{k_1 + \Delta k/2}\!\!dq_1 \int_{k_2 - \Delta k/2}^{k_2 + \Delta k/2}\!\! dq_2 \int_{k_3 - \Delta k/2}^{k_3 + \Delta k/2}\!\! dq_3\ q_1 q_2 q_3 = \frac{8\pi^2}{k_f^6} k_1 k_2 k_3 \Delta k^3\,.
\label{Ntr_closed}
\ee
In these derivations, we implicitly assumed that all values of $q_1$, $q_2$ and $q_3$ within the integration limits do form a closed triangle corresponding to a value of $\mu$ also within the integration bounds. 

Such assumption is not satisfied by all measured configurations. In fact, it breaks (by a large margin) for flattened triangle bins, i.e. those with $k_1+k_2=k_3$, or for those configurations where the values of $k_1$, $k_2$ and $k_3$ (the $k$-bin centers) cannot even form a closed triangle. These triangle bins do include closed fundamental triangles in them, that is with $\qv_{123}=0$, that we want to keep in our analysis. We refer to these configurations, with some abuse of language, as ``open'' triangle bins.\footnote{It also breaks for other configurations, such as isosceles squeezed triangles. But we've checked that this gives a small correction in those other cases.}

\begin{figure}
    \begin{subfigure}{.5\textwidth}
        \centering
        \includegraphics[width=.8\linewidth]{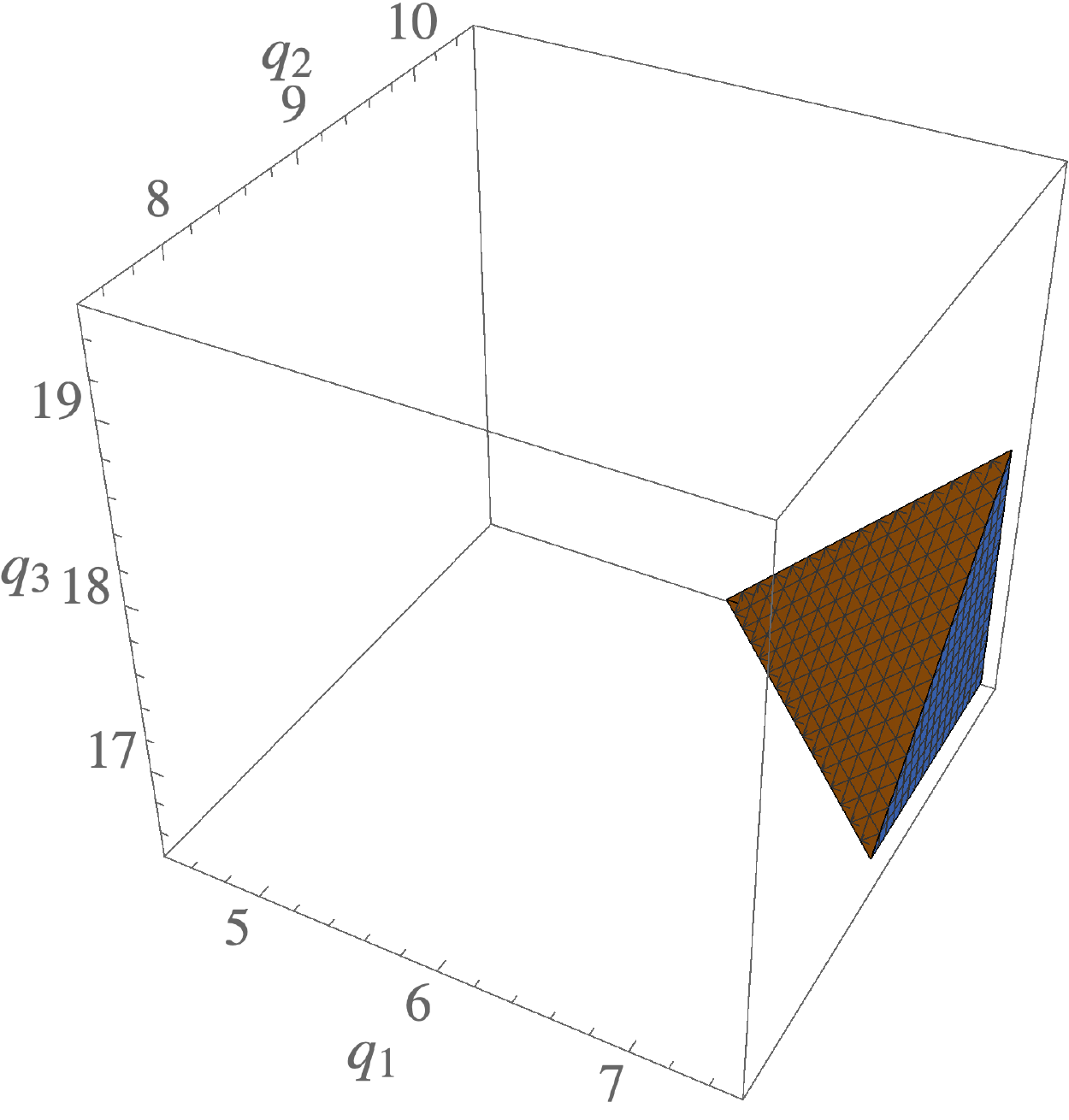}
    \end{subfigure}
    \begin{subfigure}{.5\textwidth}
        \centering
        \includegraphics[width=.8\linewidth]{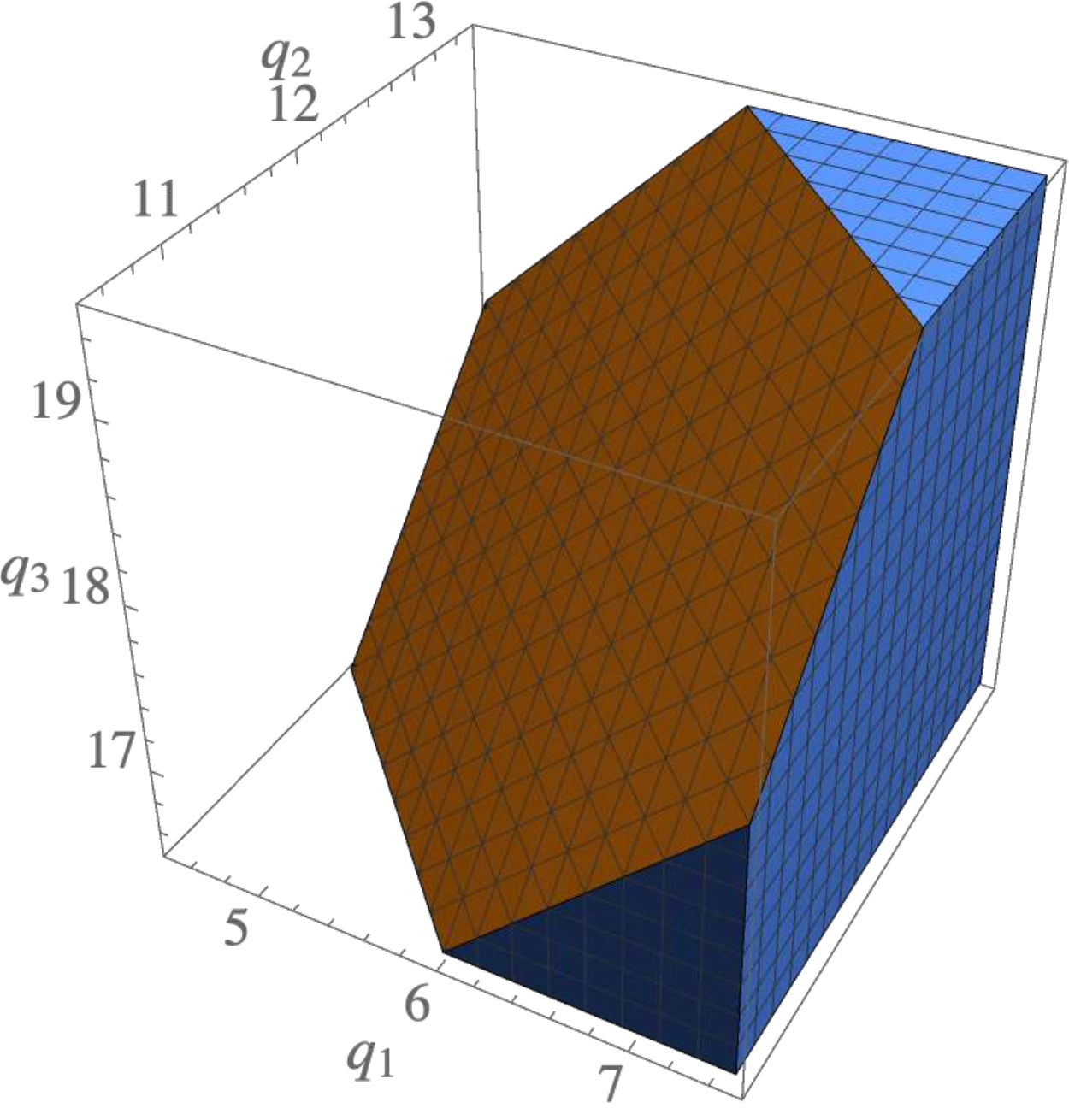}
    \end{subfigure}
    \caption{Regions of a typical bin in the magnitudes of the momenta $q_1$, $q_2$, $q_3$ with a given bin center $k_1$, $k_2$, $k_3$, where the triangle condition is satisfied $q_3 < q_1 + q_2$. Axes are in units of $k_f$. {\em Left:} A typical ``open'' bin for which $k_3 = k_1 + k_2 + \Delta k$. {\em Right:} A typical ``flattened'' bin for which $k_3 = k_1 + k_2$. Note that in these cases, integrating over the whole cube gives an $\mathcal{O}(1)$ error. Here we chose $k_1 = 6 k_f$, $k_2 = 12 k_f$, and $\Delta k = 3 k_f$, but all such configurations give identical plots. }
    \label{fig:cubes}
\end{figure}

For flattened and open triangles, the triangle condition can only be satisfied by the values of $q_1$, $q_2$ and $q_3$ in the regions depicted in Figure~\ref{fig:cubes}. The integral can still be performed analytically over those regions, giving
\be
N_{tr}^{\text{flattened}} = \frac{\pi ^2}{192 k_f^6}  \Delta k^3 \left[5 \Delta k^3+104 \left(k_1^2+k_2 k_1+k_2^2\right) \Delta k+768 k_1 k_2 \left(k_1+k_2\right)\right]\,,
\label{Ntr_open}
\ee
and
\be
N_{tr}^{\text{open}} = \frac{\pi ^2}{1152 k_f^6} \Delta k^3 \Big[24 k_1^2 \left(3 \Delta k+8 k_2\right) +\Delta k \left(72 k_2 \Delta k+17 \Delta k^2+72
   k_2^2\right)\Big]\,.
\label{Ntr_flattened}
\ee
In order to simplify these expressions we used $k_3 = k_1 + k_2$ for flattened triangles, and $k_3 = k_1 + k_2 + \Delta k$ for open triangles.

\begin{figure}
    \begin{subfigure}{.5\textwidth}
        \centering
        \includegraphics[width=\linewidth]{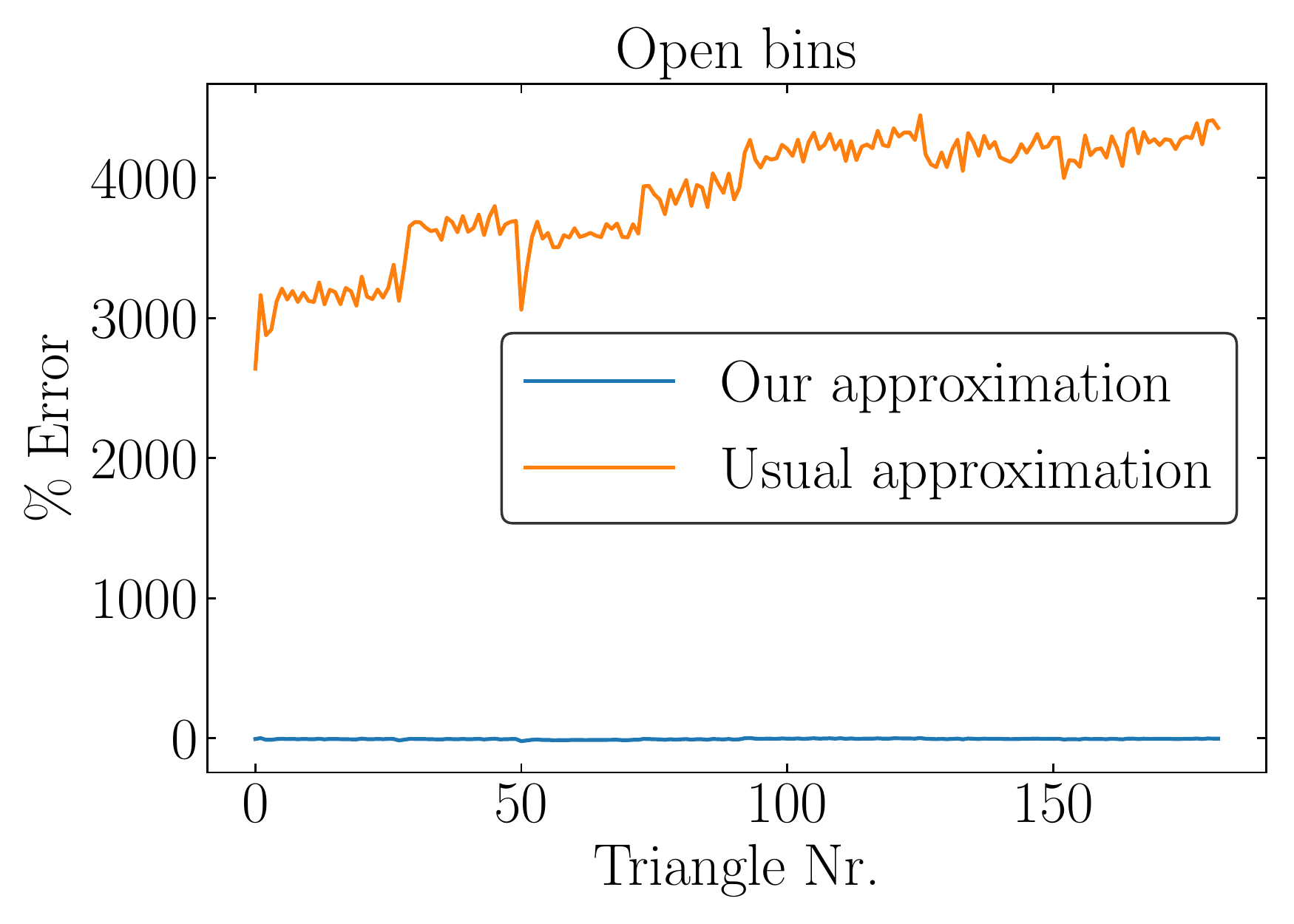}
    \end{subfigure}
    \begin{subfigure}{.5\textwidth}
        \centering
        \includegraphics[width=\linewidth]{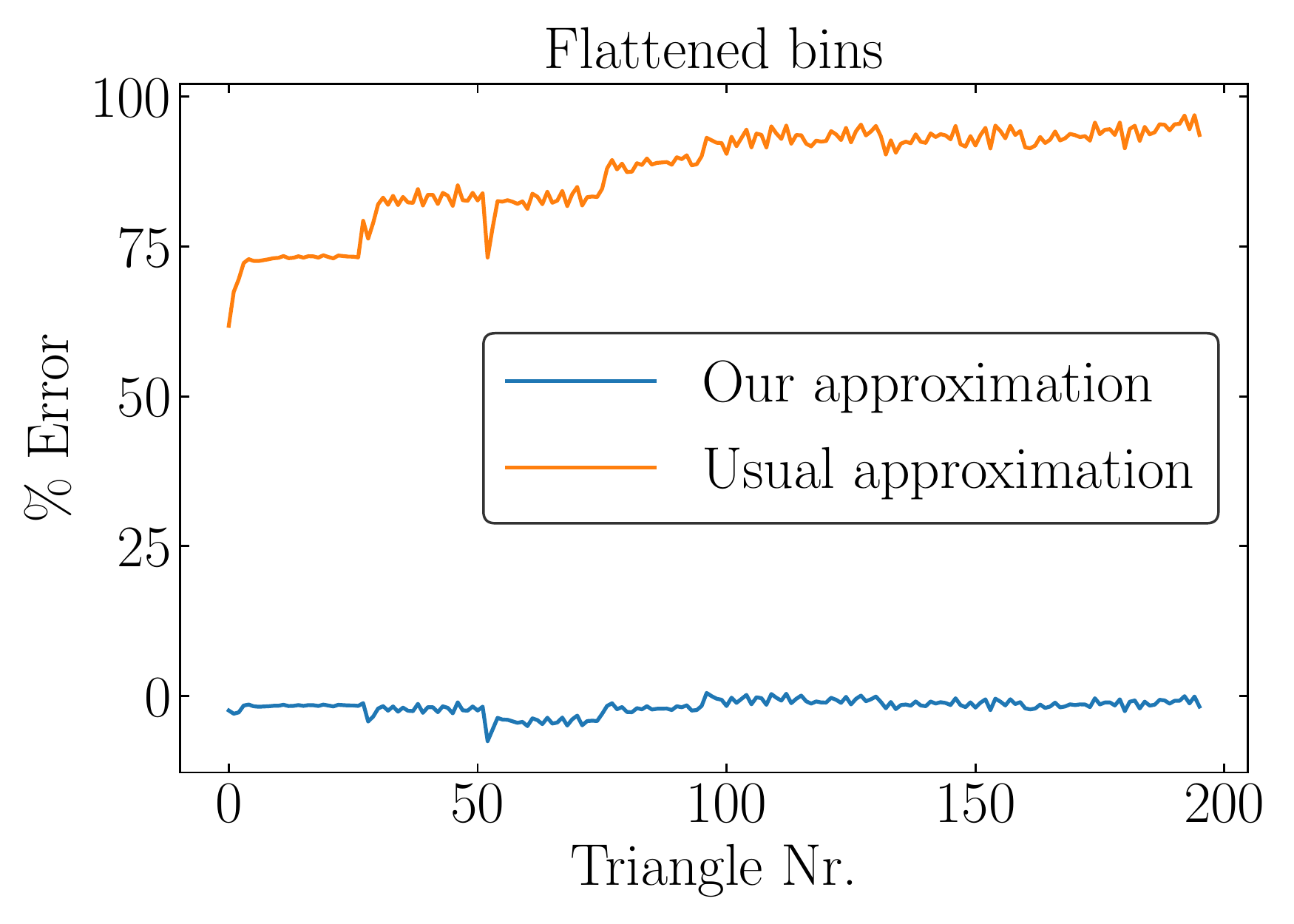}
    \end{subfigure}
    \caption{Comparison between analytical approximations and the exact sum for the number of triangles contributing to the bispectrum estimator in a bin. We present the percentage error incurred by using said analytical approximations. Here, our approximation is given by Eqs.~\eqref{Ntr_open} and~\eqref{Ntr_flattened} for ``open'' and ``flattened'' bins, respectively. This is usually approximated by Eq.~\eqref{Ntr_closed} instead. All triangle bins with sides between $3k_f$ and $84k_f$ were considered. {\em Left:} Fractional error for ``open'' bins. {\em Right:} Fractional error for ``flattened'' bins.}
    \label{fig:ntr}
\end{figure}

We plot a comparison between our approximations for $N_{tr}$ and the exact sum in Figure~\ref{fig:ntr}. We see that using equation~\eqref{Ntr_closed} for open or flattened configurations gives an error $\gtrsim \mathcal{O}(100\%)$. Our approximations for flattened bins in equations~\eqref{Ntr_flattened}, and the usual approximation of equation~\eqref{Ntr_closed} for closed bins, give an error $\lesssim 5\%$. Our approximation for open bins, equation~\eqref{Ntr_open} gives an error $\lesssim 15\%$.

Similar considerations hold for the sums appearing in the expression for the bispectrum covariance. We explicitly computed the sums only for triangles that share the longest wavelength mode $k_1$. We see that they are the most relevant in our case. From eq.~(\ref{sigma_definition}) we have
\begin{align}
N_{tr}^{i} N_{tr}^{j} \Sigma_{ij}^{11} &\simeq \frac{16\pi^3}{k_f^9} \delta^K_{k_1^a, k_1^b} \int_{k_1 - \Delta k/2}^{k_1 + \Delta k/2}\!\! dq_1 \int_{k_2^a - \Delta k/2}^{k_2^a + \Delta k/2} \!\! dq_2^a \int_{k_3^a - \Delta k/2}^{k_3^a + \Delta k/2} \!\! dq_3^a
\nn \\ & \times \int_{k_2^b - \Delta k/2}^{k_2^b + \Delta k/2} \!\! dq_2^b \int_{k_3^b - \Delta k/2}^{k_3^b + \Delta k/2} \!\! dq_3^b\,q_2^a\, q_3^a\, q_2^b\, q_3^b 
\nn \\ & \times \int_{-1}^1 d\mu_a\ \delta_D\bigg[\mu_a - \frac{(q_3^a)^2 - q_1^2 - (q_2^a)^2}{2q_1 q_2^a}\bigg] 
\nn \\ & \times
 \int_{-1}^1 d\mu_b\ \delta_D\bigg[\mu_b - \frac{(q_3^b)^2 - q_1^2 - (q_2^b)^2}{2q_1q_2^b}\bigg]\,.
\end{align}
The region of integration is again given by Figure~\ref{fig:cubes}, with $q_1$ in the intersection of the two allowed regions. Integrating over the full range is a good approximation for typical bins, which simply gives
\be
N_{tr}^{i} N_{tr}^{j} \Sigma^{11}_{ij} = \frac{16\pi^3}{k_f^9} k_2^a k_3^a k_2^b k_3^b \Delta k^5\,.
\label{sigma_closed}
\ee
As in the previous case,  using equation \eqref{sigma_closed} to approximate sums involving flattened, or open bins gives an error $\gtrsim \mathcal{O}(1)$. The results of the integration over the appropriate regions for flattened and open triangular bins are given in Appendix~\ref{appendix_integrals}. 

\section{The covariance of squeezed bispectrum configurations}
\label{sect:theory}

As already mentioned, we focus our attention on squeezed triangular configurations. For such triangles, in fact, we expect the non-Gaussian contribution to their covariance to be  dominant. 
For these configurations, it is possible to obtain an approximate, but satisfactory, model of their covariance matrix from direct measurements of the power spectrum and bispectrum of a given distribution, without the need to fit for any free parameter. This exercise illustrates relevant aspects of the modeling of the bispectrum covariance, which is helpful for its full analytical description. 

In this section, we quantify the relevance of the various non-Gaussian contributions to the covariance of the bispectrum in the squeezed limit. We then  simplify some of them using response functions \cite{Chiang:2015pwa, Barreira:2017fjz}.

\subsection{Order of magnitude estimates}

In this section we provide an order of magnitude estimate of the  relevance of the non-Gaussian contributions to the covariance matrix w.r.t. the Gaussian one. For simplicity, we assume each higher-order correlator to be described by its tree-level expression in perturbation theory, even at small scales where this approximation breaks down. We validate these results with simulation measurements in the next section. 

\paragraph{Power Spectrum.}

In this case, we have only one non-Gaussian contribution depending on the trispectrum $\bar{T}(k_i,k_j)$. We compare it to the Gaussian diagonal contribution for $k_i=k_j=k$, so that $\bar{T}(k,k)\sim P^3(k)$. We have then
\be
\frac{C^{P, (T)}}{C^{P,(PP)}} \sim \frac{k_f^3 N_k}{2}\frac{\bar{T}(k,k)}{P^2(k)} \sim 2\pi\,\left(\frac{\Delta k}{k}\right)\, k^3\,P(k) \sim \left(\frac{\Delta k}{k}\right) \Delta^2(k)\,,
\ee
where $\Delta^2(k) \equiv 4\pi k^3 P(k)$ is the dimensionless power spectrum. Note that this estimate holds also for halos taking shot noise into account. Indeed, in the shot noise dominated regime, for Poisson shot noise, $\bar{T}(k,k) \sim 1/\bar{n}^3$ and $P(k) \sim 1/\bar{n}$, so that $\bar{T}(k,k) \sim P^3(k)$.

\begin{figure}[t]
\centering
\includegraphics[width=0.65\textwidth]{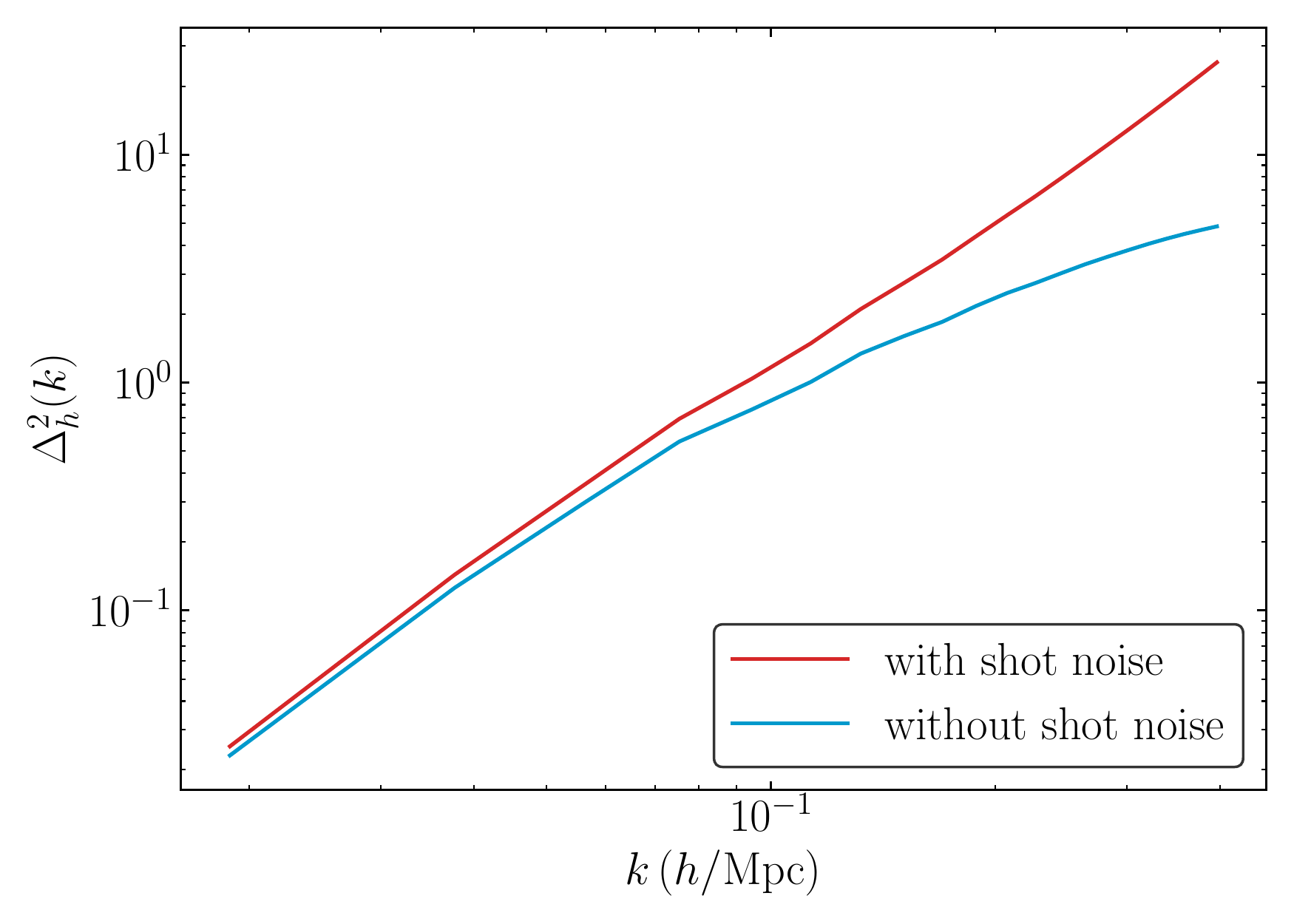}
\caption{Dimensionless halo power spectrum for the halos taken from the \quijote simulations, as described in Section\ \ref{sec:sim}.} \label{fig:DPh_Bh}
\end{figure}

For the halos we consider (see Section\ \ref{sec:sim}), this is of order $\Delta^2(k) \sim \mathcal{O}(10^{-2})$ at large scales, close to the size of the bin $k \sim \Delta k$. At small scales, e.g. $k/ \Delta k\sim 100$, we have instead $\Delta_h^2(k) \sim \mathcal{O}(10)$ (see Figure\ \ref{fig:DPh_Bh}). Therefore, this contribution is negligible at large scales but only mildly suppressed at intermediate and small scales \cite{Scoccimarro:1999kp}. On scales that are shot-noise dominated, this ratio grows like $\propto k^2$ such that the non-Gaussian term can become dominant at small enough scales and for halos with low enough number density.

\vspace{2em}

\paragraph{Bispectrum.}

In this case we have several non-Gaussian contributions, Eqs.~\eqref{appeqCBB-BB}, \eqref{appeqCBB-PT}, \eqref{appeqCBB-E}. Let us estimate the size of each with respect to the Gaussian contribution. When all scales are comparable, we set $B(k,k,k) \sim P^2(k)$. This estimate holds also when including shot noise. Indeed, in the shot noise dominated regime for Poisson shot noise $B(k,k,k) \sim 1/\bar{n}^2$ and $P(k) \sim 1/\bar{n}$. Using this in Eq.~\eqref{appeqCBB-BB} we obtain
\be
\frac{C^{B, (BB)}}{C^{B, (PPP)}} \sim k_f^3 N_{tr} \Sigma \frac{B^2(k,k,k)}{P^3(k)}\sim \left(\frac{\Delta k}{k}\right)^2 \Delta_h^2(k)\,,
\ee
where $\Sigma^{ab}_{ij} \simeq \delta_{k^i_a k^j_b}/N_{k^i_a}$ is the typical size of one of the mode-counting factors in Eq.~\eqref{appeqCBB-BB}. For the simulations and scales we consider, the contribution from $C^{B,(BB)}$ is expected to be subdominant when all scales are comparable. However, this term grows like $\propto k$ in the shot noise dominated regime, such that it can become dominant at small enough scales.

On the other hand, the bispectrum couples different scales. In the extreme case of squeezed configurations, we write $B(k_L, k_s, k_s) \sim P(k_L)P(k_s)$, where we are assuming that $P(k_L) > P(k_s)$. Again, this estimate holds even in the presence of shot noise. Indeed, when the short scale is shot-noise dominated the dominant contribution is $B(k_L, k_s, k_s) \sim P(k_L)/\bar{n}$ (see e.g.~\cite{Chan:2016ehg}). Furthermore, the mode-counting factors depend on whether the triangles correlated share the long mode or the short mode. In this way we obtain
\begin{equation}
\frac{C^{B, (BB)}}{C^{B, (PPP)}} \sim k_f^3\frac{B^2(k_L, k_s, k_s)}{P(k_L)P^2(k_s)}\left(N_{tr}\Sigma^{LL} + 4 N_{tr}\Sigma^{ss}\right)
\sim \left(\frac{\Delta k}{k_L}\right)^2 \Delta_h^2(k_L) \left(\frac{k_s^2}{k_L^2} + 4\right)\,,
\end{equation}
The first term in the parenthesis is the one proportional to $\Sigma^{11}_{ij}$ in equation~\eqref{appeqCBB}, which corresponds to a pair of triangles for which the long mode coincides. The second term in the parenthesis corresponds to all other permutations in $C^{B, (BB)}$.
Since $\Delta_h^2(k_L) \sim \mathcal{O}(10^{-2})$, this contribution to the covariance is generically suppressed. However, there is a quadratic enhancement by the ratio $k_s/k_L$ for the first term, which can easily compensate for this suppression if $k_s/k_L \gtrsim 10$. We thus see that the contribution to $C^{B,(BB)}$ for squeezed configurations that share the long mode can become dominant in the covariance. This generates off-diagonal elements in the covariance of the same order of magnitude as the elements in the diagonal, inducing a large correlation among those triangles.

The trispectrum term is analogous. When all scales are comparable $\tilde{T}(k,k,k,k,k) \sim P^3(k)$ (which again holds in the presence of shot noise), and we get from Eq.~\eqref{appeqCBB-PT}
\be
\frac{C^{B, (PT)}}{C^{B, (PPP)}} \sim k_f^3 N_{tr} \Sigma \frac{T(k,k,k,k) P(k)}{P^3(k)}\sim \left(\frac{\Delta k}{k}\right)^2 \Delta_h^2(k)\,.
\ee
For squeezed triangles we instead estimate $\tilde{T}(k_s,k_s,k_s,k_s,k_L) \sim P(k_L)P^2(k_s)$. From the explicit expression for the trispectrum in the presence of shot noise (see e.g.~\cite{Chan:2016ehg}), one can check that this estimate still holds. For example, when the short scales are shot-noise dominated $\tilde{T} \sim P(k_L)/\bar{n}^2$. We thus obtain,
\begin{multline}
\frac{C^{B, (PT)}}{C^{B, (PPP)}} \sim k_f^3\left(N_{tr}\frac{P(k_L)\tilde{T}(k_s, k_s, k_s, k_s, k_L)}{P(k_L)P^2(k_s)}\Sigma^{LL} + 4 N_{tr}\frac{P(k_s)\tilde{T}(k_s, k_L, k_s, k_L, k_s)}{P(k_L)P^2(k_s)}\Sigma^{ss}\right) \\
\sim \left(\frac{\Delta k}{k_L}\right)^2 \Delta_h^2(k_L) \left(\frac{k_s^2}{k_L^2} + 4\right)\,.
\end{multline}
Again, the contribution to $C^{B, (PT)}$ corresponding to triangles that share the long mode can be dominant in the covariance.\footnote{It is worth noticing that in previous literature these non-Gaussian terms were also calculated, either in the context of the position-dependent matter and halo power spectrum \cite{dePutter:2018jqk} or using response functions for the matter bispectrum \cite{Barreira:2019icq}. Both these analyses identified that these non-Gaussian terms are indeed large, although they did not compare with simulations, and do not discuss non-squeezed triangles.} 

For the pentaspectrum contribution, Eq.~\eqref{appeqCBB-E}, when all scales are comparable, we similarly estimate $\tilde{P}_6 \sim P^5(k)$ and obtain
\be
\frac{C^{B, (P_6)}}{C^{B, (PPP)}} \sim \frac{1}{16 \pi^2} \left(\frac{k_f}{k}\right)^6 (\Delta_h^2(k))^2\,,
\ee
For squeezed configurations we write $\tilde{P}_6 \sim P^2(k_L)P^3(k_s)$, such that
\begin{equation}
\frac{C^{B, (P_6)}}{C^{B, (PPP)}} 
\sim k_f^6 P(k_s) P(k_L) \sim \frac{1}{16 \pi^2} \frac{k_f^3}{k_s^3}\frac{k_f^3}{k_L^3} \Delta_h^2(k_s) \Delta_h^2(k_L)\,.
\end{equation}
Again, from the explicit expression of $P_6$ (e.g. from~\cite{Chan:2016ehg}), these estimates should hold even in the presence of shot-noise. In both cases, even in the shot-noise dominated regime, this contribution is suppressed.

\paragraph{Bispectrum power spectrum cross covariance.}

In order to evaluate if the cross covariance is important, let us estimate the size of the correlation coefficients. As before, we start by considering comparable scales,  using Eq.~\eqref{eq:PB_cov}
\begin{equation}
\frac{C^{PB, (PB)}}{\sqrt{C^{P, (PP)} C^{B, (PPP)}}} \sim \frac{\sqrt{2}N_{tr}^{1/2} k_f^{3/2}}{N_{k}^{1/2}}\frac{B(k, k, k)}{P^{3/2}(k)} \sim \frac{\Delta k}{k}(\Delta^2(k))^{1/2}\,.
\end{equation}
For the scales and halos we consider, this contribution is subdominant. However, in the shot noise dominated regime coefficient grows slowly as $\propto k^{1/2}$.

Let us discuss the correlation between a squeezed bispectrum $B(k_L, k_s, k_s)$ and a power spectrum evaluated at an arbitrary scale $P(k)$. We get different results depending on whether $k$ is equal to $k_s$ or $k_L$. We estimate
\begin{equation}
\frac{C^{PB, (PB)}}{\sqrt{C^{P, (PP)} C^{B, (PPP)}}} \sim \frac{\sqrt{2}N_{tr}^{1/2} k_f^{3/2}}{N_{k}^{1/2}}\frac{B(k_L, k_s, k_s)}{P^{1/2}(k_L) P(k_s)} \sim \frac{k_s}{k}\frac{\Delta k}{k_L}(\Delta^2(k_L))^{1/2}\,.
\end{equation}
When $k=k_s$ this correlation coefficient is always mildly suppressed by the dimensionless power spectrum. On the other hand, when $k = k_L$ it is enhanced by the squeezing and can easily become $\mathcal{O}(1)$. Thus, there is a large correlation between squeezed configurations and the power spectrum evaluated at the long mode.

We have assumed that the cross-covariance is dominated by $C^{PB, (PB)}$. In order to check this assumption, we compare the two contributions in Eq.~\eqref{eq:PB_cov} when all scales are comparable, estimating $\tilde{P}_5 \sim P^4(k)$. The explicit expressions for $P_5$ in the presence of shot noise can be found in~\cite{Chan:2016ehg}. We write
\begin{equation}
\frac{C^{PB, (P_5)}}{C^{PB, (PB)}} \sim \frac{ k_f^3N_k}{2}\frac{R(k, k, k, k, k)}{P(k)B(k, k, k)} 
\sim \frac{\Delta k}{k} \Delta_h^2(k)\,,
\end{equation}
This is similar to the ratio between the non-Gaussian and the Gaussian contributions to the power spectrum covariance. Thus, the tetraspectrum contribution to the cross-covariance is important at the same scales for which the trispectrum contribution is important for the power spectrum covariance. For the scales and halos we study, this contribution is subdominant. However, in the shot-noise dominated regime, this ratio scales as $\propto k^2$, with the corresponding correlation coefficient growing strongly as $\propto k^{5/2}$. 

For squeezed configurations, we again take the power spectrum to be evaluated at an arbitrary scale $k$ that can either close to $k_L$ or $k_s$. We then estimate $\tilde{P}_5(k_L, k_s, k_s, k, k) \sim P(k) P(k_L) P^2(k_s)$, which is valid in both cases. We thus get
\begin{equation}
\frac{C^{PB, (P_5)}}{C^{PB, (PB)}} \sim \frac{ k_f^3N_k}{2}\frac{\tilde{P}_5(k_L, k_s, k_s, k, k)}{P(k)B(k_L, k_s, k_s)} 
\sim \frac{\Delta k}{k} \Delta_h^2(k)\,,
\end{equation}
When $k = k_s$, this ratio is the same as before. The corresponding correlation coefficient is subdominant for the halos and scales we consider, but scales like $\propto k_s^2$ in the shot-noise dominated regime. On the other hand, when $k = k_L$, this ratio is always suppressed. Thus, the tetraspectrum can always be ignored for those configurations for which the correlation coefficient is the largest.

\subsection{The covariance in terms of response functions}\label{sec:response}

We saw in the previous section that the bispectrum covariance is dominated by the Gaussian term $C^{B, (PPP)}$ plus two contributions involving higher-order correlation functions $C^{B,(BB)}$ and $C^{B, (PT)}$. The first can be computed at non-linear scales by using the measured bispectrum or a fitting function \cite{Scoccimarro:2000ee,Gil-Marin:2011jtv,Takahashi:2019hth}. The second contribution is more challenging: trispectrum measurements are difficult (see~\cite{Sefusatti:2004xz,Bertolini:2016bmt,Gualdi:2020eag,Steele:2021lnz,Gualdi:2021yvq}), and we are not aware of fitting formulae for the trispectrum at small scales. However, in the squeezed limit they can both be written in terms of response functions. Specifically, the coupling between short wavelength modes and a long wavelength perturbation of the gravitational potential can be written as
\begin{equation}
\lim_{k_L \rightarrow 0} B(k_L, k_1, k_2) = P(k_L) \frac{\partial}{\partial \delta_L} \langle \delta({\bf k}_1)\delta({\bf k}_2)\rangle'\,. 
\end{equation}
Here, a prime denotes that the Dirac delta of momentum conservation is dropped, and $\partial/\partial \delta_L$ is the response to a change of the long-wavelength curvature.  Such a change also correlates two-point functions at different points. This gives a contribution to the trispectrum in the limit in which the sum of two momenta goes to zero (which we call an internal squeezed limit)\footnote{This response function diverges at unequal times, and this divergence is fixed by symmetry \cite{Creminelli:2013mca}. We discuss here the subleading terms in the response, which do not vanish at equal times.}
\begin{multline}
\lim_{|{\bf k}_1 + {\bf k}_2| \rightarrow 0} T(k_1, k_2, k_3, k_4, |{\bf k}_1 + {\bf k}_2|, |{\bf k}_4 + {\bf k}_4|) \\ \approx P(|{\bf k}_1 + {\bf k}_2|) \left(\frac{\partial}{\partial \delta_L}\langle \delta({\bf k}_1)\delta({\bf k}_2)\rangle'\right)\left(\frac{\partial}{\partial \delta_L}\langle \delta({\bf k}_3)\delta({\bf k}_4)\rangle'\right)\,.
\end{multline}
This contribution is dominant when the power spectrum is the largest when evaluated at $k_L = |{\bf k}_1 + {\bf k}_2|$ compared to other combinations of the momenta.\footnote{It can be checked in perturbation theory that this contribution can be rendered subdominant in the squeezed limit, with respect to other terms that were discarded, when $P(k_s) \geq P(k_L)$.} In our case, we take the long mode to be $k_L \sim 0.01\ \text{Mpc}^{-1}$ and the short modes to be 
$k_s \gtrsim 3k_L$.
It can be checked that $P(k_L) \gg P(k_s)$, such that we can use this approximation for the trispectrum.

The response functions of  short scale power spectra should, in principle, be measured in simulations. However, we can use these expressions to simplify the angle-averaged trispectrum appearing in the bispectrum covariance
\begin{align}
\tilde{T}(k_1, k_2, k_3, k_4, q) &\approx P(k_L) \left(\frac{\partial}{\partial \delta_L}\langle \delta({\bf k}_1)\delta({\bf k}_2)\rangle'\right)\left(\frac{\partial}{\partial \delta_L}\langle \delta({\bf k}_3)\delta({\bf k}_4)\rangle'\right) \frac{1}{\Delta p} \int\ dp \nonumber \\
&= \frac{1}{P(k_L)} B(k_L, k_1, k_2) B(k_L, k_3, k_4)\,,
\end{align}
where we have used the fact that the response functions do not depend on the momentum combination being integrated. Using this, $C^{B, (PT)}$ can be written as \footnote{Note that the analysis in this section holds also in the presence of shot noise.}
\begin{align}
C^{B, (PT)}_{ij} &\approx P(k_1^i)\tilde{T}(k_2^i, k_3^i, k_2^j, k_3^j, k_1^i)\Sigma^{11}_{ij} \nonumber \\
&\approx B(k_1^i, k_2^i, k_3^i) B(k_1^j, k_2^j, k_3^j)\Sigma^{11}_{ij} \nonumber \\
&\approx C^{B,(BB)}_{ij}\,.
\end{align}

\subsection{Prescription for the covariance and its inverse}\label{sec:inversion}

We conclude that the following expression for the bispectrum covariance is a good approximation for squeezed triangles 
\begin{equation}
C_{ij}^{B} \simeq \frac{\delta_{ij} \,s_B}{k_f^3 N_{tr}^i}   P(k_1^i)P(k_2^i)P(k_3^i) + \frac{2 B(k_1^i,k_2^i, k_3^i)B(k_1^j, k_2^j, k_3^j)}{N_{tr}^i N_{tr}^j}   \sum_{{\bf q}\, \in\, \{k\}} \delta_{{\bf q}_1^i + {\bf q}_2^i + {\bf q}_3^i} \delta_{{\bf q}_1^j + {\bf q}_2^j + {\bf q}_3^j} \delta_{{\bf q}_1^i + {\bf q}_1^j}\,,\label{eq:Total_BB}
\end{equation}
\noindent where as before $s_B = 6$, $2$, $1$ for equilateral, isosceles and scalene triangles, respectively, $N^i_{tr}$ is the number of fundamental triangles in the triangle bin $\{k^i_1, k^i_2, k^i_3\}$, eq. \eqref{eq:Nt}, and the sum can be approximated as described in section \ref{section_sums_approx}. This expression is expected to work reasonably well even for non-squeezed triangles, which are dominated by the Gaussian covariance. In the next section we show that the non-Gaussian terms in this expression are large even for mildly squeezed triangles for which $k_L \lesssim 3k_s$.

Furthermore, the correlation coefficient between a squeezed triangle and the corresponding long-mode power spectrum is order~$\mathcal{O}(1)$ and dominated by the $C^{PB, (PB)}$ term.

We derive equation~\eqref{eq:Total_BB} under the following assumptions:
\begin{itemize}
    \item The 6-point function can be ignored. This is reasonable when the long mode is linear. The 6-point function is expected to be important only when all scales involved are non-linear.
    \item Correlations among triangles sharing the long mode are found to be the largest ones. Other correlations, such as when all scales are short, become important only when all those short scales are deep in the non-linear regime
    \item The trispectrum appearing in these correlations can be approximated by the squeezed bispectrum. This is expected to hold as long as the triangles are squeezed and $P(k_L) > P(k_s)$, as shown in Section 3.2.
\end{itemize}

For any procedure involving constraints on a parameter, like Fisher matrix and likelihood analyses, we need to invert the covariance matrix. If, as in our case, the covariance matrix has sizable elements far outside the diagonal, the inversion operation can be tricky from a numerical point of view. The off-diagonal terms we introduce have a particular structure that allows for analytic inversion using a generalization of the Sherman-Morrison formula (see~\cite{10.1214/aoms/1177729893,10.1214/aoms/1177729698}, also used recently in a scenario similar to ours~\cite{dePutter:2018jqk}). 

Both from our expectations and from our measurements, we see that the dominant term in the cross-covariance is when the long mode of a squeezed triangle is the same as the momentum of the power spectrum. In that case, the cross-covariance can be written as a sum of outer products of vectors
\be
\left(\begin{array}{cc}\bm{C}^{P} & \sum_q {\bm a}_q {\bm b}_q^T \\ \sum_q {\bm b}_q {\bm a}_q^T & \bm{C}^{B}\end{array}\right)\,,
\ee
with the sum running over the values of the momentum $q$. The vectors are
\be
a_{q}^k = \frac{2}{N_{k}}\delta_{q, k} P(k)\,,\quad a_q^{(k_1, k_2, k_3)} = 0\,,
\ee
\be
b_q^k = 0\,,\quad b_q^{(k_1, k_2, k_3)} = \delta_{k_1, q} B(k_1, k_2, k_3)\,.
\ee
We labeled the elements of the data vector corresponding to the power spectrum with a single index representing its momentum, and those corresponding to the bispectrum with the three momenta at which it is evaluated.

Our generalization of the Sherman-Morrison formula then gives the following inverse
\be
\left(\begin{array}{cc}(\bm{C}^{P})^{-1} + \sum_q \, x\, \beta_q\,(\bm{C}^{P})^{-1} {\bm a}_q {\bm a}_q^T (\bm{C}^{P})^{-1} & -\sum_q \, x\, (\bm{C}^{P})^{-1} {\bm a}_q {\bm b}_q^T (\bm{C}^{B})^{-1} \\ -\sum_q \, x\, (\bm{C}^{B})^{-1} {\bm b}_q {\bm a}_q^T (\bm{C}^{P})^{-1} & (\bm{C}^{B})^{-1} + \sum_q\, x\,\alpha_q\,(\bm{C}^{B})^{-1} {\bm b}_q {\bm b}_q^T (\bm{C}^{B})^{-1}\end{array}\right)\,,
\ee
where
$$
\alpha_q = \bm{a}_q^T (\bm{C}^{P})^{-1} {\bm a}_q\,,\quad \beta_q = \bm{b}_q^T (\bm{C}^{B})^{-1} {\bm b}_q\,,\quad x=1/(1-\alpha_q\,\beta_q),
$$
and we used the diagonal approximation for ${\bf C}^{P}$ and the fact that ${\bf C}^{B}$ in Eq.~\eqref{eq:Total_BB} is block-diagonal. We numerically invert the bispectrum covariance ${\bm C}^{B}$, and use this formula for the inverse of the total covariance.

\section{Comparison with N-body simulations}\label{sec:sim_meas}
In this section we provide a detailed comparison between theoretical models of the power spectrum and bispectrum covariance discussed in the previous section and measurements from numerical simulations. We analyze catalogs of dark matter halos in a cubic box at fixed redshift $z=0$. Our setup allows for a comparison where the shot-noise contribution is relevant, even if its level is not realistic for typical galaxy redshift surveys. Working in this simplified, but controlled, scenario allows us to verify the accuracy of the model in Sect.~\ref{sec:inversion} with little in the way of observational systematic errors.

\subsection{N-body simulations}\label{sec:sim}

We use the publicly available suite of simulations \quijote\footnote{Information on \quijote simulations can be found at \href{https://quijote-simulations.readthedocs.io/en/latest/}{https://quijote-simulations.readthedocs.io/en/latest/} and on the reference paper~\cite{Villaescusa-Navarro:2019bje}.}, run using the \textsc{Gadget}-3 code~\cite{Springel:2005mi}. We consider a subset of $2377$ realizations from the fiducial cosmology (see Table \ref{tab:sims}), which are run on cubic boxes of $1000\, $Mpc $/h$ side length with $512^3$ particles in them. The initial conditions are set at $z_i=127$ using the code \textsc{2LPTic}~\cite{Crocce:2006ve}. The linear transfer function is obtained using the Boltzmann code \textsc{CAMB}~\cite{Lewis:1999bs}.  The halo catalogs are generated using a Friends-of-Friends (FoF) algorithm with a linking  length $\lambda= 0.2 $. We require that halos are constituted by a minimum of $50$ particles, implying a number density in each box of $\bar n \sim 5\cdot 10^{-6}\, (h/{\rm Gpc})^3$. 
In Table \ref{tab:sims} we summarize the specifications and the fiducial cosmology for \quijote.
\begin{table}
\centering
\resizebox{\columnwidth}{!}{%
\begin{tabular}{ccccccccccc}\hline
    Name & $n_s$ & $h$ & $ \Omega_b$ & $\Omega_m$ & $\sigma_8$ & \# &$N_\mathrm{p}^{1/3}$&$L_{\rm box}$ & $V_\mathrm{tot}$ &$m_\mathrm{p}$ \\
    & & & & & & sims & &$[\Mpc]$ & $[\cGpc]$ &$[10^{10} \,h^{-1}M_\odot]$  \\
    \hline\hline
    \quijote & $0.9624$&$0.6711$&$0.049$&$0.3175$&$0.834$&$2377$ & $512$ & $1000$ & $2377$ & $65.6$ \\
    \hline
\end{tabular}%
}
\caption{Cosmological and structural parameters for the \quijote simulations.}
\label{tab:sims}
\end{table}

\subsection{Measurements and binning strategy}\label{sec:grids}

We measure the matter and halo power spectra and bispectra using the estimators Eqs. \eqref{eq:estP} and \eqref{eq:estB}. We implement a fourth-order density interpolation and the interlacing scheme described in~\cite{Sefusatti:2015aex}. 
Bins have width of $\Delta k = 3 k_f$, with $k_f \simeq 0.006 \, h/$Mpc. 
We provide bispectrum measurements based on estimates of the halo number density on two different grids  of different sizes, as described  and summarized in Table \ref{tab:binning}:
\begin{itemize}
    \item From a small grid with a linear size of $256$ , we measure the power spectrum and {\em all} bispectrum configurations up to $k_{\rm max} = 0.528\, h/$Mpc. 
    We call these measurements \qone.
    \item From a larger grid of linear size $450$ we only measure the bispectrum on squeezed triangles, since measuring all the triangles would have required  an exceedingly  large amount of memory. The selected triangles are composed of long modes in the range $0.0018\, h/$Mpc to $0.075\, h/$Mpc, while the short modes are between $0.641\, h/$Mpc and $0.942\, h/$Mpc, \footnote{Note that, for such short modes, the shot noise dominates in our setup, although that might not be the case for realistic galaxy surveys.}  for a total of $4$ bins for long modes and $16$ bins for short modes. We refer to these measurements as \qtwo.
\end{itemize} 

Note that we do not subtract shot noise from any of these measurements. This allows us to make use of the theoretical expressions for the covariance in section~\ref{sect:theory} assuming that all correlators in such expressions include the usual shot-noise contributions. As pointed out in \cite{Smith:2008ut}, and extended for the bispectrum case in \cite{Sugiyama:2018yzo}, if shot noise is subtracted from the measurements then the non-Gaussian contributions are reduced to some extent.

\begin{table}
\centering
\resizebox{0.7\columnwidth}{!}{%
\begin{tabular}{cccccc}\hline\hline
    Name & $N_{\rm grid}$ & $k_{\rm min} (h/$Mpc) & $k_{\rm max} (h/$Mpc) & \# bins & \# triangles \\
    \hline
    \qone  & $256$ & $0.018$&$0.528$&$28$ & $2513$\\
    \hline
    \qtwo  & $450$  & $0.018$&$0.942$&$50$ & $254$\\
    \hline
\end{tabular}}
\caption{Binning strategy for \quijote simulations. All bins have width $\Delta k = 3 k_f$, where $k_f \simeq 0.006 \, h/$Mpc. For the \qtwo sets of measurements, only squeezed triangles are measured. The triangles we consider are composed of long modes in the range $0.009\, h/$Mpc to $0.075\, h/$Mpc, while the short modes are between $0.641\, h/$Mpc and $0.942\, h/$Mpc.}
\label{tab:binning}
\end{table}

\subsection{Results}

We estimate the covariance for halo power spectrum $P$, bispectrum $B$ and their cross-covariance directly from the simulations using Eq. \eqref{eq:defcov}. 
We notice that measurements of the covariance from a limited number of realizations suffer from correlated numerical noise. Since it is correlated, this noise can be easily confused with additional structure in the matrix. An illustration and a brief discussion of this is found in Appendix~\ref{app:noise}. Our goal is to describe the covariance to an accuracy of roughly $\sim 20\%$. The model of Sect.~\ref{sec:inversion} is able to achieve this goal, as we show from the comparison to N-body measurements we present in this section. This leads us to believe that the numerical error is also below this threshold.

The model we are plotting in this section contains the following terms:
\begin{itemize}
	\item For the power spectrum covariance we only consider the first term in Eq. \eqref{eq:cpp}, $C^{P,(PP)}$. We expect a small contribution from the term $C^{P,(T)}$ at large scales. At small scales, this term is not negligible, but the power spectrum is not the main focus of this work. For a recent study of the power spectrum covariance, we refer the reader to \cite{Wadekar:2019rdu} and references within. 
	\item For the bispectrum covariance we use the result in Eq. \eqref{eq:Total_BB}.
	We call the first term ``PPP'' and the second one ``2BB'', where the factor of 2 accounts for the ``PT'' term as explained in Section \ref{sec:response}.
	\item For the bispectrum-power spectrum cross covariance we take into account only the first term in Eq. \ref{eq:PB_cov} as we expect the  5-point function to have a small contribution. 
\end{itemize}
For all these predictions, we used the halo power spectrum and bispectrum measured from the simulations themselves. We have also tested results using perturbation theory predictions in place of the measured ones, finding good agreement at large scales. When entering the non-linear regime for the most squeezed triangle, the agreement degrades significantly.  An alternative could be using a nonlinear model or fitting function calibrated on simulations \cite{Scoccimarro:2000ee, Gil-Marin:2011jtv, Takahashi:2019hth}.

In order to compare the model with measurements, we start by plotting the variance. We then compare the off-diagonal terms of the covariance by plotting the correlation matrix. Finally, we check the inverse covariance by computing $\chi^2$ values using our approximation. 

\subsubsection{Bispectrum variance}

\begin{figure}[t]
\includegraphics[width=0.49\textwidth]{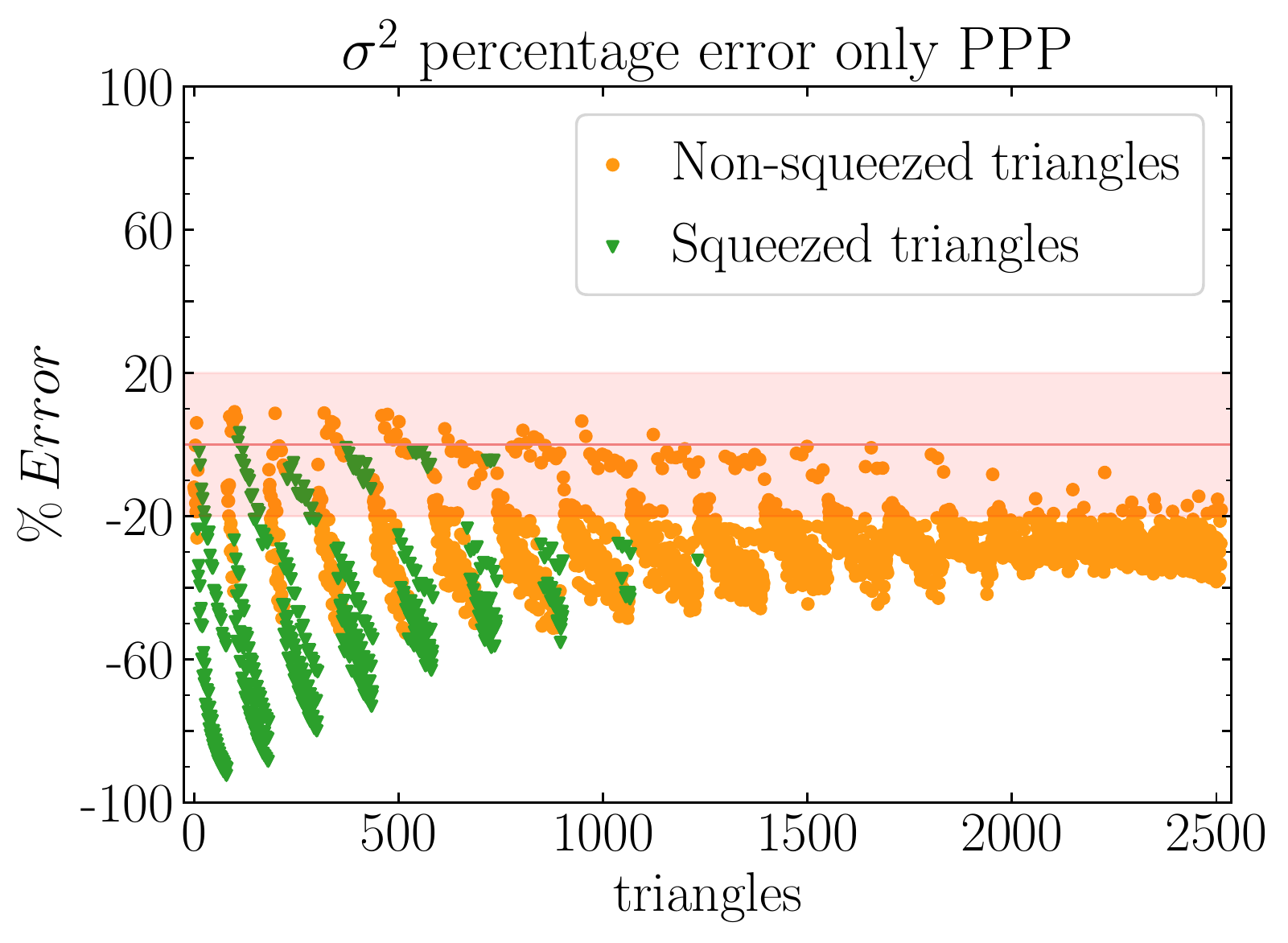}
\includegraphics[width=0.49\textwidth]{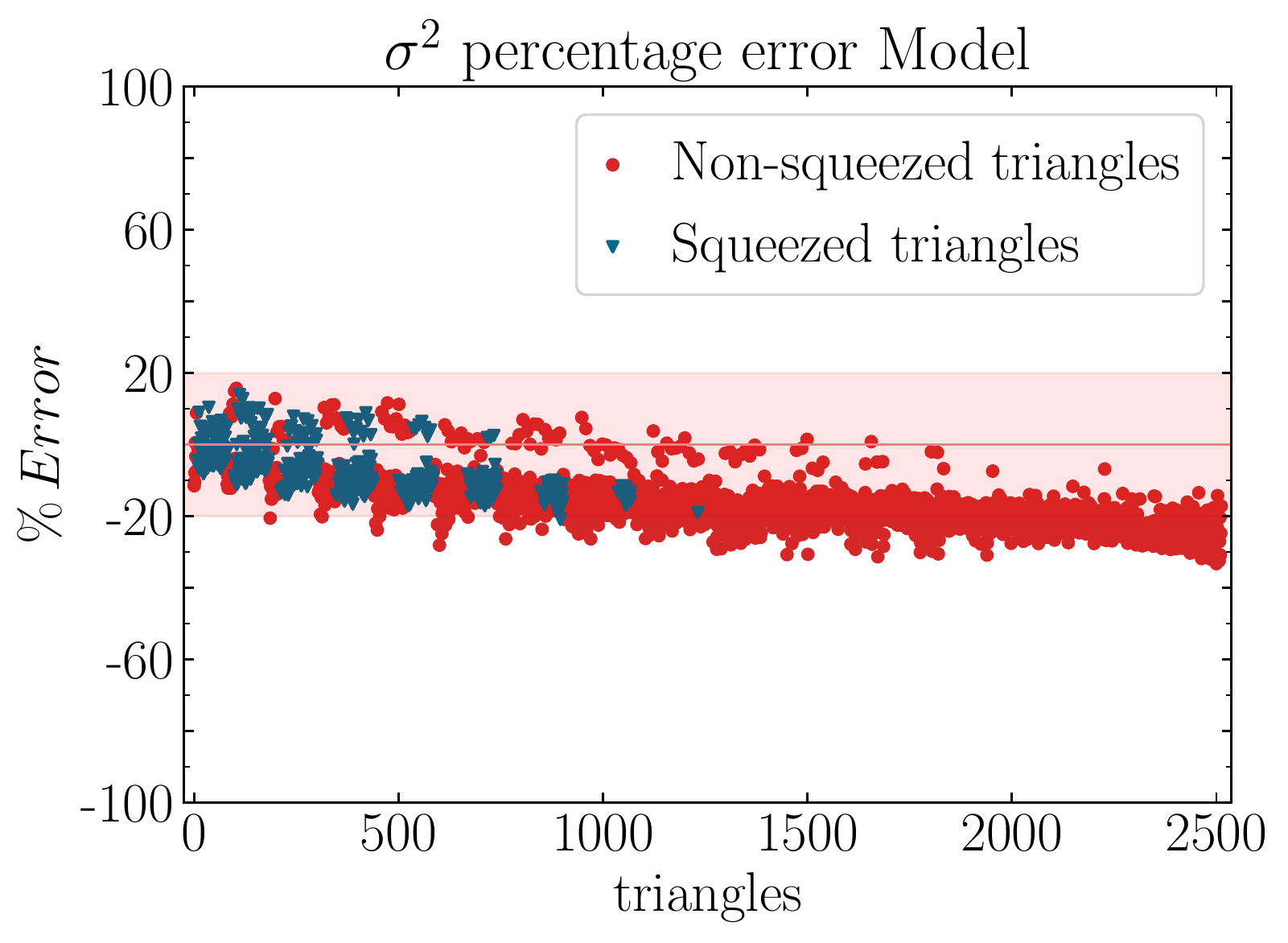}\caption{{\em Left}: Percentage error between simulation measurements and only ``PPP'' contribution to the theoretical bispectrum variance. {\em Right}: Percentage error between simulation measurements and ``Model'' bispectrum variance from \eqref{eq:Total_BB}. Both the power spectrum and bispectrum used to compute eq. \eqref{eq:Total_BB} are measured directly on the simulations. Squeezed triangles (such that the long mode is at least $3$ times the short mode) are marked with a different color and symbol.  These estimates clearly show that using a Gaussian model for the bispectrum variance, i.e. only including the ``PPP'' contribution, leads to errors up to $100\%$.} \label{fig:Full_Variance}
\end{figure}

\begin{figure}[t]
\includegraphics[width=0.49\textwidth]{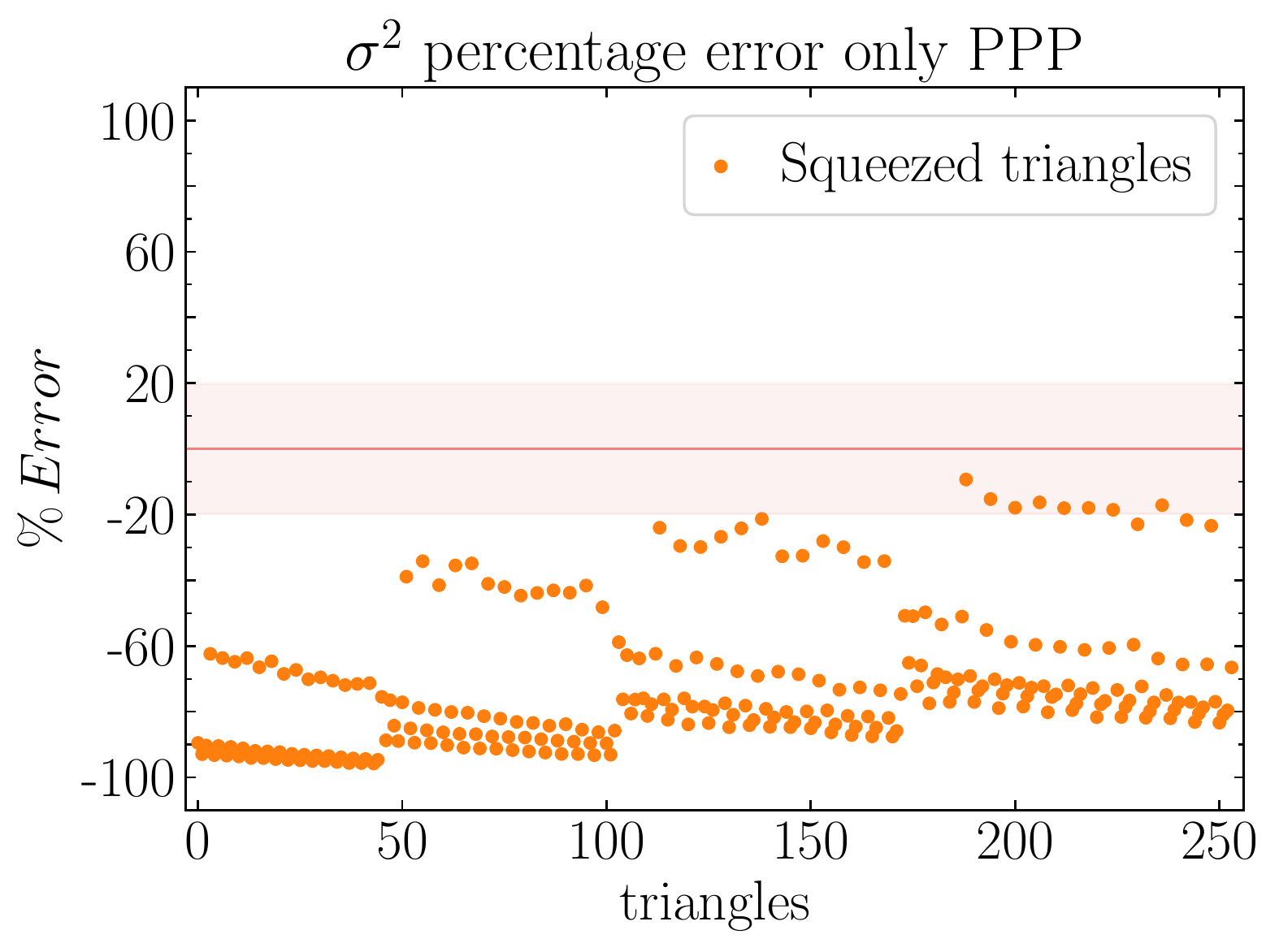}
\includegraphics[width=0.49\textwidth]{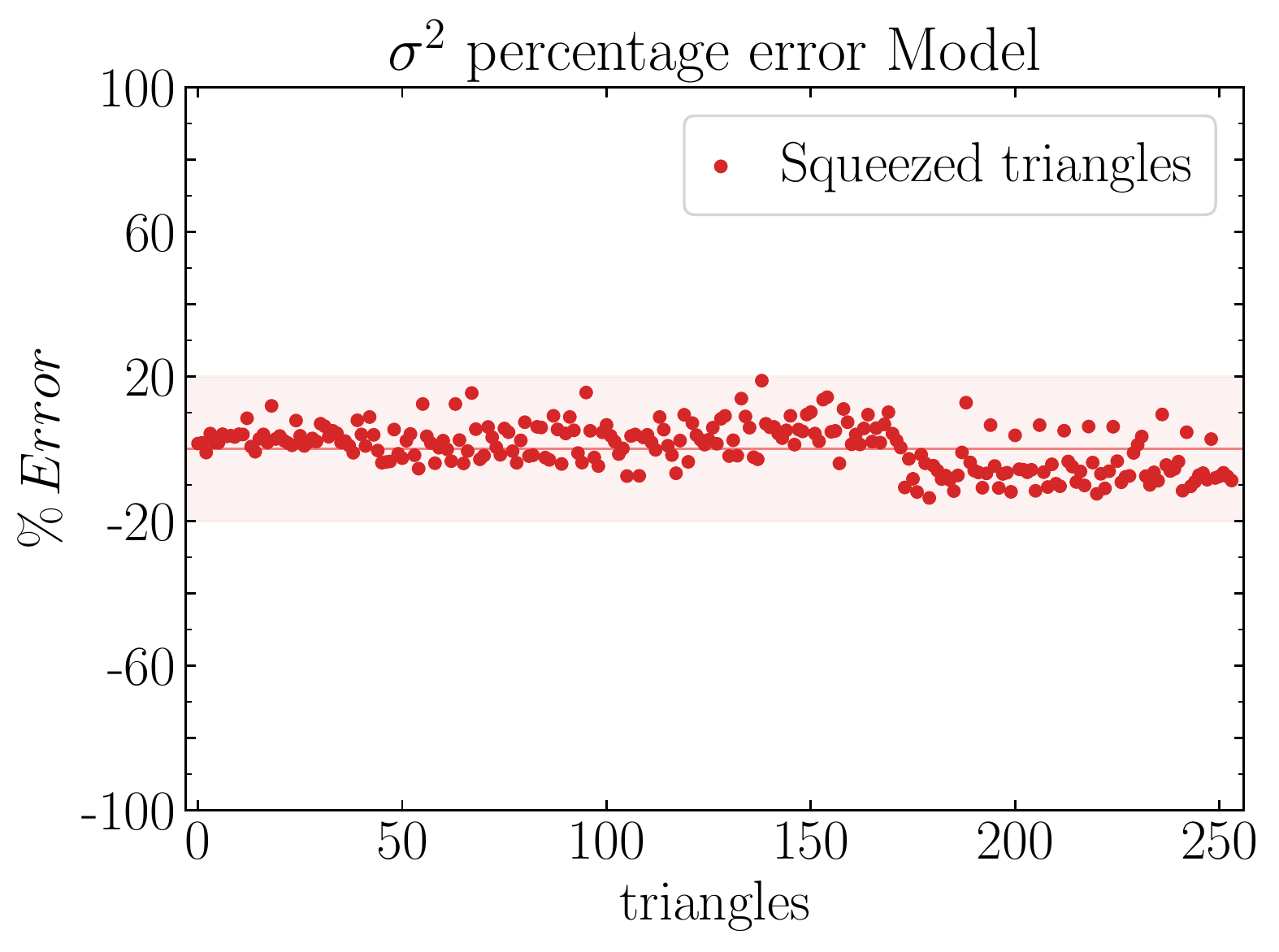}\caption{Percentage error for the bispectrum variance for \qtwo grid. We compare the variance obtained from N-body simulations with the theoretical variance considering only ``PPP'' term and the ``Model'' from \eqref{eq:Total_BB}. Note that all triangles for this grid are squeezed. {\em Left}: Percentage error between simulation measurements and ``PPP'' covariance. {\em Right}: Percentage error between simulation measurements and ``Model'' theoretical covariance. As for the previous plot, a Gaussian model for the bispectrum variance, i.e. only including the ``PPP'' contribution, clearly fails at predicting the variance for squeezed triangles.}\label{fig:Full_Variance_Squeezed}
\end{figure}

In Figures \ref{fig:Full_Variance} and \ref{fig:Full_Variance_Squeezed} we show the percentage error between the measurements and the theoretical variance of the bispectrum. We compare our results by considering only the Gaussian approximation ``PPP'' and the model of Section~\ref{sec:inversion}, for the \qone and \qtwo grids respectively.

In Figure~\ref{fig:Full_Variance}, we present the percentage error of the variance between measurements and the theoretical prediction considering only the ``PPP'' term (left) and the model (right) for the full set of triangle configurations in the \qone grid (containing squeezed and non-squeezed configurations). For triangles that are not squeezed, especially at large scales, the ``PPP'' term is expected to dominate. For squeezed triangles, non-Gaussian terms become relevant.  Indeed, Figure~\ref{fig:Full_Variance} clearly shows that  only including
the “PPP” contribution leads to errors up to $100\%$
Using our model of Eq. \eqref{eq:Total_BB}, we recover the measured variance to within a $\lesssim 40\%$ error. We notice that the error is largest at smaller scales (higher triangle index), where non-Gaussian terms beyond the ones we consider are expected to become important according to the estimates of Section~\ref{sect:theory}. We also highlight those triangles that are squeezed, that are described by our approximation within a $20\%$ error.

In Figure~\ref{fig:Full_Variance_Squeezed} we present the percentage error of the variance for the \qtwo grid. Since the triangles measured are very squeezed, our approximation works very well. The percentage error is within $20\%$ even for very small non-perturbative scales. The Gaussian approximation fails to approximate the variance by more than $\mathcal{O}(1)$. Even if the short scales are shot-noise dominated, the non-Gaussian terms are crucial in computing the variance.
    
    
\begin{figure}
\includegraphics[width=0.5\textwidth]{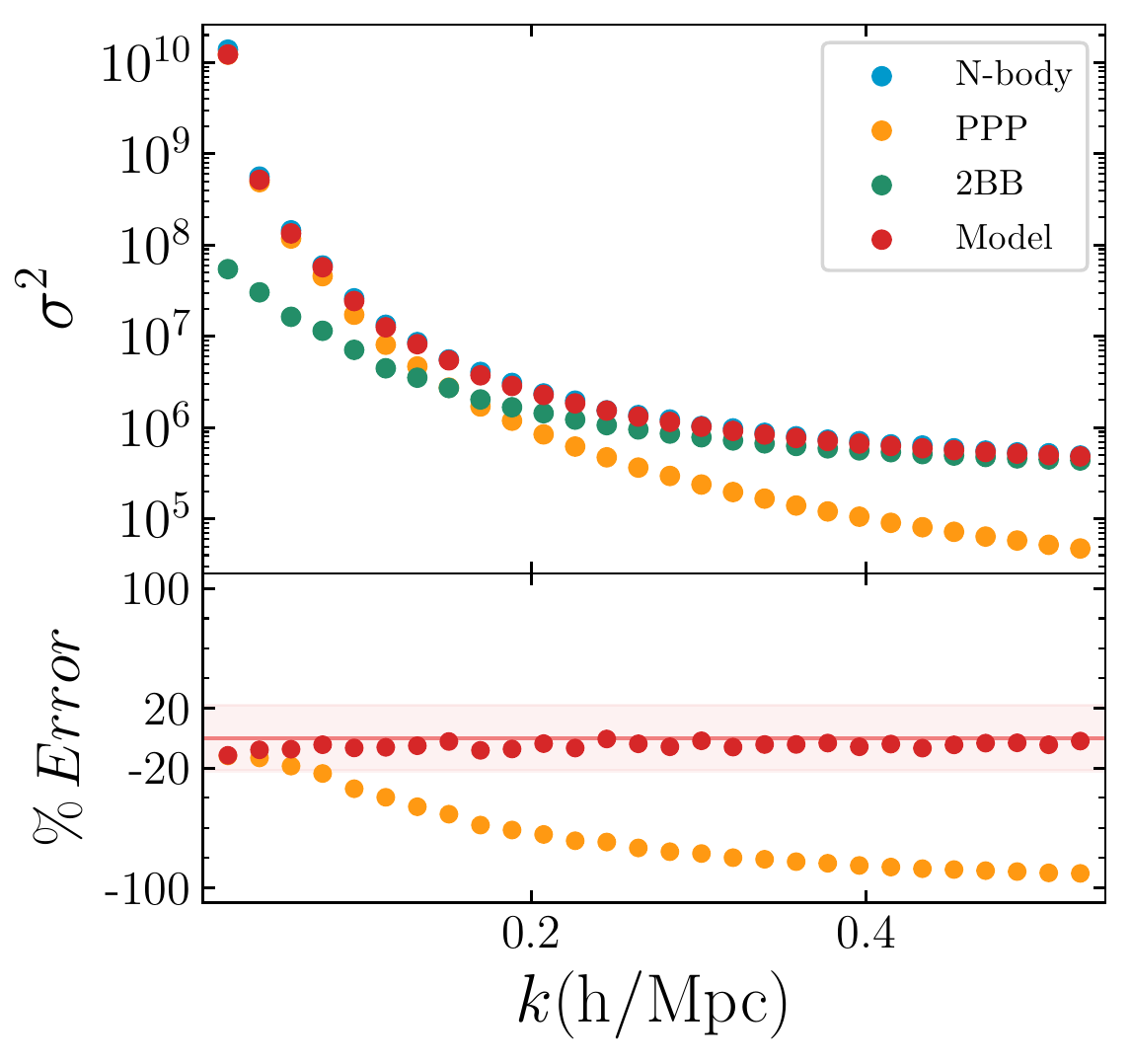}
\includegraphics[width=0.5\textwidth]{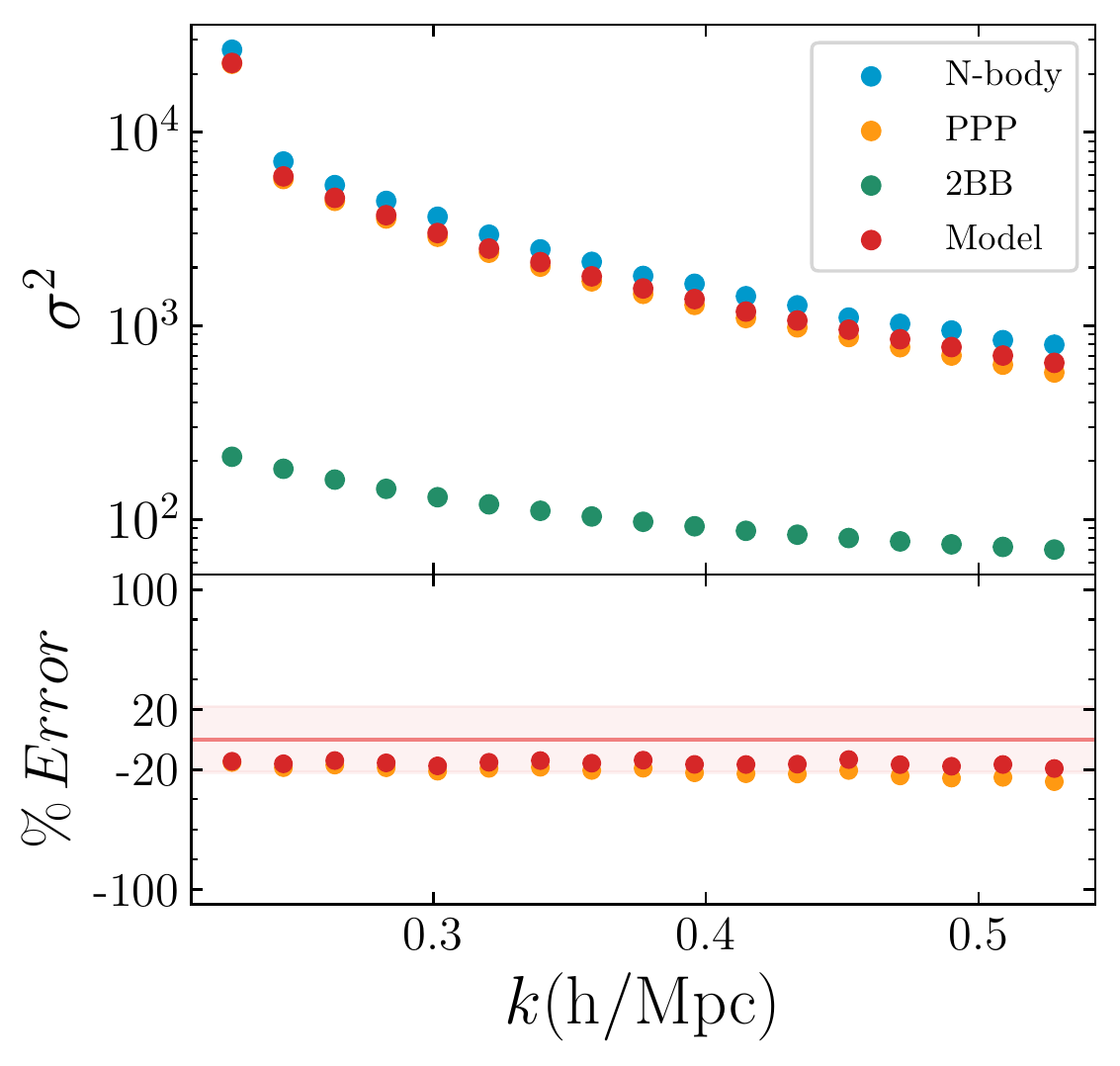}
\caption{\textbf{Bispectrum variance and percentage error between simulation measurements and our theoretical prediction, considering only PPP term (orange) and Model (red)}. {\em Left}: Squeezed triangles, where we fix the long mode to $q = 0.018\, h\,\text{Mpc}^{-1}$  as a function of one of the two remaining sides that satisfy the condition $k_2=k_3$. {\em Right}: Generic triangles, where we fix one side to $k =0.226\, h\,\text{Mpc}^{-1}$  as a function of one of the two remaining sides that satisfy the condition $k_2=k_3$. Measurements are performed using the \qone grid. These plots show that our model of Eq. \eqref{eq:Total_BB} predicts the bispectrum variance to within $20\%$ for any given triangle configuration.} \label{fig:varianceQ1}
\end{figure}

For ease of visualization, we focus on a subset of triangles in Figure \ref{fig:varianceQ1}. We consider two types of triangles: The left panel shows the bispectrum variance and percentage error for squeezed triangles, where we fix the long mode to $q=0.018 \text{h Mpc}^{-1}$ and the remaining sides satisfy the condition $k_2=k_3=k$. The error here is within the $20\%$ range. 
We observe that as the triangle becomes squeezed (toward the right part of the Figure), the ``2BB'' term dominates over the ``PPP'' term.
In the right panel we consider triangles that are not squeezed, with one of the sides fixed at $k_1=0.226\text{ h Mpc}^{-1}$ and the other two satisfying the condition $k_2=k_3=k$. 
In this case the error is somewhat larger than before. The Gaussian ``PPP'' is a good approximation in all the range considered, and the error is slightly larger than $20\%$ for a few points.

\begin{figure}
\includegraphics[width=0.49\textwidth]{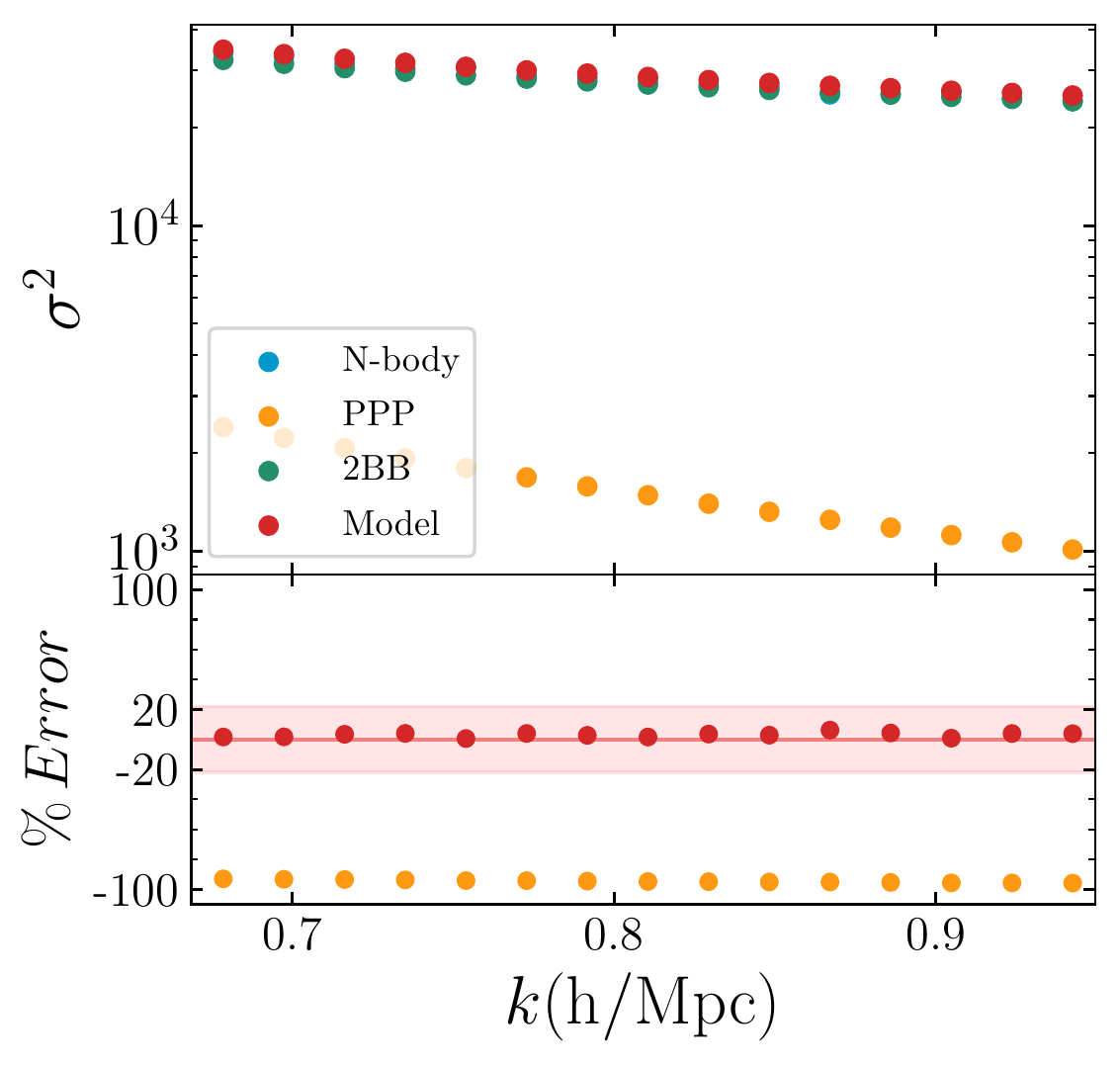}
\includegraphics[width=0.49\textwidth]{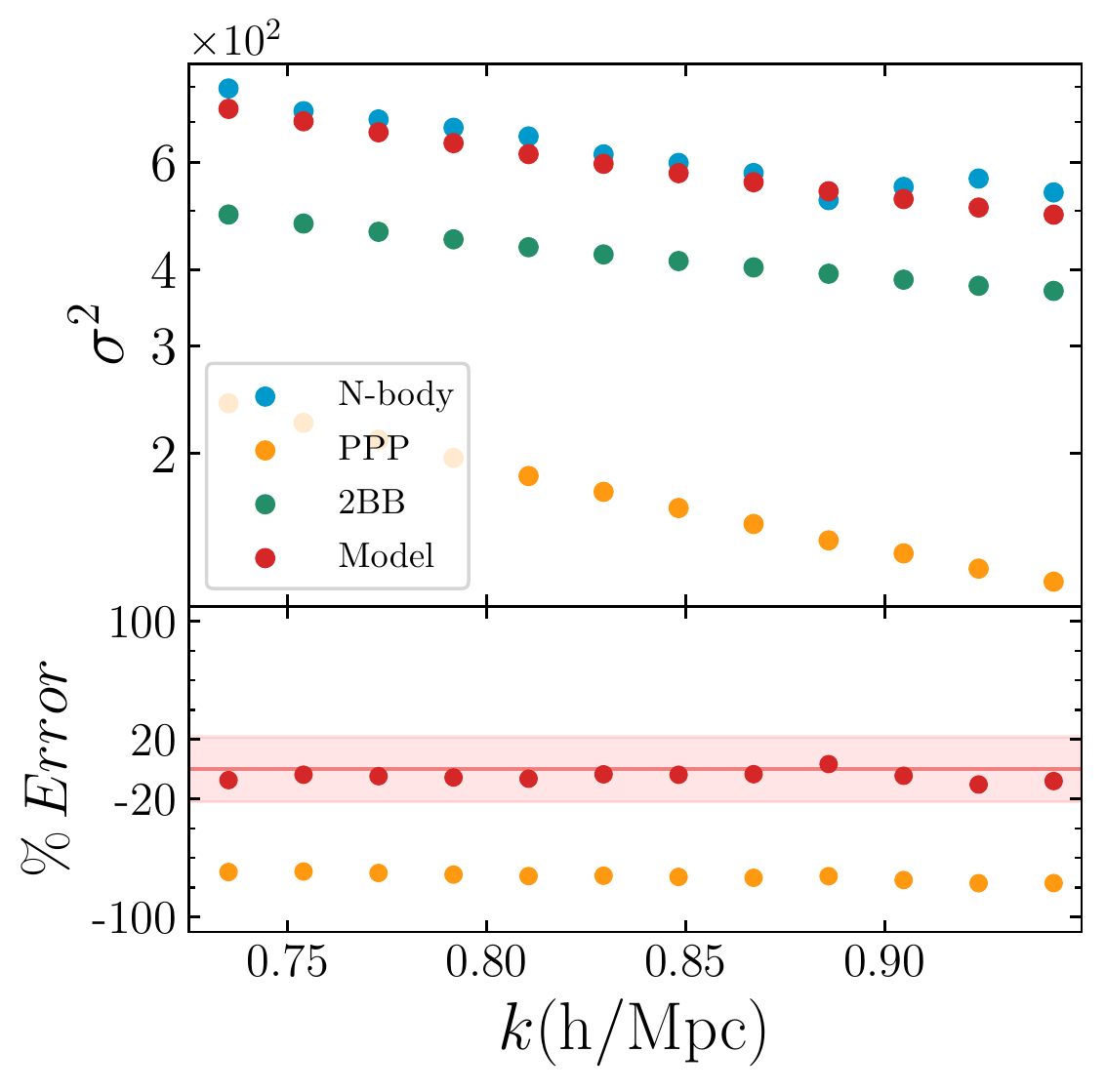}
\caption{\textbf{Bispectrum variance and percentage error between N-body simulations and theoretical prediction, considering only PPP term (orange) and Model (red)}. {\em Left}: Very squeezed triangles, where we fix the long mode to $q = 0.018\, h\,\text{Mpc}^{-1}$ as a function of the two remaining sides that satisfy the condition $k_3 =k_2+q$ (``flattened'' triangles). {\em Right}: Less squeezed triangles, where we fix the long mode to  $q =0.075\, h\,\text{Mpc}^{-1}$  as a function of the two remaining sides that satisfy the condition $k_3 =k_2+q$ (``flattened'' triangles). Measurements are performed using the \qtwo grid. These plots show that our model of Eq. \eqref{eq:Total_BB} predicts the bispectrum variance to within $20\%$ even for very squeezed configurations.} \label{fig:varianceQ2}
\end{figure}

In Figure~\ref{fig:varianceQ2}, we show similar plots for very squeezed triangles involving highly non-linear scales in the \qtwo grid. We see that, as expected, the model works very well, while the Gaussian approximation fails to order $\mathcal{O}(1)$. In this Figure, we show ``flattened'' configurations for which the usual approximation to the mode-counting factor $\Sigma_{ij}$ fails. We are able to recover the measured covariance using the approximation described in section~\ref{section_sums_approx}.

    
We confirm that our approximations recover the variance measured from simulations within a $20\%$ error for squeezed triangles involving scales up to $k \lesssim 1\, h\,\text{Mpc}^{-1}$. For these triangles, the non-Gaussian ``2BB'' term is dominant, and the Gaussian approximation fails. For non-squeezed triangles, we used the larger scales frequently considered in the literature, and find an agreement of $40\%$. For these configurations, as seen in Figure~\ref{fig:varianceQ1}, the non-Gaussian terms we keep are subdominant, and this thus quantifies how good the Gaussian approximation is.

\subsubsection{Correlation matrix  and off-diagonal terms}

It is useful to plot the correlation matrix of the halo power spectrum and bispectrum, defined as
\begin{equation}
    r_{ij} = \frac{C_{ij}}{\sqrt{C_{ii}C_{jj}}},
\end{equation}
where $C_{ij}$ is the covariance matrix as defined in Eq. \eqref{eq:fullmat}. This allows us to highlight the importance of off-diagonal elements. 

\begin{figure}
\centering
\includegraphics[width=0.45\textwidth]{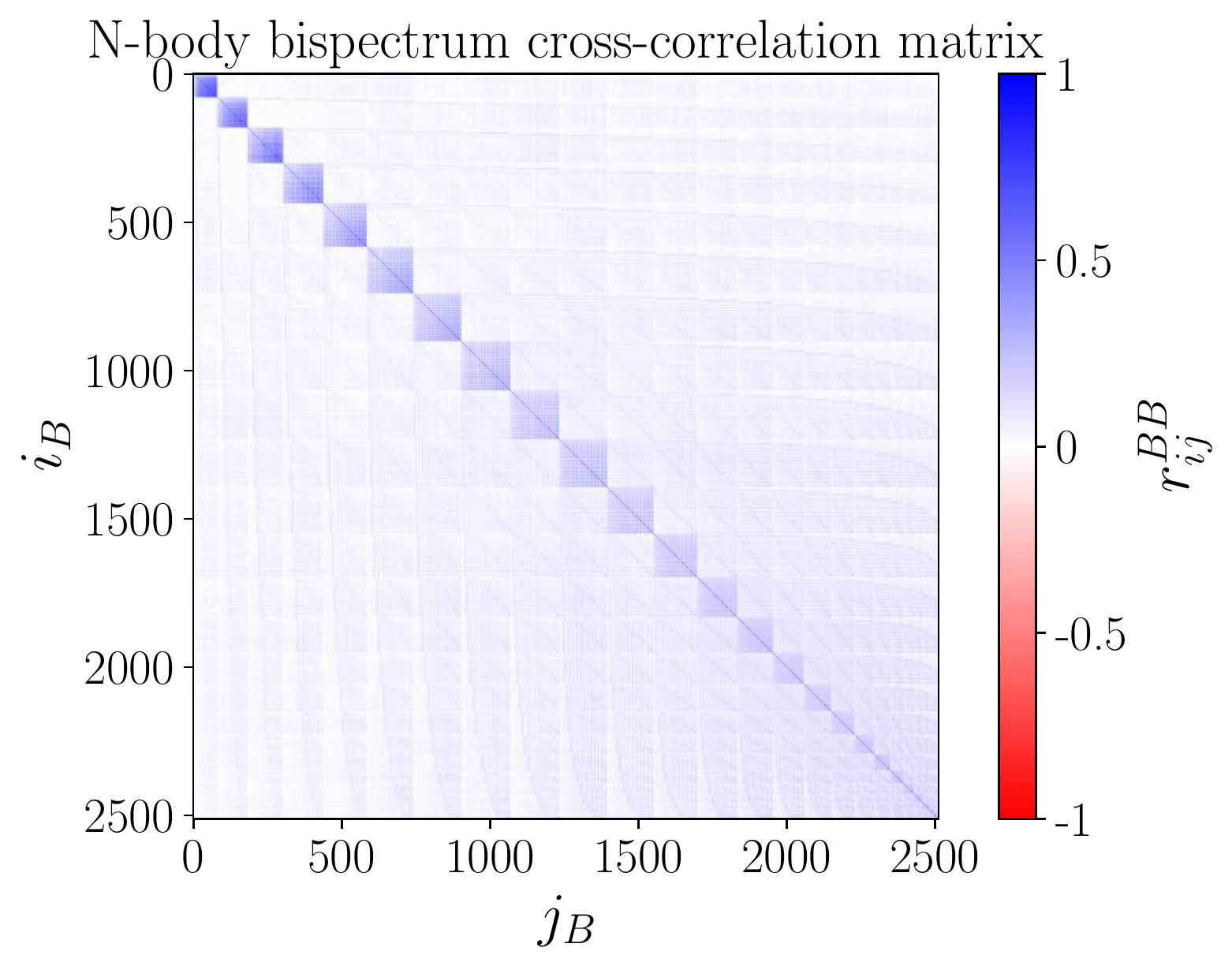}
\includegraphics[width=0.45\textwidth]{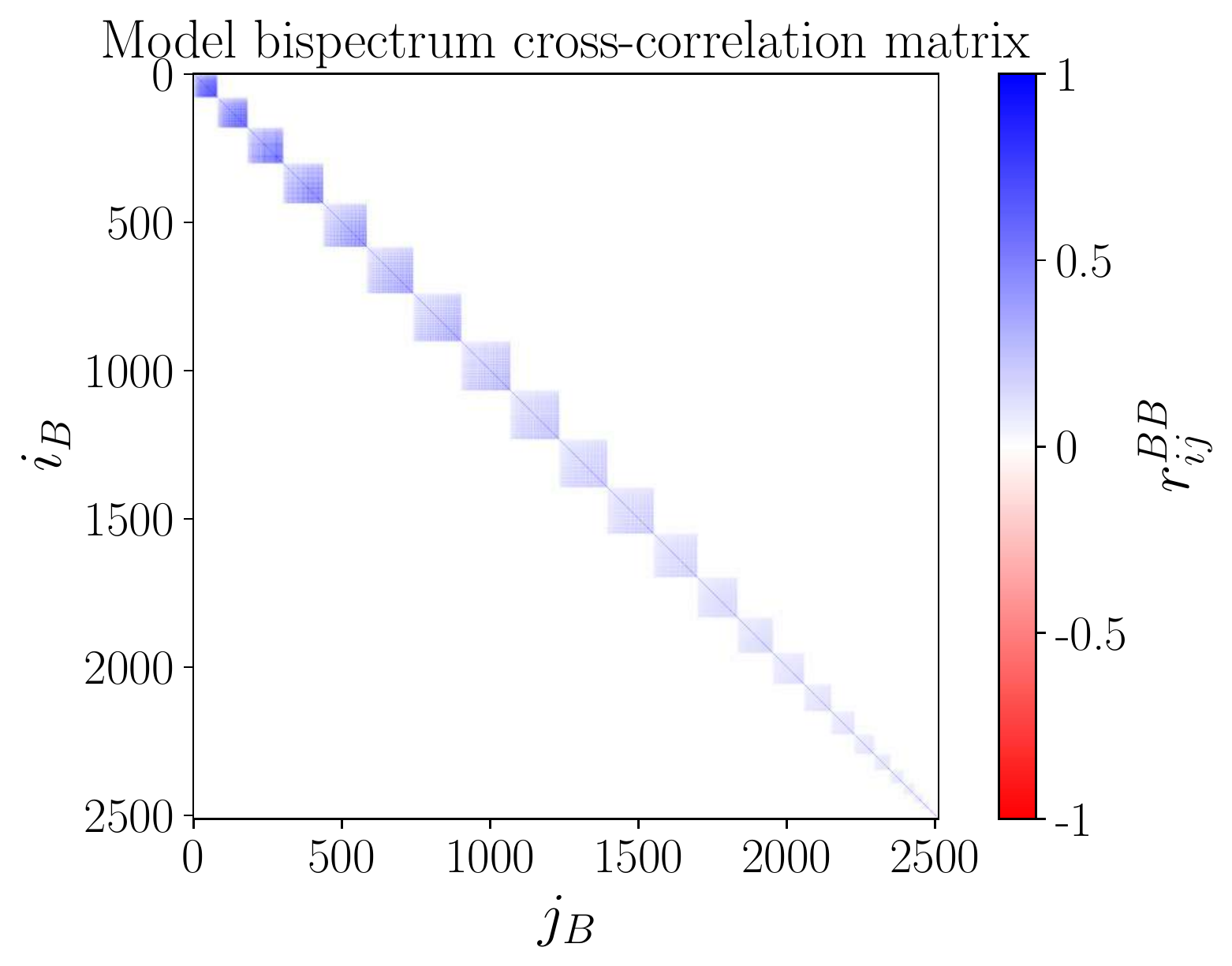}
\includegraphics[width=0.45\textwidth]{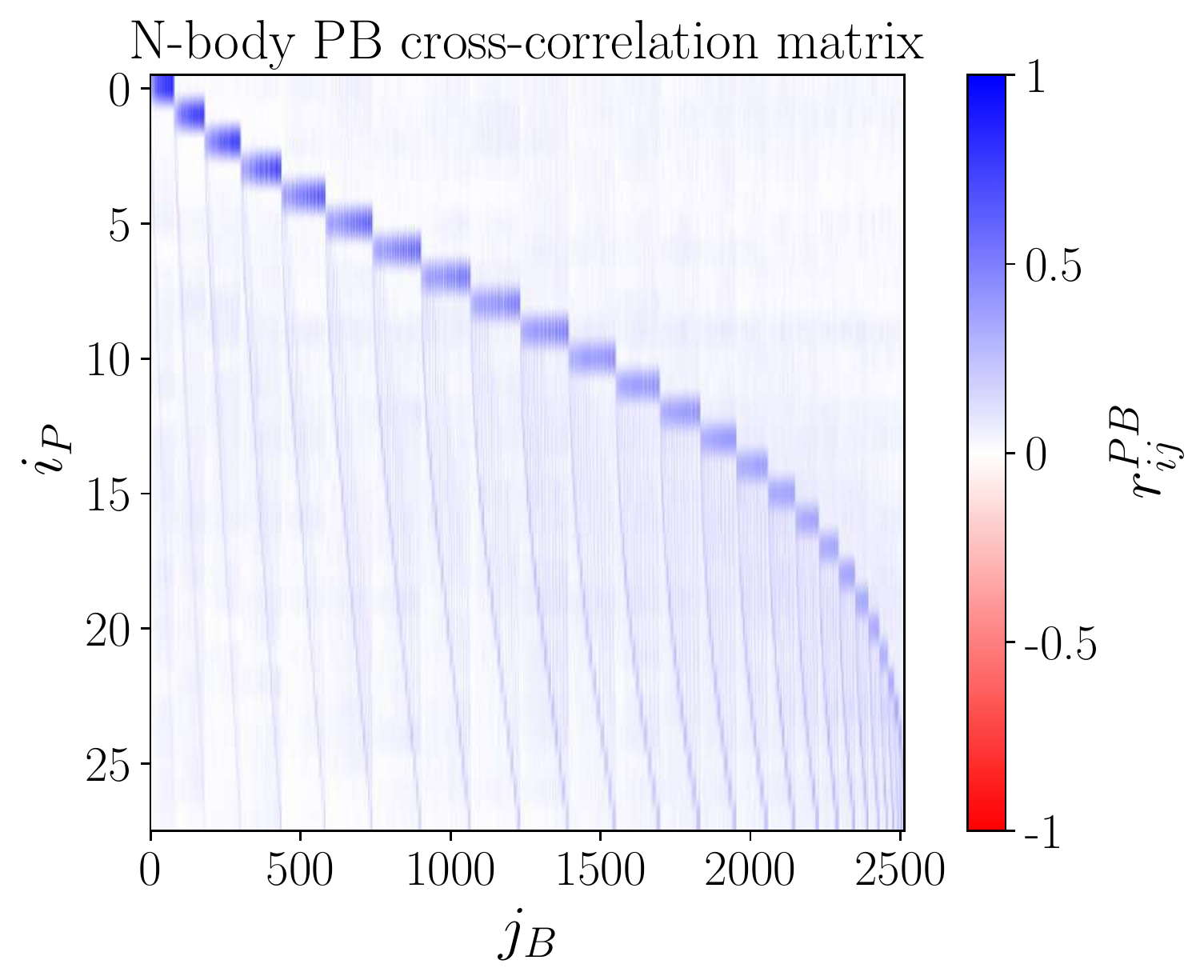}
\includegraphics[width=0.45\textwidth]{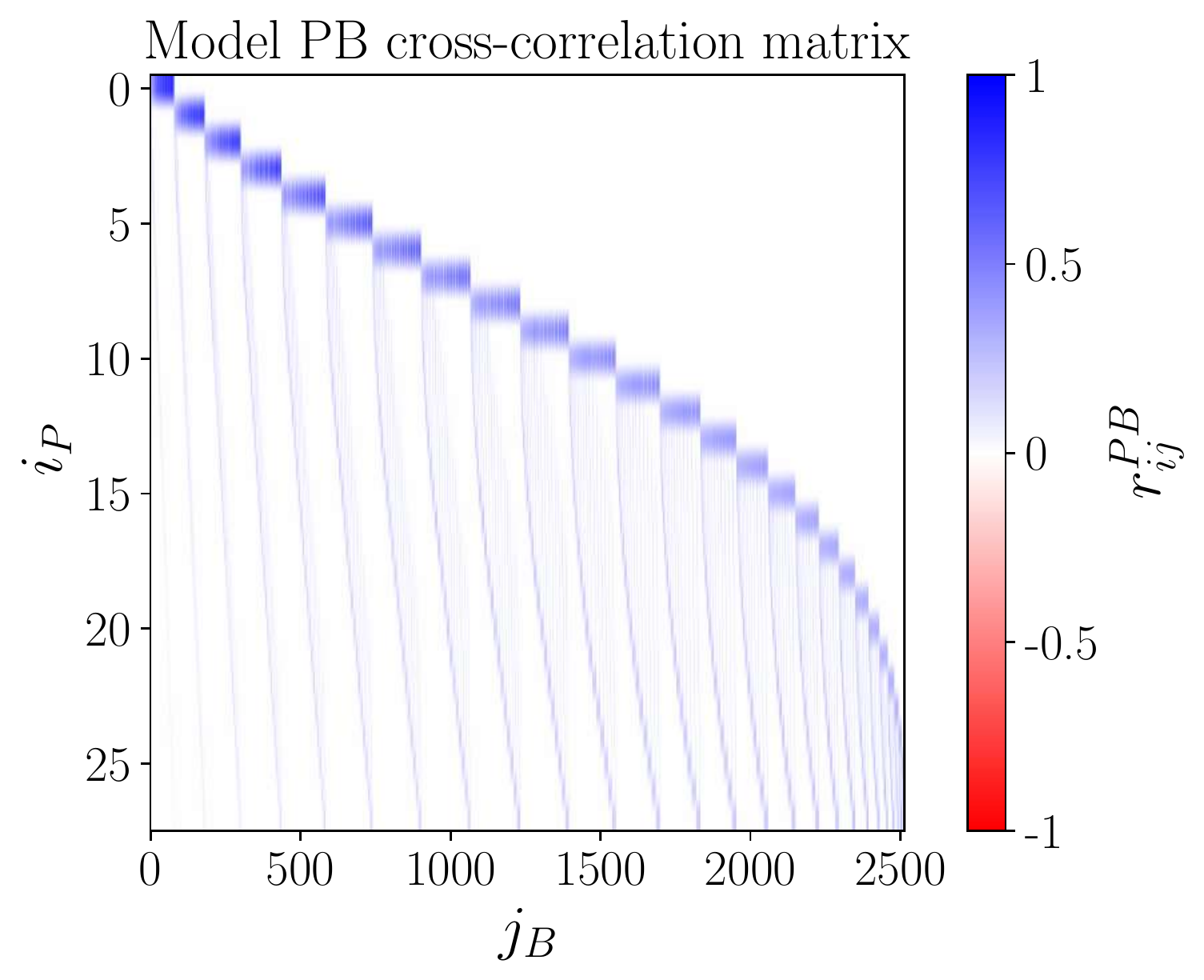}
\includegraphics[width=0.45\textwidth]{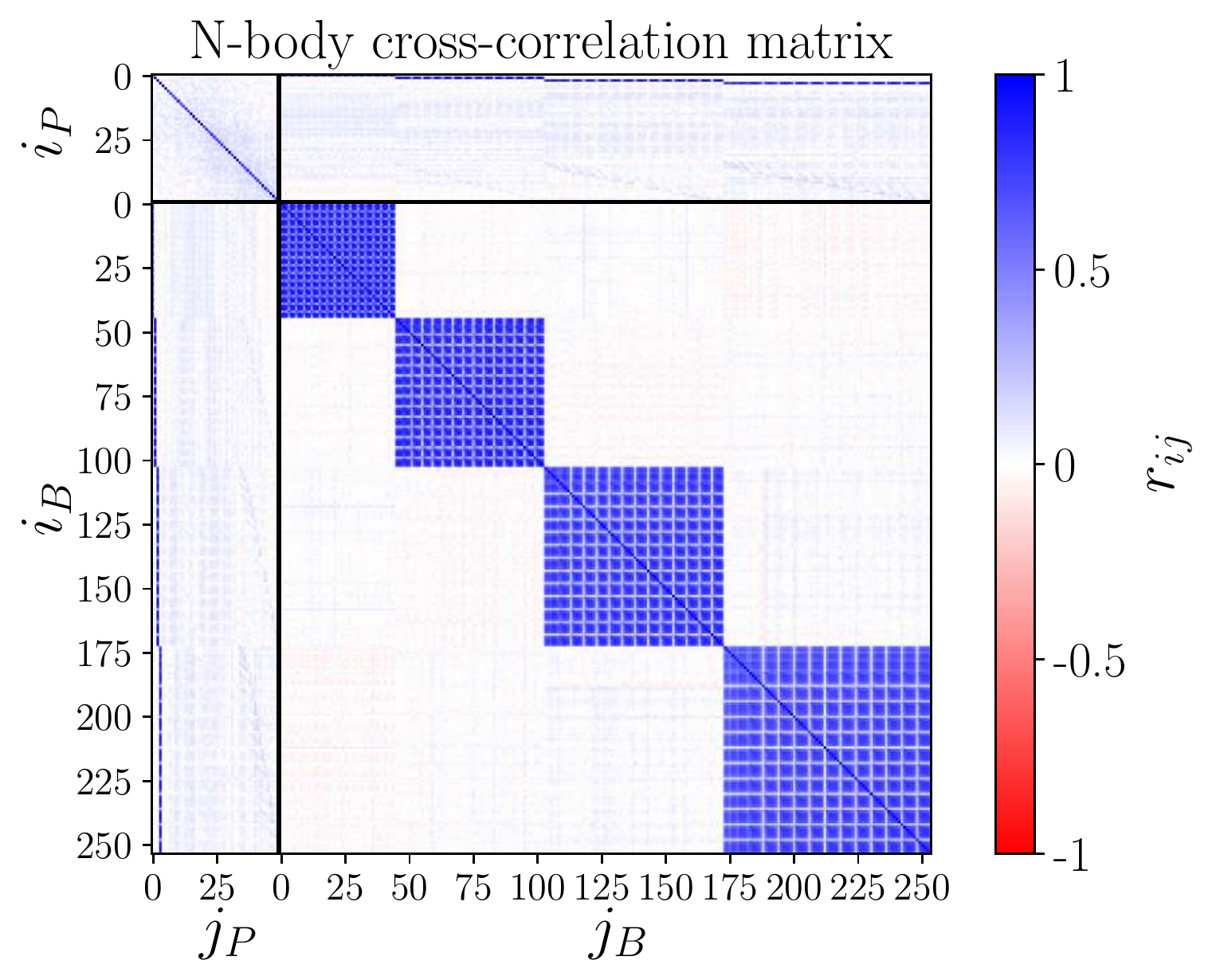}
\includegraphics[width=0.45\textwidth]{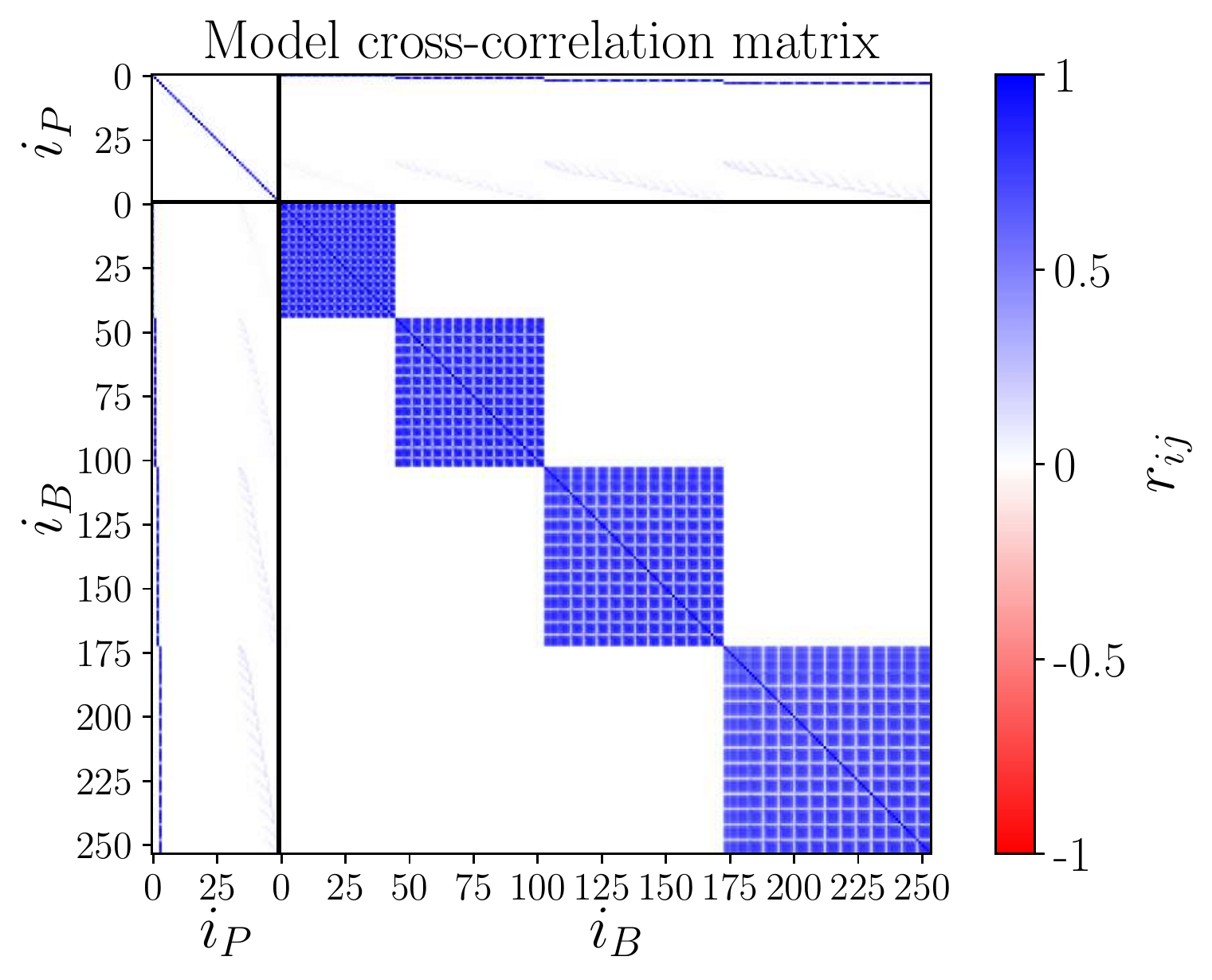}
\caption{{\em Top:} Cross-correlation matrix $r^{B}_{ij}$ of the halo bispectrum covariance from simulation measurements (left) and the model (right) in the \qone grid. {\em Middle:} Cross-correlation matrix $r^{PB}_{ij}$ of the halo bispectrum power spectrum cross covariance from simulation measurements (left) and model (right) using \qone grid. {\em Bottom:} simulation measurements (left) and the model (right), using the \qtwo grid. For both panels, the upper-left quadrant shows the power spectrum  cross-correlation matrix $r^{PP}_{ij}$, the botton-right quadrant shows the halo bispectrum cross correlation $r^{B}_{ij}$, and the off-diagonal quadrants show the cross correlation between power spectrum and bispectrum $r_{ij}^{PB}$.  Note that colors are saturated where correlations are close to one. This comparison shows qualitatively how our model captures the largest off-diagonal contributions of the cross power-spectrum-bispectrum and bispectrum covariance.  Differences between N-body and model in the $r^{PP}_{ij}$ quadrant are related to the fact that we do not model the non-Gaussian covariance of the power spectrum, as the focus of this work is in the bispectrum and cross-bispectrum covariance.} \label{fig:CM}
\end{figure}

In Figure~\ref{fig:CM} we show the simulation measurements and theoretical cross correlation matrix for the halo bispectrum covariance and the power spectrum bispectrum cross covariance  for the \qone and \qtwo grids.
The indexes $i_P, j_P, i_B, j_B$ in these figures refer to elements in the data vector. We ordered the bispectra by their smaller momentum first, then its next-to-smallest, and then the largest. In  this way, the data vector is such that triangles that share their smallest momentum are situated close to each other. We see large off-diagonal elements both in the simulation measurements and in our model. These correspond to triangles that share a long mode. Perhaps not surprisingly, they show the largest values in the correlation matrix. For the cross-covariance, there are large correlations between squeezed triangles and large-scale power spectra.
We can see that our model is in qualitative agreement with the covariance measured on simulations. 

\begin{figure}
\includegraphics[width=0.48\textwidth]{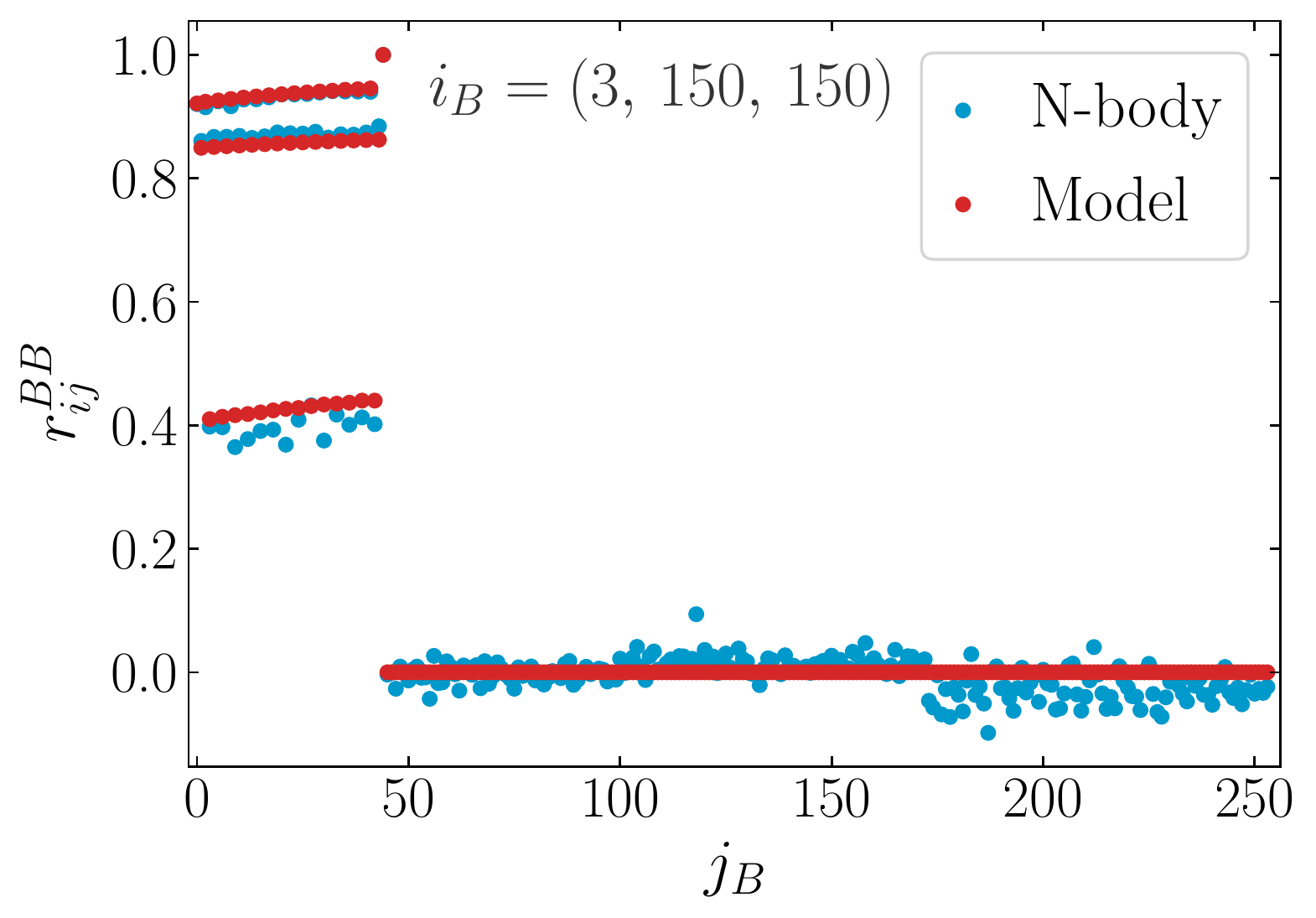}
\includegraphics[width=0.49\textwidth]{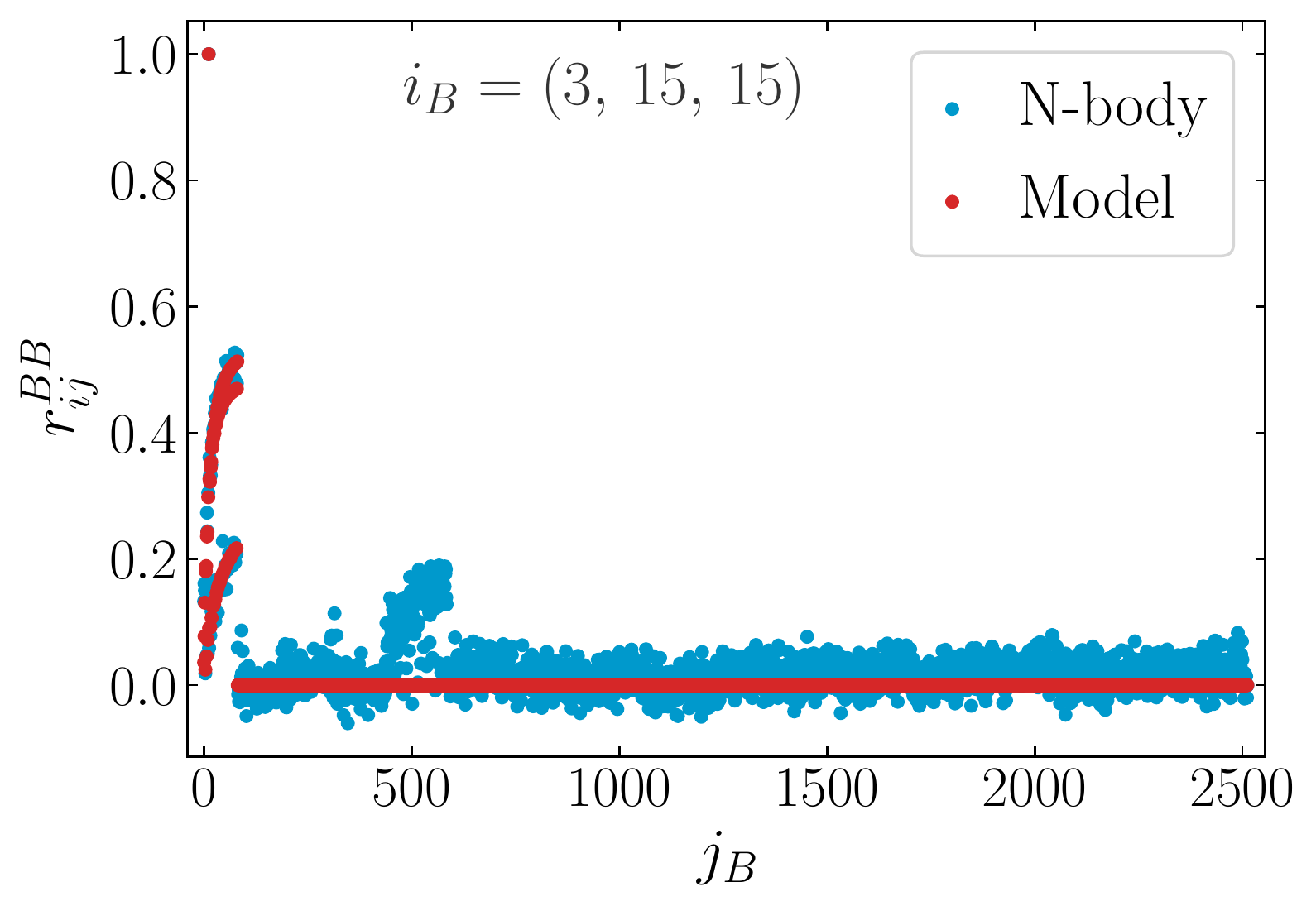}
\includegraphics[width=0.49\textwidth]{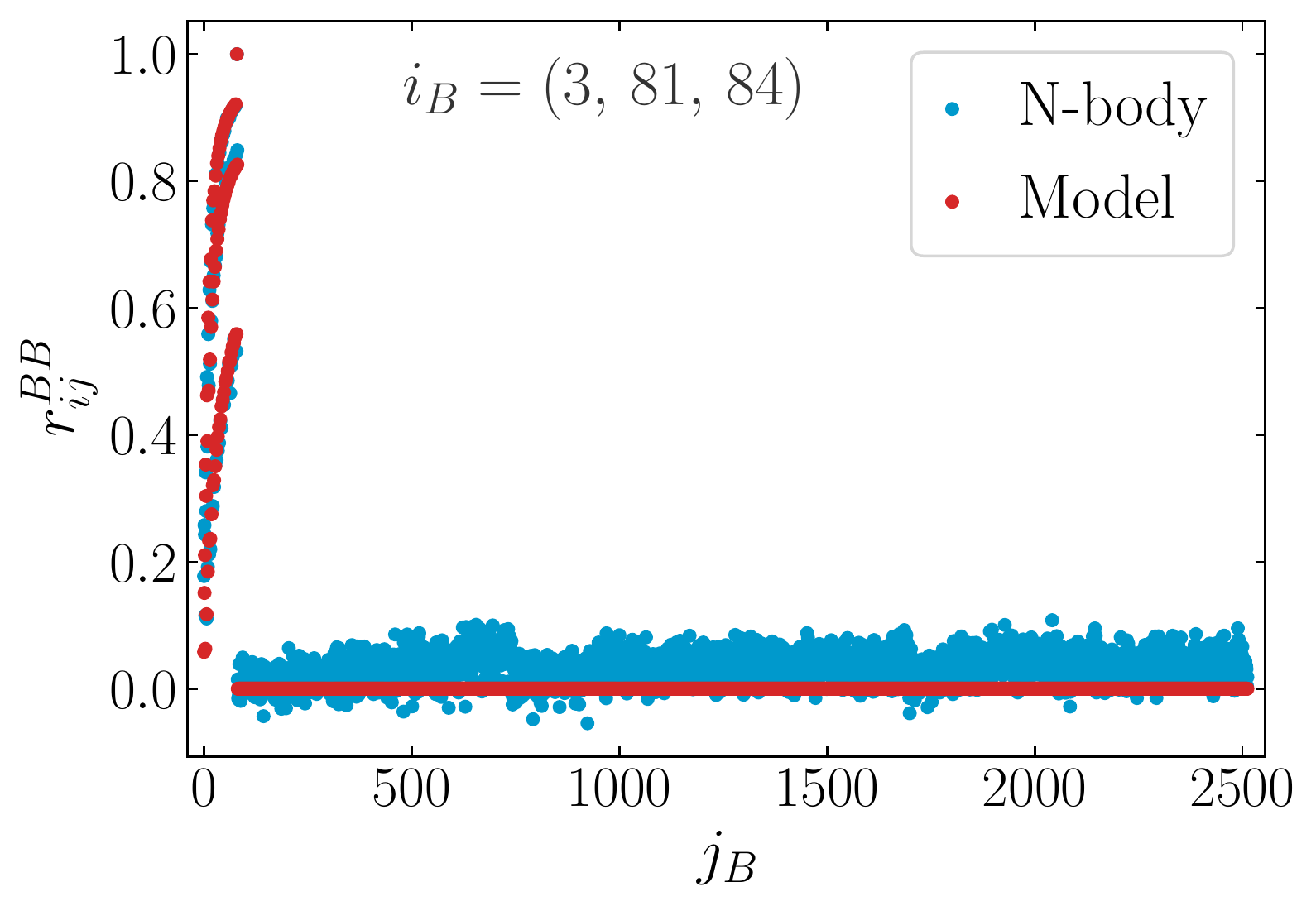}
\includegraphics[width=0.49\textwidth]{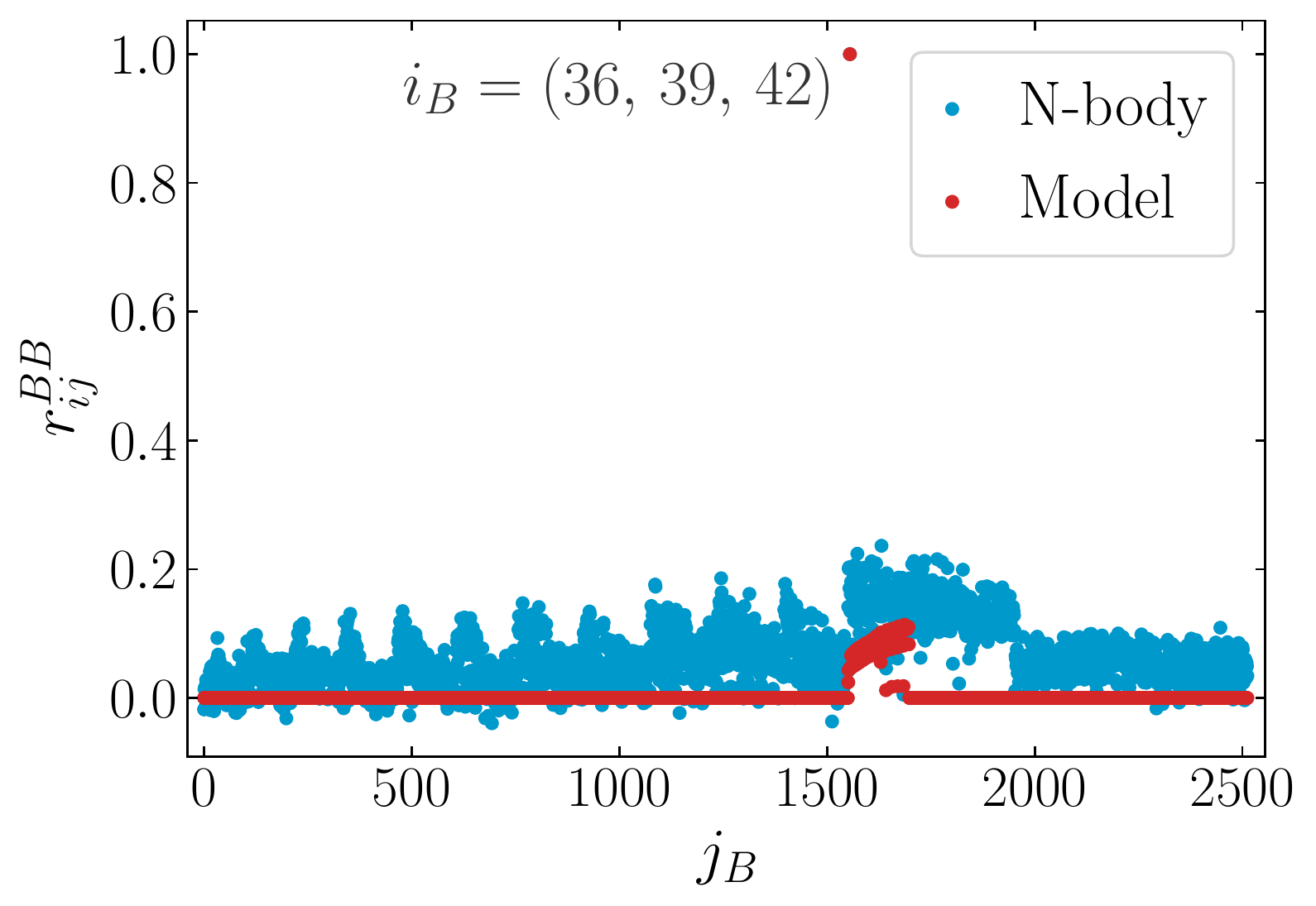}
\caption{\textbf{Rows of the bispectrum correlation matrix}. In each plot, there is a point with $r=1$ corresponding to the correlation of a triangle with itself. {\em Upper-left panel}: very squeezed configuration for the \qtwo grid. {\em Upper-right panel}: squeezed configuration for the \qone grid. {\em Bottom-left panel}: Flat squeezed configuration for the \qone grid. {\em Bottom-right panel}: Non-squeezed configuration for the \qone grid. These plots show the agreement of our model of Eq. \eqref{eq:Total_BB} with off-diagonal terms of the bispectrum covariance.} \label{fig:BB_rows}
\end{figure}

 In order to better  quantify the model accuracy on the off-diagonal elements of the correlation matrix, we plot several rows of $r^{B}_{ij}$ in Figure~\ref{fig:BB_rows}. These figures show very squeezed configurations (left panels), a mildly squeezed configuration (upper-right panel), and a non-squeezed configuration (lower-right panel). For squeezed configurations, the simulation measurements agree well with the theoretical model, which captures the dominant structure in the matrix. We notice that even when the triangles do not share one of the momenta, the bispectrum covariance is systematically above or below zero by a small amount. For these configurations, the ``PPP'', ``BB'' and ``PT'' terms are zero. This correlation can either be due to correlated noise due to the numerical uncertainty in our measurement (see Appendix~\ref{app:noise}) or a manifestation of the  6-point function. We notice that such contributions are, with few exceptions, of order $10\%$ in the correlation matrix.
We show a non-squeezed configuration in the lower-right panel. We see that, as expected, there are no large correlations between this triangle and others (they are all below $\sim 20\%$). The additional structure can be due to terms we ignore, such as other $BB$ or $PT$ terms in Eq.~\ref{appeqCBB-E}, the 6-point function, or  numerical uncertainty. In the upper-right panel we show a mildly squeezed configuration. We see that it has a large correlation with triangles sharing the long-wavelength mode (towards the left of the plot). We seem to be missing the correlation with the short-wavelength modes, showing at $j_B \sim 500$, but it is $\lesssim 20\%$.

\begin{figure}
\includegraphics[width=0.49\textwidth]{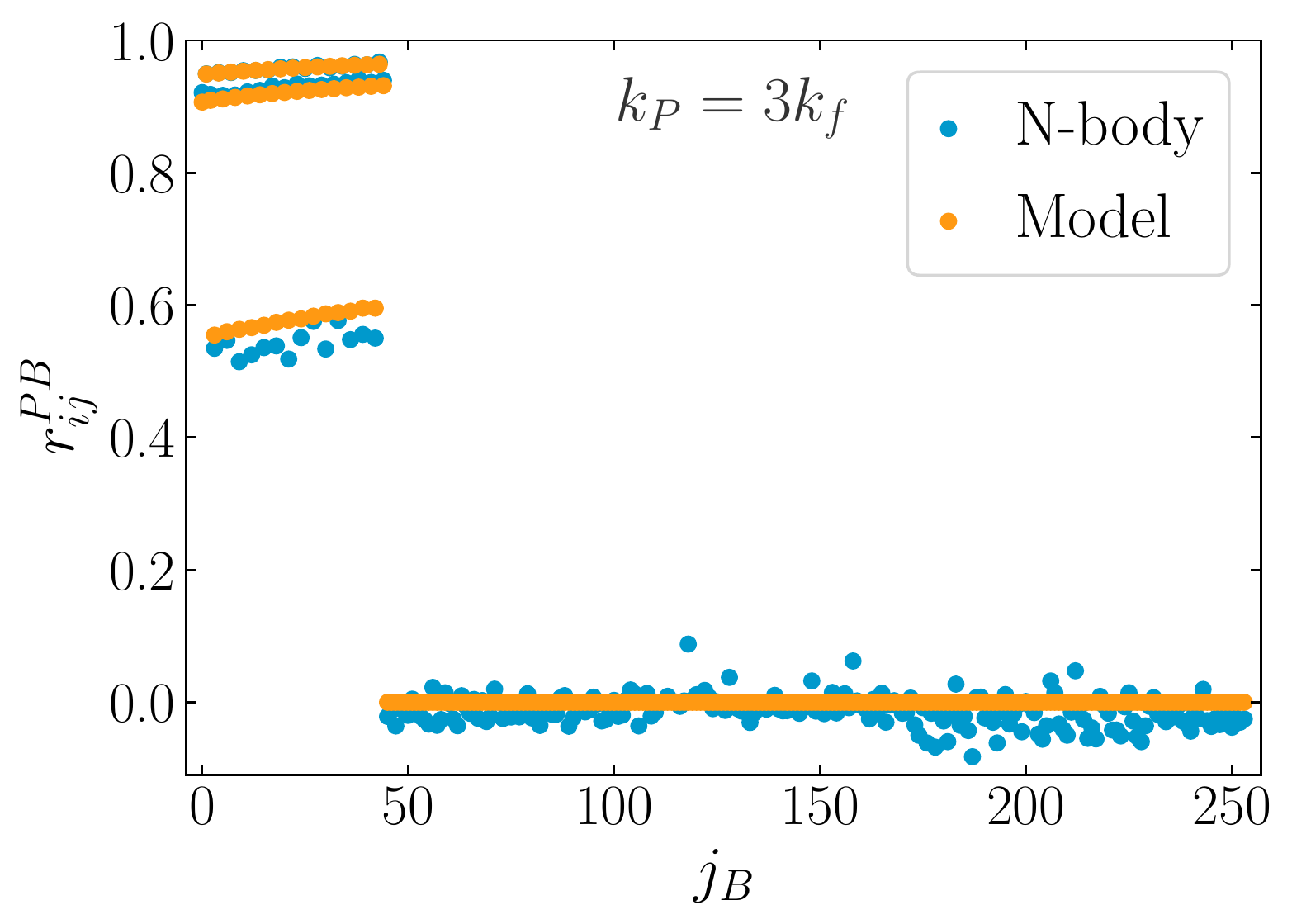}
\includegraphics[width=0.5\textwidth]{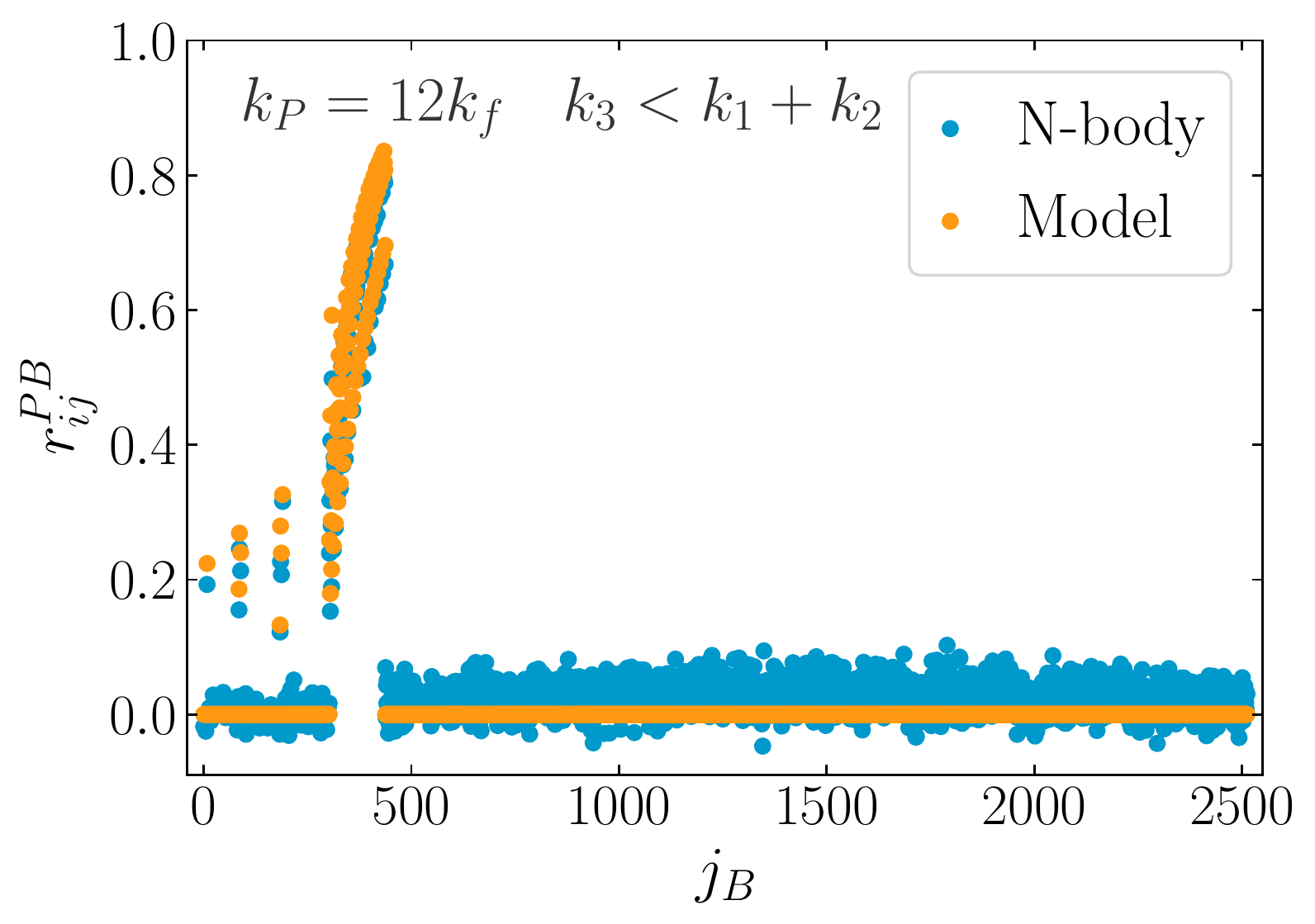}
\caption{\textbf{Rows of the bispectrum-power spectrum cross covariance correlation matrix}. {\em Left panel}: Very squeezed triangles for the \qtwo grid. {\em Right panel}: Triangles from the \qone grid.  These plots show the agreement of our model of Eq. \eqref{eq:Total_BB} with off-diagonal terms of the cross bispectrum-power spectrum covariance.}
\label{fig:PB_rows}
\end{figure}

We plot some rows of the power spectrum-bispectrum cross-correlation matrix $r_{ij}^{PB}$ in Figure~\ref{fig:PB_rows}. We see again that our model captures the structure of the cross-correlation. It is large when the power spectrum is evaluated at the long-wavelength mode of the bispectrum. The correlation is systematically different from zero even when the power spectrum does not share a mode with the bispectrum. Since the ``PB'' contribution is zero for these cases, this can be due to numerical uncertainty (see Appendix~\ref{app:noise}) or the  $5$-point function.

This comparison corroborates the model described in Section~\ref{sec:inversion}. It describes the correlation matrix with $\sim 10\%$ accuracy, and its main features are verified:
\begin{enumerate}
	\item In the bispectrum covariance sub-matrix, the ``PPP'' term is not the only sizable contribution to the diagonal for squeezed triangles. In fact, it gets corrected by up to $O(1)$ factors because of the ``BB'' and ``PT'' terms.
	\item Squeezed triangles sharing the same long mode in the bispectrum covariance are highly correlated, such that its off-diagonal component is not negligible.
	\item Squeezed triangles are very correlated with power spectra sharing the same long mode. As such, the cross-covariance between the power spectrum and the bispectrum is not negligible.
\end{enumerate}

\subsection{A $\chi^2$ test for the inverse covariance}

We have been comparing the measured covariance with the prescription of Eq.\ \eqref{eq:Total_BB}, in order to prove that triangles are correlated and this correlation is expected from theory. However, what is used in Fisher forecasts and likelihood analyses is actually the precision matrix, i.e. the inverse of the covariance matrix. An accurate numerical inversion of the N-body covariance requires a large number of realizations. With the number of realizations we are using this inverse is very noisy (keeping only squeezed triangles) or even impossible to compute (keeping all triangles).

\begin{figure}[h]
\includegraphics[width=0.49\textwidth]{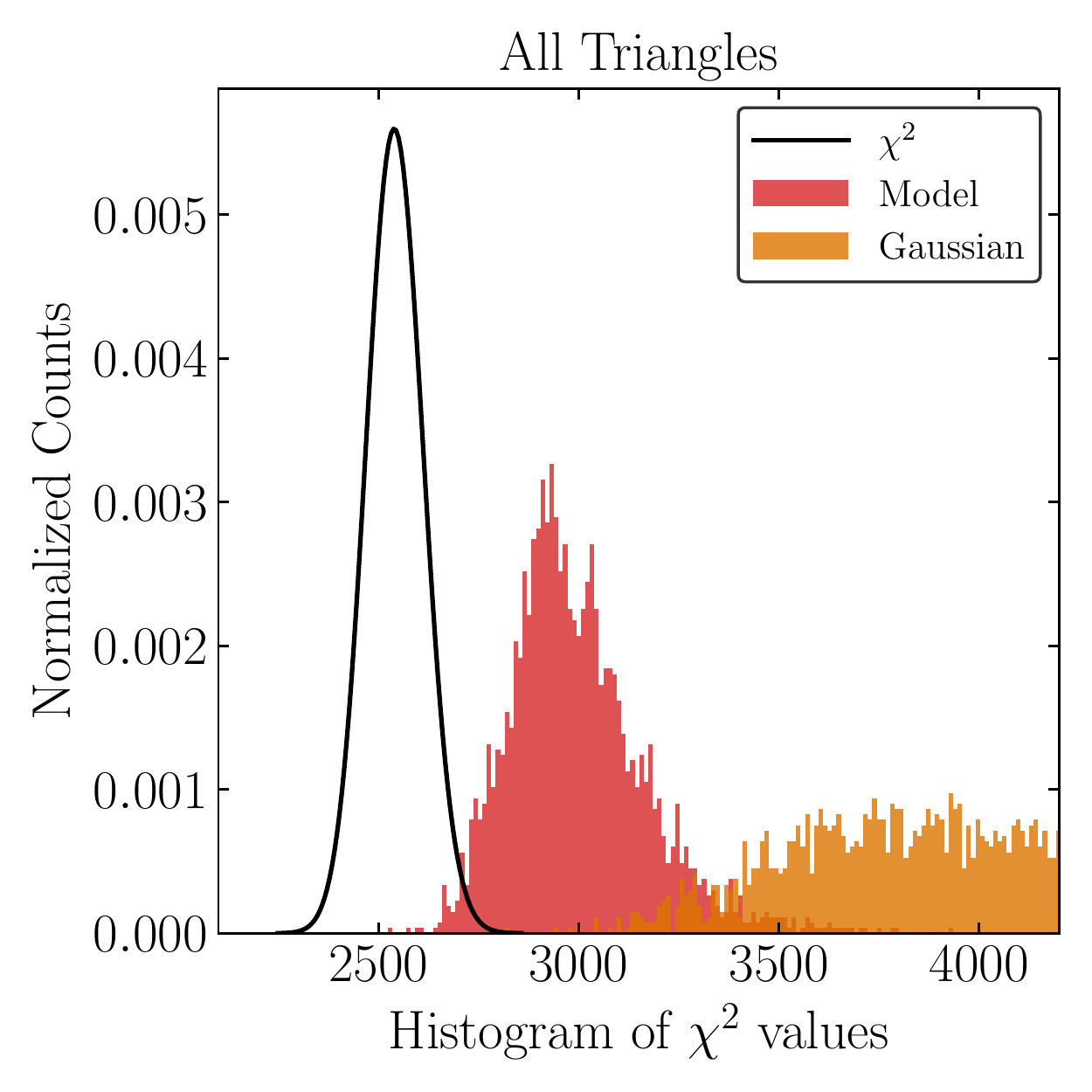}
\includegraphics[width=0.49\textwidth]{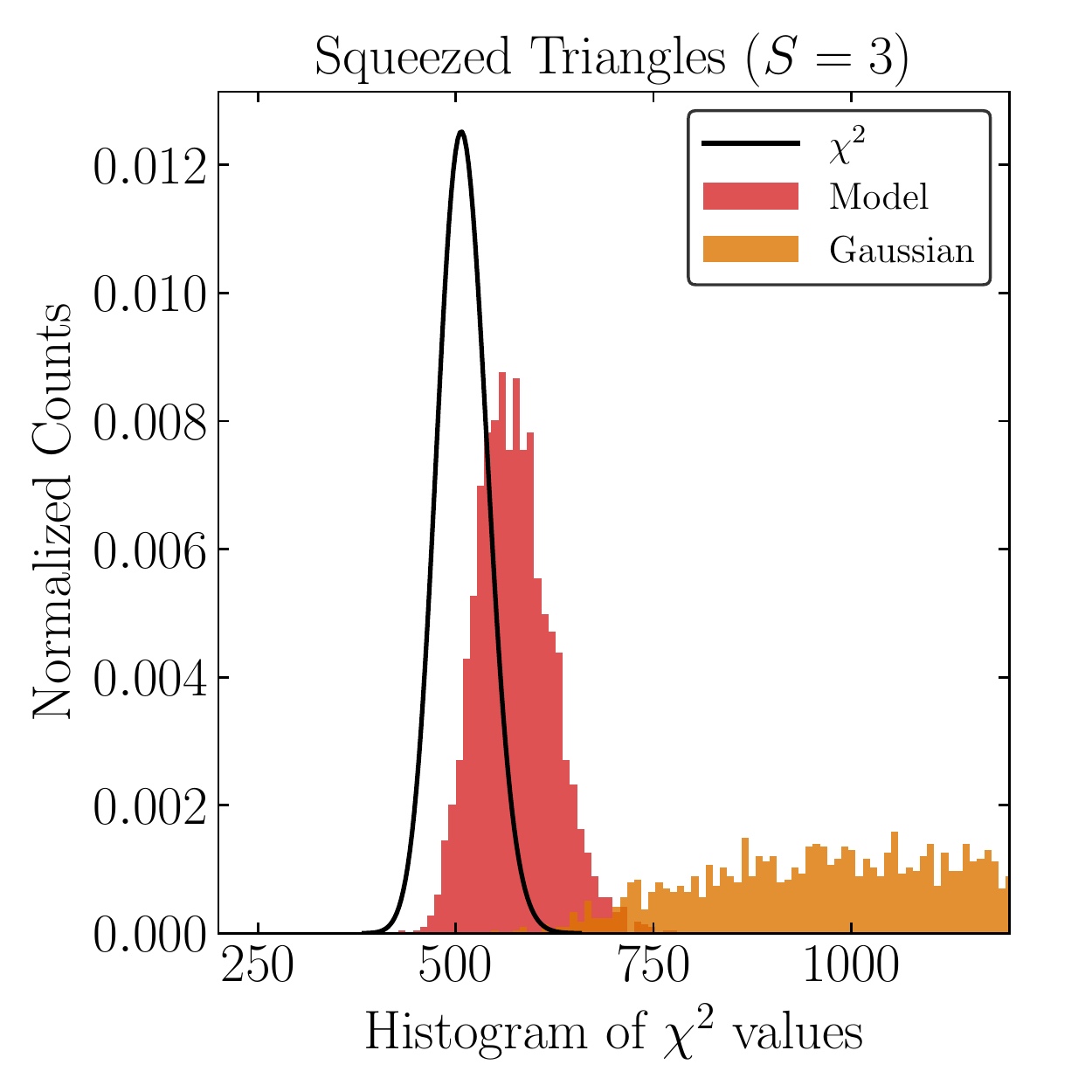}
\includegraphics[width=0.49\textwidth]{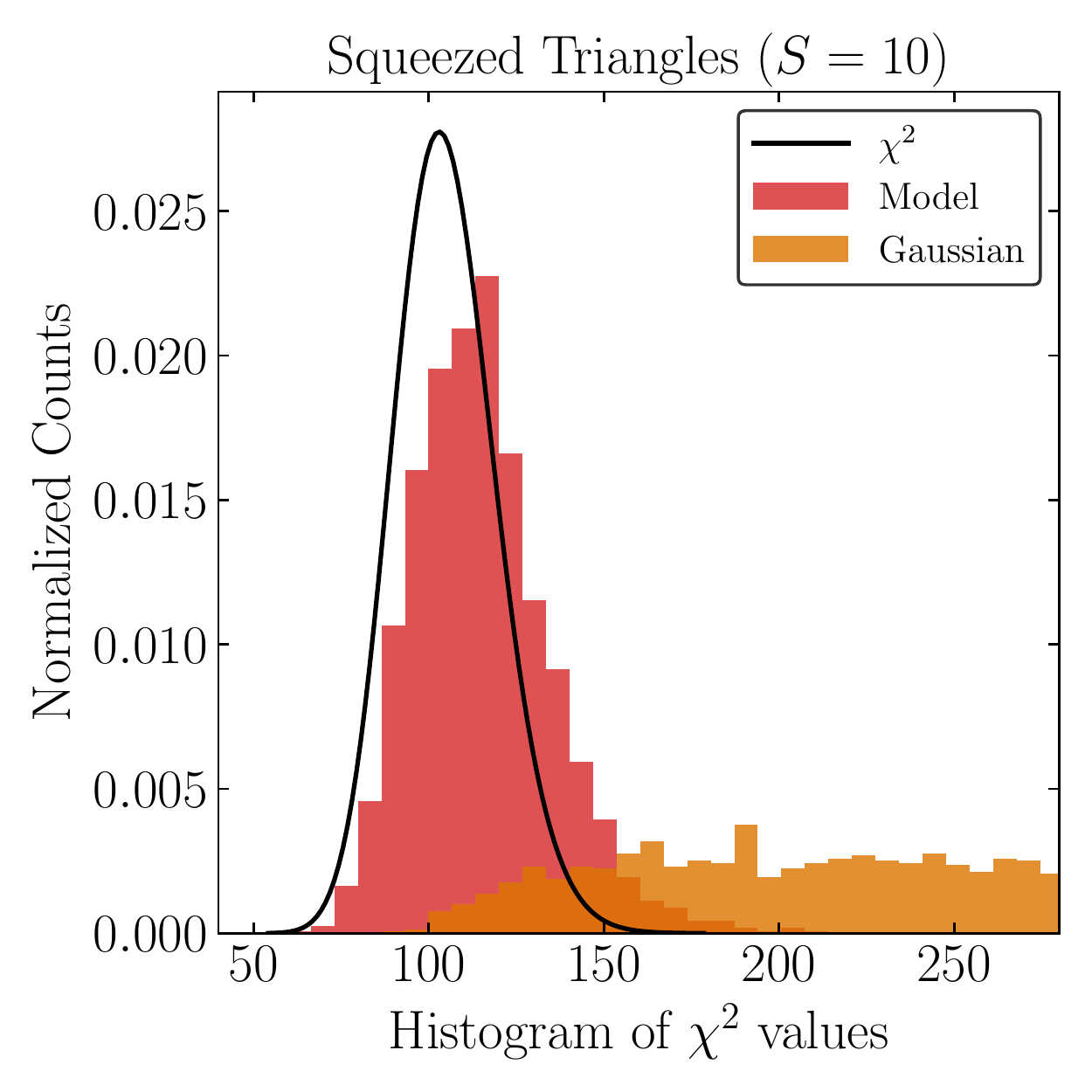}
\includegraphics[width=0.49\textwidth]{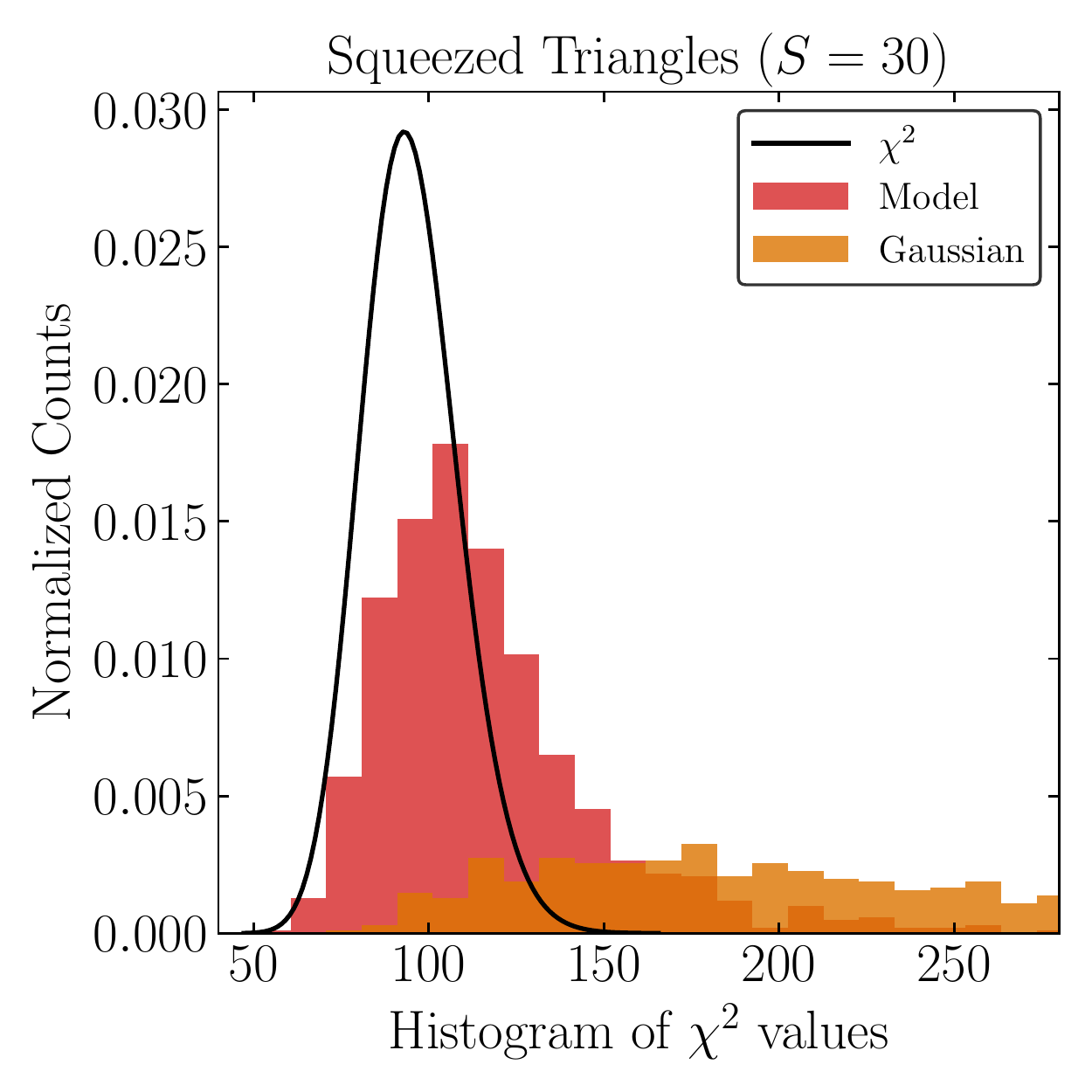}
\caption{{\bf Histogram of $\chi_{th}^2$ values using the theoretical covariance} as defined in Eq.~\eqref{chi2th}. \emph{Top Left}: All triangles for the \qone grid. \emph{Top Right:} Squeezed triangles such that the long mode is at least $3$ times smaller than the short modes for the \qone grid. \emph{Bottom Left:} Squeezed triangles such that the long mode is at least $10$ times smaller than the short modes for the \qone grid. \emph{Bottom Right:} Squeezed triangles such that the long mode is at least $30$ times smaller than the short modes for the \qtwo grid.  This test shows that our model is much closer to a $\chi^2$ distribution than considering a Gaussian covariance. As expected, the agreement of our model improves more and more as we restrict to squeezed triangles. For all configurations, the shift of the mean of the model with respect to the the expected $\chi^2$ is around $20\%$, which is what we find from comparing our model to simulations.}\label{fig:chi2}
\end{figure}

In order to check the accuracy of the precision matrix obtained from the model of the covariance in Section~\ref{sec:inversion}, we instead compute\footnote{Another check on the reliability of the inverse can be done by performing the so-called \emph{half-inverse test}, first introduced in \cite{Slepian:2015hca}, which we outline in Appendix \ref{app:hitest}. We thank an anonimous referee for suggesting this test.}
\be\label{chi2th}
\chi_{th,i}^2 =  (D_i-D_{\rm mean}) {\bf C}^{-1}_{th} (D_i-D_{\rm mean})^T\,,
\ee
where $D_i = (P_i, B_i)$ is the vector of measurements for the $i$-th realization, $D_{\rm mean}$ is the mean over $2377$ realizations and ${\bf C}_{th}$ is the theoretical covariance, inverted as described in section \ref{sec:inversion}. 
 If the theoretical covariance is close to the true covariance, $\chi^2_{th}$ should follow a $\chi^2$ distribution with degrees of freedom equal to the dimension of $D_i$.\footnote{We are thankful to Anže Slosar for suggesting this check.}

In Figure~\ref{fig:chi2} we plot this check for the configurations in the \qone, and \qtwo grids. We show the histogram of $\chi_{th}^2$ values computed with Eq.~\eqref{chi2th}.  Different colors correspond to using Eq.~\eqref{eq:Total_BB} (Red) or using the Gaussian diagonal covariance (orange), and are compared to the expected $\chi^2$ (black solid line). The panels show the $\chi^2$ values for all triangles, a squeezing factor of $3$ and a squeezing factor of $10$ for the \qone grid, and a squeezing factor of $30$ for the \qtwo grid.

When keeping all triangles, we see that the Gaussian approximation to the covariance fails at reproducing the expected distribution. On the other hand, our prescription produces a distribution that is about $20\%$ off from the expected central $\chi^2$ value. For squeezed triangles, as expected, our distribution fares even better, while the failure of the Gaussian approximation is even more evident. This is a clear indication that correlations among triangles cannot be neglected.

\section{ Impact on parameter constraints}
\label{sect:fisher}

In the previous section, we showed that one cannot neglect all off-diagonal terms in the bispectrum covariance, nor in the cross power-spectrum-bispectrum covariance. As a consequence, it would be important to assess the impact of these terms on parameter constraints. As a test, in this section we estimate the Fisher information content with and without our corrections on local primordial non-Gaussianity. In this context, constraints on local non-Gaussianity are expected to be affected significantly by our findings, given that the bispectrum information is crucial to break degeneracies~\cite{MoradinezhadDizgah:2020whw}.

\subsection{Local Primordial non-Gaussianity}

Interactions among fields taking place during inflation produce a deviation from Gaussianity in the primordial curvature perturbations that seed the formation of structure. Knowing about these interactions is paramount for understanding fundamental physics at very high energies, hence the search for so called primordial non-Gaussianity is one of the major goals in cosmological searches. While primordial perturbations are nowadays constrained to be very close to Gaussian, there is still room for interesting primordial non-Gaussianity to be discovered, and large scale structure surveys are expected to be at the forefront of such an effort~\cite{Meerburg:2019qqi}. Here, we consider primordial non-Gaussianity of the local type, which is modeled as
\begin{equation}\label{eq:locpng}
    \phi(\xv) = \phi_{\rm G}(\xv) + \fnl\left(\phi^2_{\rm G}(\xv) - \langle \phi_{\rm G}^2\rangle \right) +\mathcal O\left(\phi_{\rm G}^3\right) ,
\end{equation}
being $\phi$ the curvature perturbation, $\phi_{\rm G}$ a Gaussian random field, and $\fnl$ parametrizes the amplitude of non-Gaussianity. The primordial bispectrum generated by $\phi$ peaks  for  squeezed  configurations. That is why we expect our findings to be particularly important for this parameter. 
The modeling of how the primordial signature of $\phi$ is imprinted in galaxy clustering has been extensively studied in the literature, both for the power spectrum (\cite{Dalal:2007cu,Matarrese:2008nc,Slosar:2008hx}, see~\cite{Biagetti:2019bnp} for a recent review and references therein) and for the bispectrum~\cite{Baldauf:2010vn,Sefusatti:2010ee,Sefusatti:2011gt,Yokoyama:2013mta,Tasinato:2013vna,Dizgah:2015kqi,Hashimoto:2015tnv,Tellarini:2016sgp,Hashimoto:2016lmh,Yamauchi:2016wuc,DiDio:2016gpd,Chiang:2017vsq,An:2017rwo,MoradinezhadDizgah:2018ssw,dePutter:2018jqk,MoradinezhadDizgah:2018pfo,Barreira:2020ekm,MoradinezhadDizgah:2020whw,Barreira:2021ueb}. 

\subsection{N-body simulations with non-Gaussian initial conditions}

We are only interested in the qualitative, rather than quantitative, impact of our terms on $\fnl$ constraints. Thus, we use simulations with and without primordial non-Gaussianity to infer the change in the power spectrum and bispectrum as a function of $\fnl$, rather than modeling it using perturbation theory.  We take advantage of the \eos\footnote{Information on the \eos  suite is available at \href{ https://mbiagetti.gitlab.io/cosmos/nbody/eos/}{https://mbiagetti.gitlab.io/cosmos/nbody/eos/}. For a recent publication using the same subset, see~\cite{Biagetti:2020skr}.} simulations, run using the \textsc{Gadget}-2 \cite{Springel:2005mi}. The suite consists of several realizations run with Gaussian and non-Gaussian initial conditions of the local and equilateral type. For our analysis, we use halo catalogs from $10$ realizations in the \textsf{G85L}, \textsf{NG250L} and \textsf{NG250mL} sets, which are initialized with local type primordial non-Gaussianity $\fnl = 0$, $+250$ and $-250$, respectively. These simulations are following the evolution of $1536^3$ dark matter particles in a periodic cubic box with a side length of $2000\,\,$Mpc$/h$. The initial conditions are set at $z_i=99$ by displacing each particle using second order Lagrangian perturbation theory with the publicly available code \textsc{2LPTic}~\cite{Crocce:2006ve} for Gaussian initial conditions, and its modification \textsc{2LPTicNG} \cite{Scoccimarro:2011pz} for non-Gaussian initial conditions. The linear transfer function is obtained using the Boltzmann code \textsc{CLASS}~\cite{Blas:2011rf}.  The halo catalogs are generated using the public code Rockstar~\cite{Behroozi:2011ju}, identifying candidate halos with a Friends-of-Friends (FoF) algorithm~\cite{Davis:1985rj} with a linking  length $\lambda= 0.28 $. We require that halos are constituted by a minimum of $50$ particles. 
We compute the power spectrum and bispectrum in an analogous grid to \qone (see Table \ref{tab:binning}), where both $k_{\rm min}$ and $k_{\rm max}$ are half the size, since the \eos box size is twice the one of \quijote simulations.  We call these measurements \textsf{E1}.

\subsection{Fisher matrix forecasts}

\begin{figure}[t]
\includegraphics[width=0.5\textwidth]{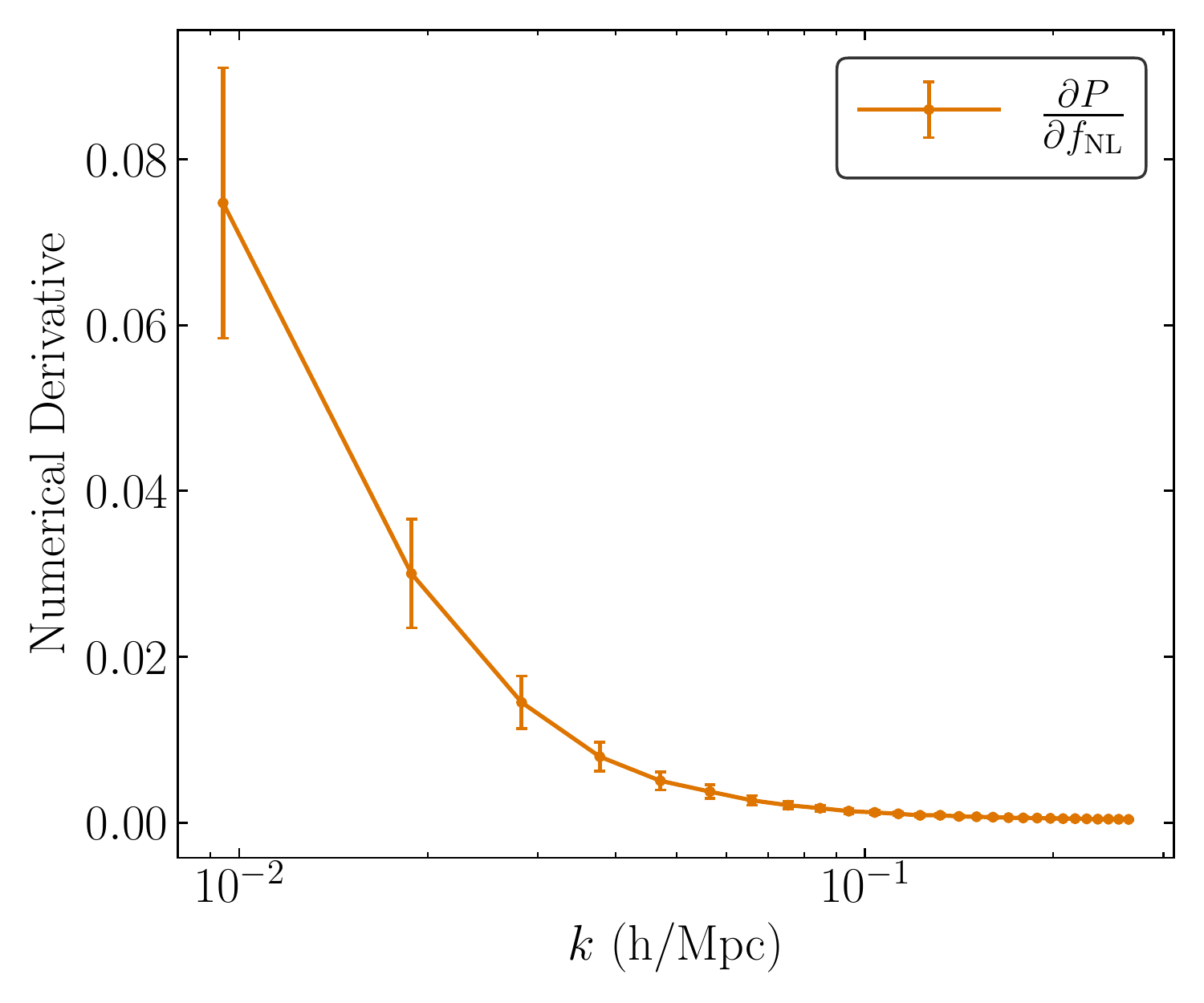}
\includegraphics[width=0.5\textwidth]{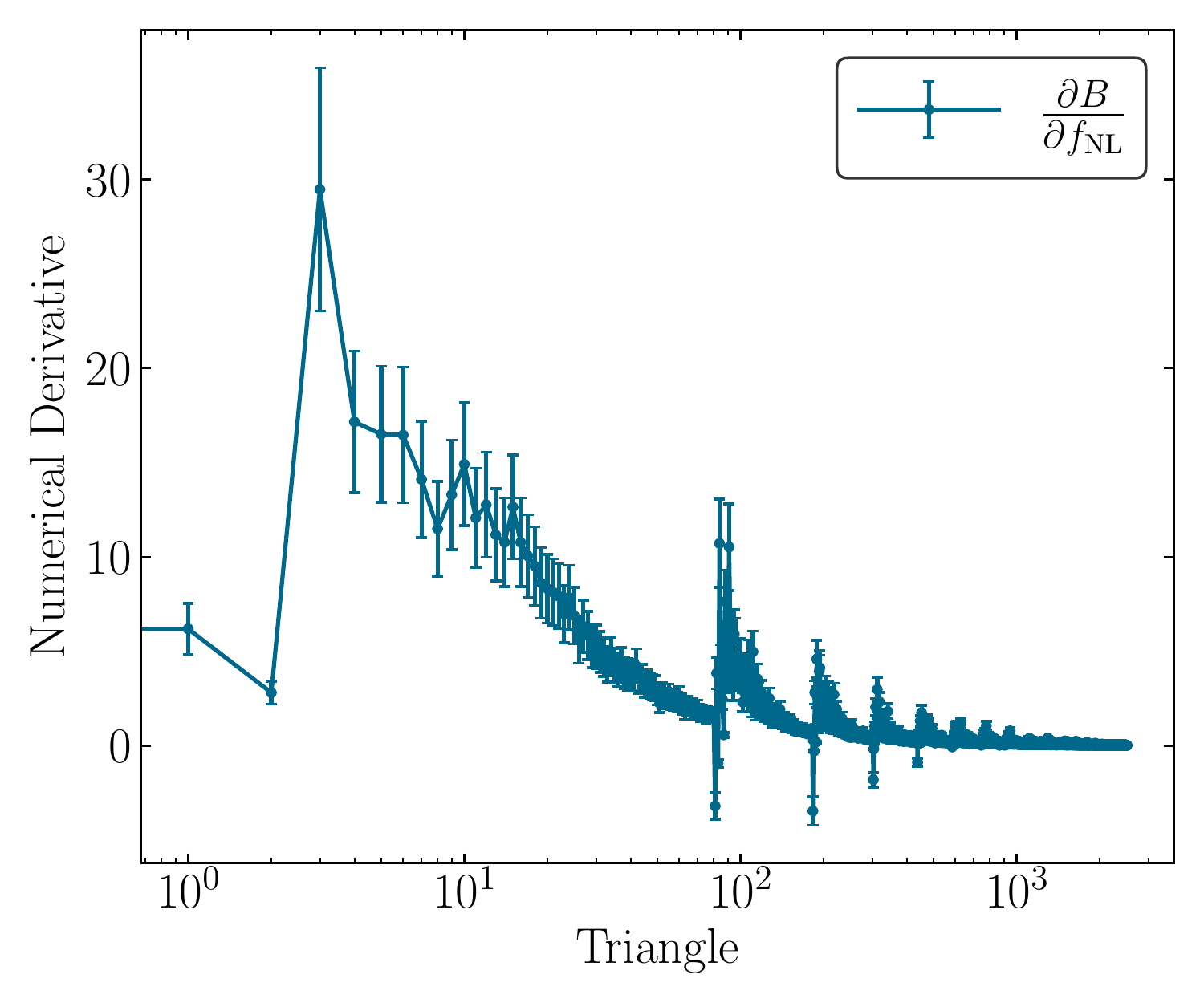}
\caption{\textbf{Numerical derivative of power spectrum and bispectrum}. {\em Left}: Power spectrum numerical derivative with respect to $\fnl$ as a function of wavenumber $k$ for the \textsc{E1} measurements. Error bars indicate standard deviation from the mean of $10$ realizations. {\em Right}: Bispectrum numerical derivative with respect to $\fnl$ for all the triangles in the \textsc{E1} measurements. Error bars indicate standard deviation from the mean of $10$ realizations.} \label{fig:derivatives}
\end{figure}

We want to determine the Fisher matrix for the parameter $\fnl$ using our predictions for the theoretical covariance matrix of the power spectrum and bispectrum. We use the simulations with primordial non-Gaussianity to infer the numerical derivative of the power spectrum and bispectrum with respect to $\fnl$. The Fisher matrix is defined as
\begin{equation}
    F_{\fnl} = \frac{\partial \mathbf{D}}{\partial \fnl} \cdot \mathbf{C}^{-1} \cdot \frac{\partial \mathbf{D}^T}{\partial \fnl}
\end{equation}
being $\mathbf{D} = (P,B)$ the vector of power spectrum and bispectrum values as a function of $k$ and $\mathbf{C}$ the full covariance matrix as defined in Equation~\eqref{eq:fullmat}.\footnote{ One might be worried that by neglecting non-Gaussian covariance terms in the $P$ covariance we are underestimating the gain due to the bispectrum. However, for the specific case of local-type non-Gaussianity, most of the constraining power in the power spectrum comes from the large scales, because of the scale dependent bias effect \cite{Dalal:2007cu,Slosar:2008hx,Matarrese:2008nc}. Therefore we do not expect a strong impact in our estimations due to neglecting non-Gaussianity in the power spectrum covariance. A similar discussion applies for the cross power spectrum-bispectrum covariance: we do include terms where the power spectrum is evaluated at large scales, since these are the sizable terms, while the small scale power spectrum terms have a lower impact on the uncertainty.}  The derivative is performed numerically as
\begin{equation}
    \frac{\partial \mathbf{D}}{\partial \fnl} = \frac{\mathbf{D}(\textsf{NG250L}) - \mathbf{D}(\textsf{NGm250L})}{2 \bar f_{\rm NL}},
\end{equation}
where $\bar f_{\rm NL} = 250 $ and $\mathbf{D}(\textsf{X})$ is the vector of $P$ and $B$ values as measured for the $\textsf{X}$ simulation using the \eone grid. We show the derivatives of the power spectrum and bispectrum as a function of wavenumber in Figure~\ref{fig:derivatives}. 

\begin{figure}[t]
\centering
\includegraphics[width=0.8\textwidth]{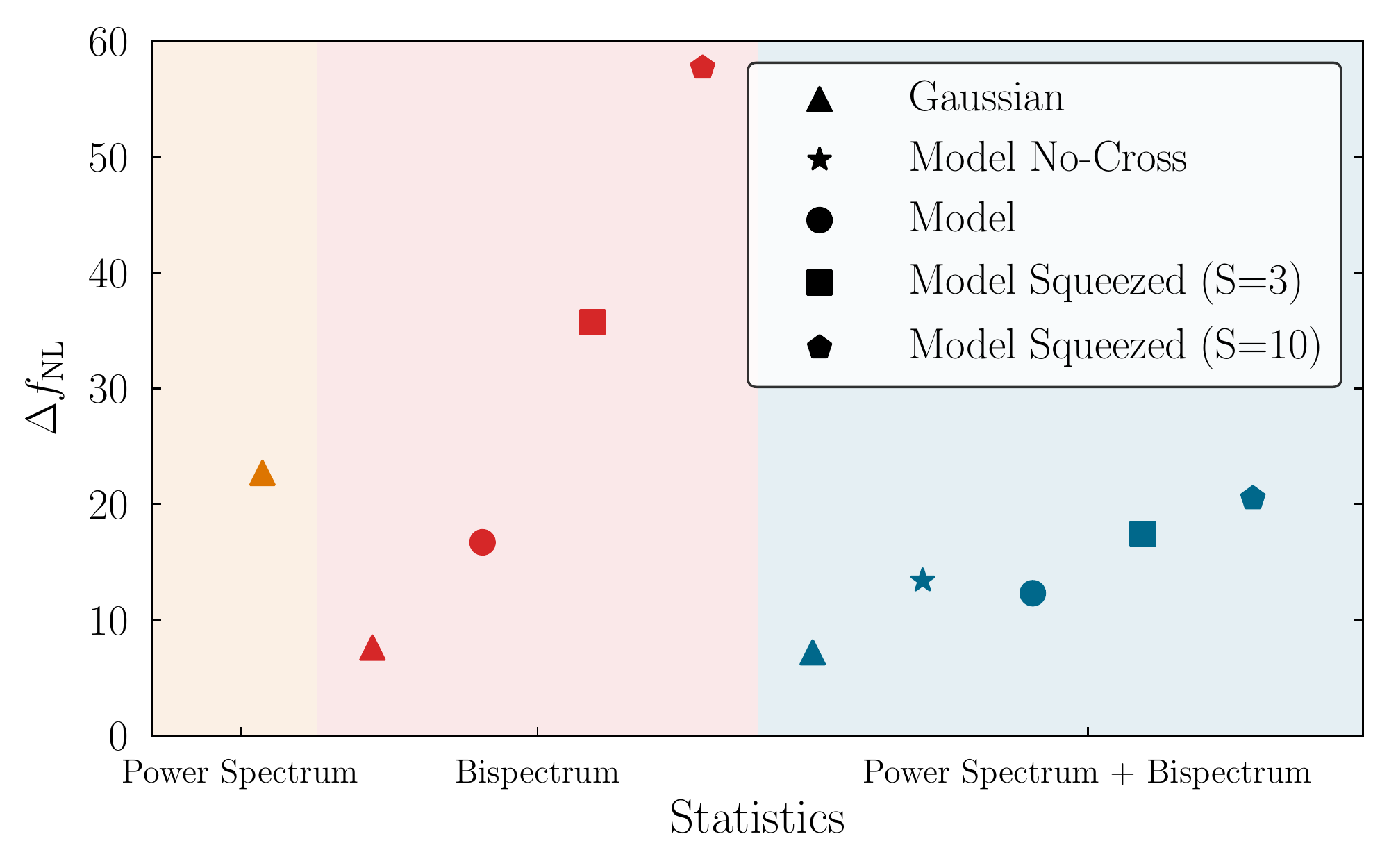}
\caption{\textbf{Predicted uncertainties ($1$-$\sigma$) on $\fnl$}. Power spectrum (Orange), bispectrum (Red) and joint power spectrum-bispectrum (Blue) uncertainties for different prescriptions we took for the covariance matrix. } \label{fig:results}
\end{figure}

The predicted uncertainty on $\fnl$ is given by $\Delta \fnl = 1/F_{\fnl}^{1/2}$. We compute $\Delta \fnl$ for four different combinations of terms in the covariance matrix
\begin{itemize}
    \item {\bfseries \textsf{Gaussian}}: Numerical diagonal covariance matrix 
    \item {\bfseries \textsf{Model}}: The full covariance matrix.
    \item {\bfseries \textsf{Model Squeezed (S=N)}}: The full covariance matrix, where we include only squeezed triangles with a squeezing $N k_L < k_S$ for two values of $N = 3$ and $10$.
    \item {\bfseries \textsf{Model No-Cross}}: Covariance matrix setting the P-B cross-covariance to zero, but including off-diagonal terms in the bispectrum covariance
\end{itemize}
We show the results of the Fisher matrix analysis in Figure \ref{fig:results} and we quote all uncertainties in a Table \ref{tab:results}. As expected, including the bispectrum off-diagonal terms has a strong impact on constraints, degrading the constraint on $\fnl$ by more than a factor of $2$ when considering the bispectrum alone and similarly for the joint power spectrum-bispectrum constraints. This result is in agreement with what found in \cite{Barreira:2020ekm} using a perturbative model. The cross power spectrum-bispectrum covariance does not have a sizable impact on the joint constraint, improving it only by $\sim 10\%$. 

\begin{table}[t]
\centering
\begin{tabular}{cccc}
    \hline\hline
    $\mbox{}$ & P  & B & P+B\\
    \hline\hline
    {\bfseries \textsf{Gaussian}} & $22.8 $ & $7.6$ & $7.2$ \\
    \hline
    {\bfseries \textsf{Model No-Cross}} & $22.8$ & $16.7$ & $13.45$ \\
    \hline
    {\bfseries \textsf{Model}} & $22.8$ & $16.7$ & $12.5$\\
    \hline
    {\bfseries \textsf{Model Squeezed (S=3)}} & $22.8$ & $35.7$ & $17.4$\\
    \hline
    {\bfseries \textsf{Model Squeezed (S=10)}} & $22.8$ & $57.7$ & $20.5$\\
    \hline\hline
\end{tabular}%
\caption{$1$-$\sigma$ uncertainties on $\fnl$ from the four different covariance matrices considered. Each column represents constraints from the power spectrum alone, the bispectrum alone and the joint power spectrum-bispectrum, respectively. The covariance matrix is on the set \textsf{G85L}, which has $\fnl=0$.  The table (and related figure \ref{fig:results}) shows that the Gaussian covariance underestimates the uncertainty on $\fnl$ by a factor of $\approx 2$ with respect to our model covariance, eq. \eqref{eq:Total_BB}. Interestingly, including  the cross bispectrum-power spectrum covariance in the model slightly improves the constraint over a model where this cross-covariance is neglected.}
\label{tab:results}
\end{table}

\section{Conclusions}
\label{sec:conclusions}

We studied the halo bispectrum covariance with a robust numerical estimate obtained from a large set of over 2,000 \quijote simulations. We compared such estimate with an approximate theoretical model using response function arguments. We thus identified the regime where non-Gaussian terms are most important in the bispectrum covariance: squeezed configurations. We pay particular attention to the accurate evaluation of numerical factors related to mode counts in the power spectrum and bispectrum estimators. In addition to the complete measurement of all triangular configurations at large scales, we provide measurements for squeezed configurations where short-wavelength mode extends deep into the nonlinear regime, $k\sim 1\kMpc$, for the first time. 

We can summarize our findings as follows:
\begin{itemize}
    \item The Gaussian approximation for the bispectrum variance does not work when squeezed triangular configurations are involved (see e.g. Figures~\ref{fig:varianceQ1} and \ref{fig:varianceQ2}). Non-Gaussian contributions can exceed by an order of magnitude the Gaussian one, as already pointed out by \cite{dePutter:2018jqk} for a different (though related) observable, and \cite{Barreira:2019icq} for the matter bispectrum case. 
    \item For the same reason, off-diagonal contributions to the bispectrum covariance are particularly large, with correlation coefficients of $\mathcal{O}(1)$, for squeezed configurations. Even mildly squeezed triangles (where $k_L \lesssim 3 k_s$) are highly correlated when they share the same long mode. 
    \item Off-diagonal contributions to the power spectrum-bispectrum cross-covariance can also be large for squeezed triangles correlated to the power spectrum of the long mode. This leads to correlation coefficients of order one for the full $P+B$ covariance matrix. This was pointed out by \cite{dePutter:2018jqk, Barreira:2019icq}, who, however, did not compare with simulations.
    \item We roughly estimate the size of each contribution to the covariance. We then simplify the largest terms using response function arguments. We thus find a simple prescription, Eq.~\eqref{eq:Total_BB}. This is similar to that adopted by \cite{Barreira:2020ekm, Alkhanishvili:2021pvy}, who included the factor of $2$ in front of the $BB$ term, only as a way to estimate the size of the $PT$ term. We also give a simple prescription to invert the full covariance matrix.
    \item Our implementation of Eq.~\eqref{eq:Total_BB}, taking advantage of direct measurements of the power spectrum and bispectrum from simulations shows an accuracy better than 20\%, not restricted to squeezed triangles. For triangles closer to equilateral, and at small scales, this degrades but does not exceed 40\% (see Figure~\ref{fig:Full_Variance}). 
    \item It is crucial, in order to obtain an accurate theoretical description of the full $P+B$ covariance, to carefully evaluate the mode-counting factors that enter its expression. We provide an analytic approximation to the exact sums over the grid wave-numbers.  
    \item We explicitly quantify how a fully non-Gaussian prescription for the bispectrum covariance can affect constraints on parameters such as the amplitude of local primordial non-Gaussianity $\fnl$. This case is particularly interesting, since most of the information on $\fnl$ is encoded in squeezed configurations \cite{Sefusatti:2011gt}. We find that the uncertainty on $\fnl$ almost doubles with respect to the Gaussian diagonal covariance case. The power spectrum-bispectrum cross-covariance, on the other hand, can reduce it by $\sim 10 \%$.
\end{itemize}

Clearly, an alternative implementation of our theoretical prescription can be given in terms of perturbation theory (e.g. \cite{Eggemeier:2018qae}) or some fitting function (e.g. \cite{Scoccimarro:2000ee, Gil-Marin:2011jtv, Takahashi:2019hth}), since it should capture nonlinear corrections at small scales. We stress that Eq.~\eqref{eq:Total_BB} constitutes a relatively simple extension of the Gaussian variance. It provides a much better approximation to the full covariance that can find application to likelihood and Fisher analyses (e.g. \cite{Alkhanishvili:2021pvy, Gualdi:2021yvq, Eggemeier:2021cam, Ivanov:2021kcd}) or to tests of consistency relations \cite{dePutter:2018jqk, Esposito:2019jkb, Marinucci:2019wdb, Marinucci:2020weg}. These results could be affected by additional non-Gaussian contributions to the covariance, although in the realistic setting of an actual redshift survey the difference might be less relevant than in our ideal case (e.g. \cite{Wadekar:2019rdu} for the power spectrum case).

Of course, several improvements can be considered. For instance, it would be straightforward to include shot-noise contribution to the 5- and 6-point correlation functions \cite{Sugiyama:2019ike}. We leave this for future work, along with the extension to redshift-space, preferring to highlight here the range of validity of the simplest non-Gaussian model. Furthermore, we can use this covariance to assess the power of very squeezed configurations to constrain $\fnl$, see e.g.~\cite{dePutter:2018jqk,Esposito:2019jkb}.

\section*{Acknowledgments}

The authors would like to thank An\v{z}e Slosar for useful comments. We would also like to thank Francisco Villaescusa-Navarro and the whole \quijote team for making the simulation suite available.
M.B acknowledges support from the Netherlands Organization for Scientific Research (NWO), which is funded by the Dutch Ministry of Education, Culture and Science (OCW) under VENI grant 016.Veni.192.210. M.B. also acknowledges support from the NWO project ``Cosmic origins from simulated
universes'' for the computing time allocated to run a subset of the
\eos simulations on \textsc{Cartesius}, a supercomputer that is part of the Dutch National Computing Facilities.
L.C. was supported by ANID scholarship No.21190484 and is supported by "Beca posgrado PUCV, termino de tesis, 2021".
J.N. is supported by FONDECYT grant 1211545, ``Measuring the Field Spectrum of the Early Universe''. E.S. is partially supported by the INFN INDARK PD51 grant.

\appendix

\section{Integrals for open and flat triangles}
\label{appendix_integrals}

We list here the coefficients $\Sigma^{11}_{ij}$ appearing in the expression for the covariance \eqref{appeqCBB} (for the terms satisfying $k_1^i = k_1^j$). In section~\ref{section_sums_approx} we outlined how they can be computed. In all these expressions we take $k_3 > k_2 > k_1$, and we used $k_3 = k_1 + k_2$ for flattened configurations, and $k_3 = k_1 + k_2 + \Delta k$ for open configurations.

The coefficient for the covariance between {\bf two flattened triangles} is
\begin{align}
N_{tr}^{i} N_{tr}^{j} \Sigma^{11}_{ij} &= \frac{\pi ^3 \Delta k^5}{2903040 k_f^9} \bigg[144 \Delta k^2 \left(867 k_1 \left(k^i_2+k^j_2\right)+867 \left((k^i_2)^2+(k^j_2)^2\right)+1532 k_1^2\right) \nn \\
&\phantom{= \frac{\pi ^3 \Delta k^5}{2903040} \bigg[}+1572480 k_1 \Delta k \left(k_1 \left(k^i_2+k^j_2\right)+(k^i_2)^2+(k^j_2)^2\right) \nn\\
&\phantom{= \frac{\pi ^3 \Delta k^5}{2903040} \bigg[}+14176512 k^i_2 \left(k_1+k^i_2\right) k^j_2 \left(k_1+k^j_2\right)+6347 \Delta k^4\bigg]\,.
\end{align}

The coefficient for the covariance between {\bf two open triangles} is
\begin{align}
N_{tr}^{i} N_{tr}^{j} \Sigma^{11}_{ij} &= \frac{\pi ^3}{10080 k_f^9} \bigg[\frac{3}{32} \Delta k^8 \left(632 \left(k^i_2+k^j_2\right)+457 k_1\right)\nn\\
&\phantom{= \frac{\pi ^3}{10080} \bigg[}+ \frac{1}{4} \Delta k^7 \left(601 k_1 \left(k^i_2+k^j_2\right)+3 \left(336 k^i_2 k^j_2+79(k^i_2)^2+79 (k^j_2)^2\right)+132 k_1^2\right)\nn\\
&\phantom{= \frac{\pi ^3}{10080} \bigg[}+7 \Delta k^6 \left(13 k_1^2 \left(k^i_2+k^j_2\right)+k_1 \left(72 k^i_2 k^j_2+13 (k^i_2)^2+13 (k^j_2)^2\right)+36 k^i_2 k^j_2
\left(k^i_2+k^j_2\right)\right)\nn\\
&\phantom{= \frac{\pi ^3}{10080} \bigg[}+252 k^i_2 \left(k_1+k^i_2\right) k^j_2 \Delta k^5 \left(k_1+k^j_2\right)+\frac{8027 \Delta k^9}{576}\bigg]\,.
\end{align}

The coefficient for the covariance between {\bf a flattened triangle ($i$) and an open triangle ($j$)} is
\begin{align}
N_{tr}^{i} N_{tr}^{j} \Sigma^{11}_{ij} &= \frac{\pi ^3 \Delta k^5}{5806080 k_f^9} \bigg[54 \Delta k^3 \left(663 k_1+488 k^j_2\right)\nn\\
&\phantom{= \frac{\pi ^3 \Delta k^5}{5806080} \bigg[} + 144 \Delta k^2 \left(939 k_1 k^j_2+288 k_1^2+2549 k^i_2 k_1+2549 (k^i_2)^2+183 (k^j_2)^2\right) \nn\\
&\phantom{= \frac{\pi ^3 \Delta k^5}{5806080} \bigg[} +4032 \Delta k
   \left(27 k_1 (k^j_2)^2+3 \left(9 k_1^2+128 k^i_2 k_1+128 (k^i_2)^2\right) k^j_2+143 k_1 k^i_2 \left(k_1+k^i_2\right)\right) \nn\\
&\phantom{= \frac{\pi ^3 \Delta k^5}{5806080} \bigg[}+1548288 k^i_2 \left(k_1+k^i_2\right) k^j_2 \left(k_1+k^j_2\right)+6253 \Delta k^4\bigg]\,.
\end{align}

The coefficient for the covariance between {\bf a closed triangle ($i$) and an open triangle ($j$)} is
\be
N_{tr}^{i} N_{tr}^{j} \Sigma^{11}_{ij} = \frac{1}{240 k_f^9} \pi ^3 k^i_2 k^i_3 \Delta k^5 \left(10 \Delta k \left(3 k_1+8 k^j_2\right)+80 k^j_2 \left(k_1+k^j_2\right)+19 \Delta k^2\right)\,.
\ee

The coefficient for the covariance between {\bf a closed triangle ($i$) and a flattened triangle ($j$)} is
\be
N_{tr}^{i} N_{tr}^{j} \Sigma^{11}_{ij} = \frac{1}{12 k_f^9} \pi ^3 k^i_2 k^i_3 \Delta k^5 \left(96 k^j_2 \left(k_1+k^j_2\right)+13 k_1 \Delta k\right)\,.
\ee

\section{Cross-correlation matrix with correlated noise}
\label{app:noise}

\begin{figure}[H]
\includegraphics[width=0.5\textwidth]{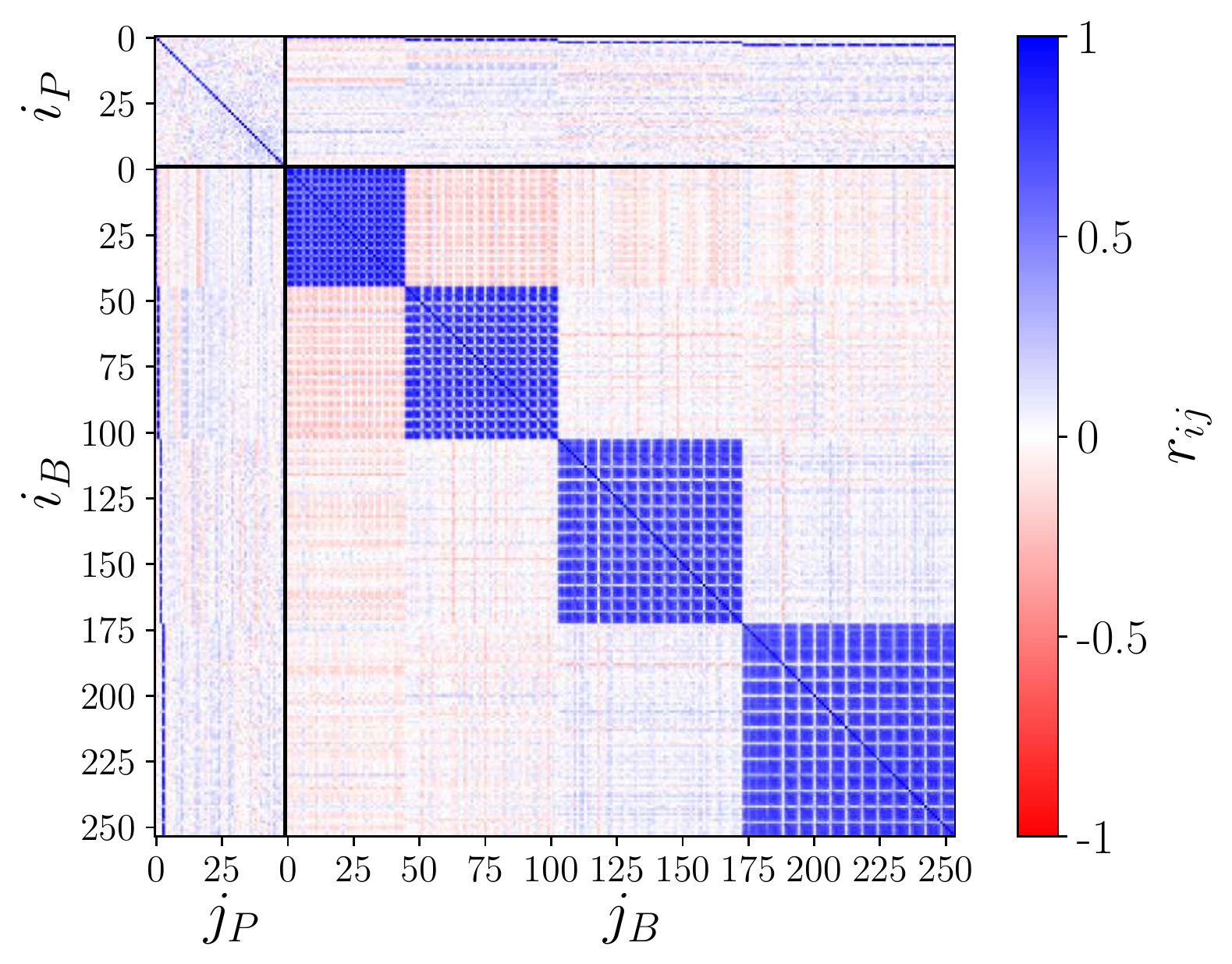}
\includegraphics[width=0.5\textwidth]{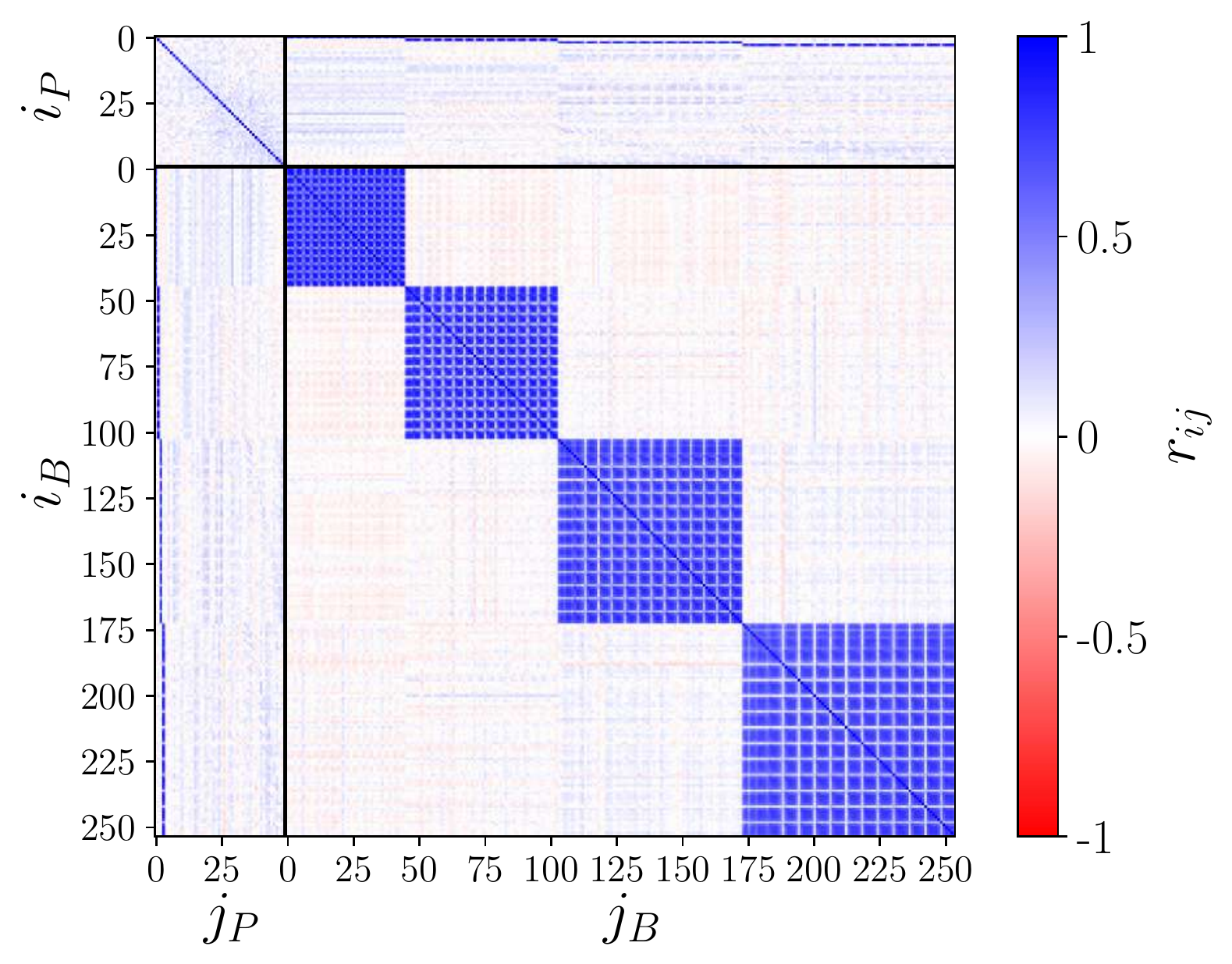}
\includegraphics[width=0.5\textwidth]{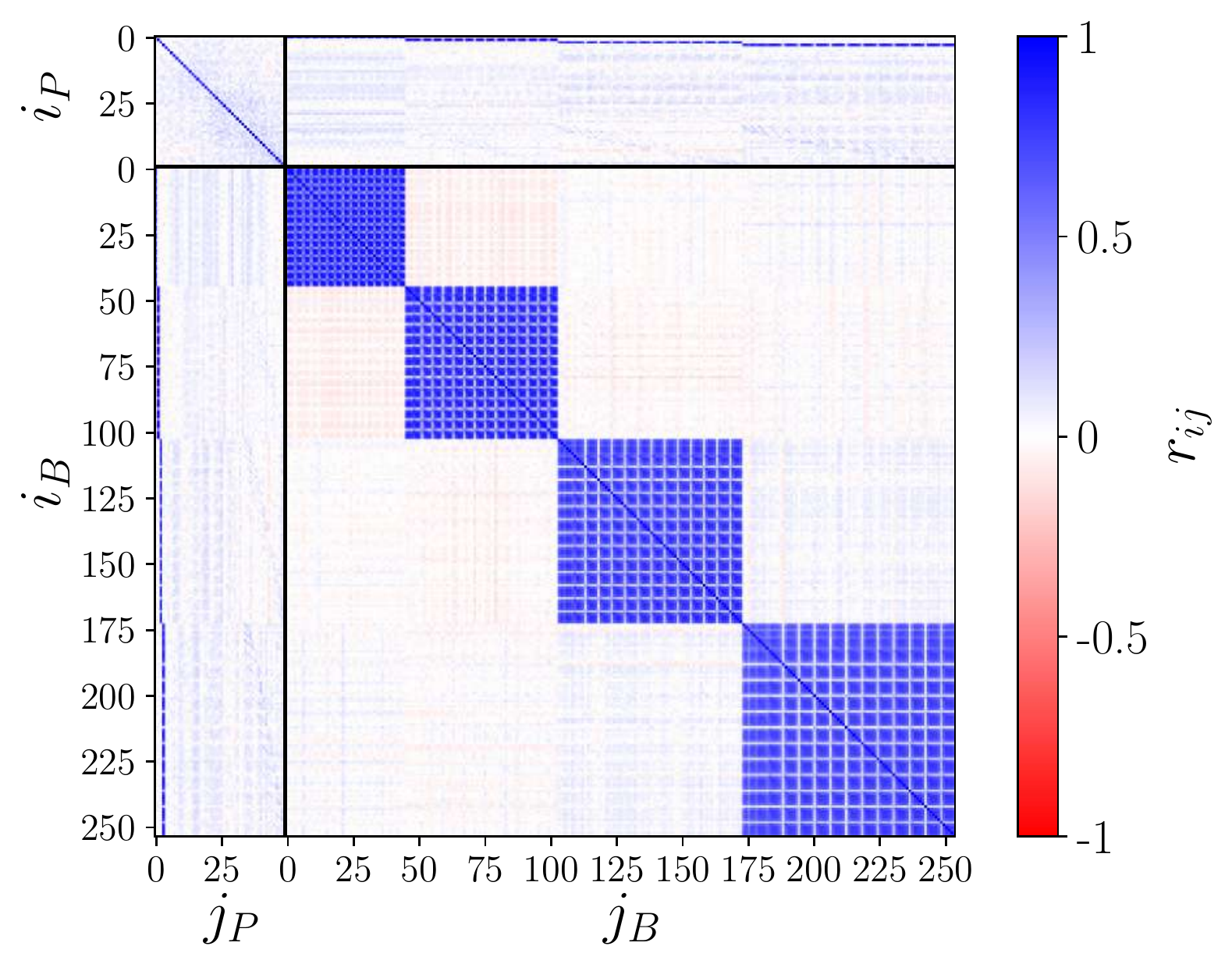}
\includegraphics[width=0.5\textwidth]{Final_Plots/Full_correlation_matrix_data_1000_t.pdf}
\caption{\textbf{Model cross correlation matrix for Q2 set}, as in Figure~\ref{fig:CM}, considering different numbers of realizations. Upper left: 100 realizations. upper right: 300 realizations. Bottom left: 500 realizations. Bottom right: 1000 realizations.}\label{fig:CM__E2}
\end{figure}

Measurements of the covariance matrix need a large number of simulations. Using few simulations makes the measurement noisy and unstable, with noise that is {\em highly correlated}. We illustrate this in Figure~\ref{fig:CM__E2}. We show the correlation matrix measured from simulations using an increasingly large number of realizations. Note that for a few realizations there are large regions far from the diagonal where the matrix seems to be systematically negative (red) or systematically positive (blue) rather than randomly distributed around zero (as expected from uncorrelated noise).\footnote{One could also check whether the noise is Gaussian by performing a Kolmogorov-Smirnov test. We deserve a more thorough check of the noise for future projects.} This structure gradually disappears as the number of realizations increases.

\section{Half-inverse test}
\label{app:hitest}

\begin{figure}[H]
\centering
\includegraphics[width=0.46\textwidth]{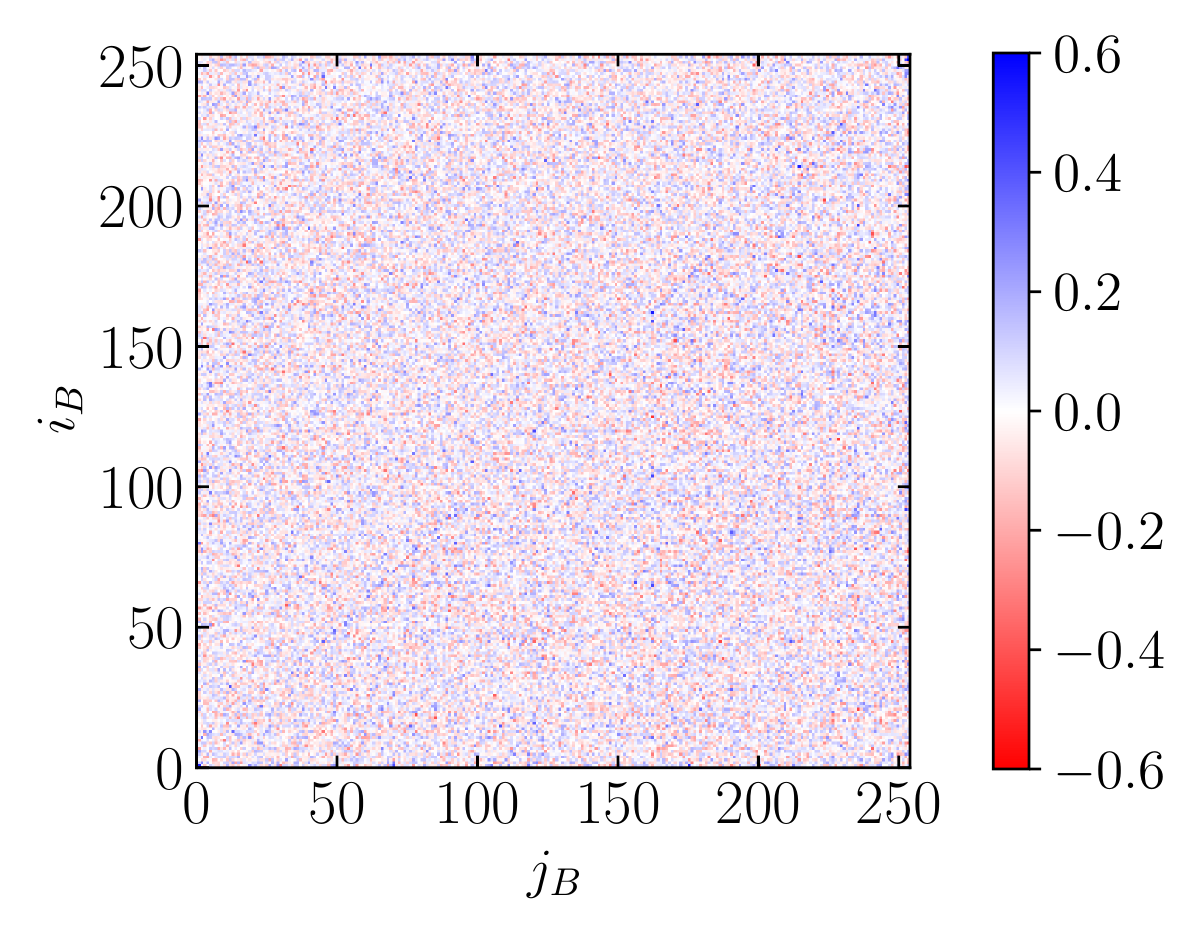}
\includegraphics[width=0.46\textwidth]{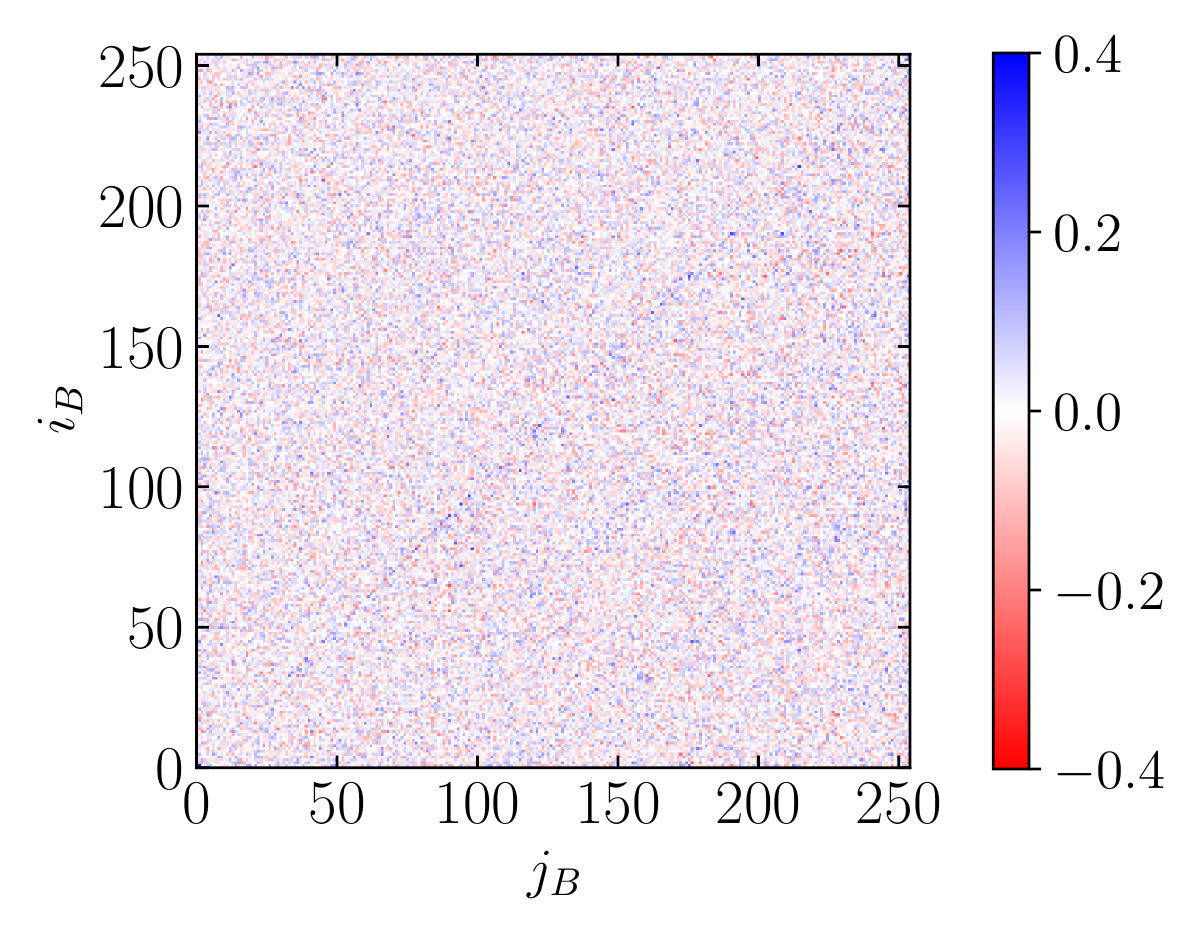}
\includegraphics[width=0.46\textwidth]{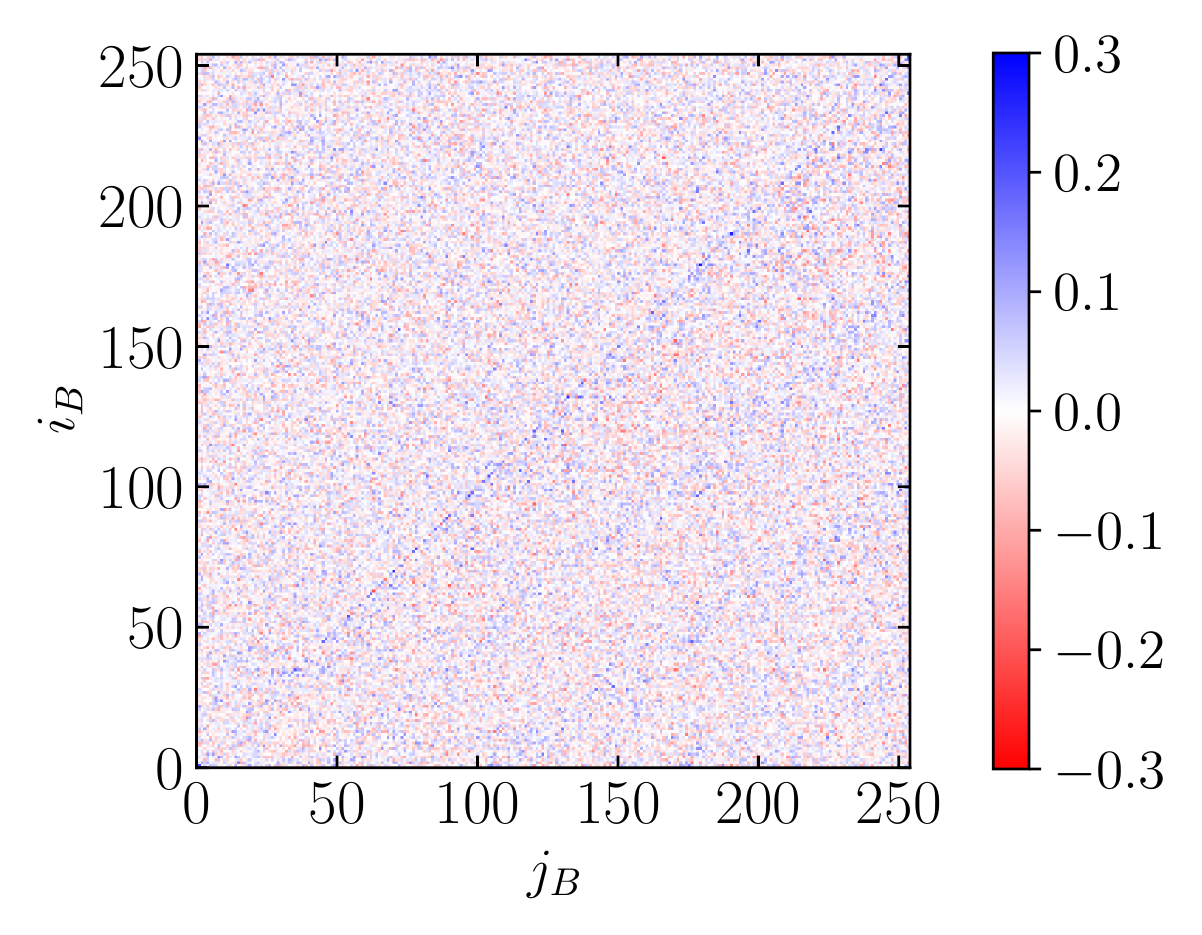}
\includegraphics[width=0.46\textwidth]{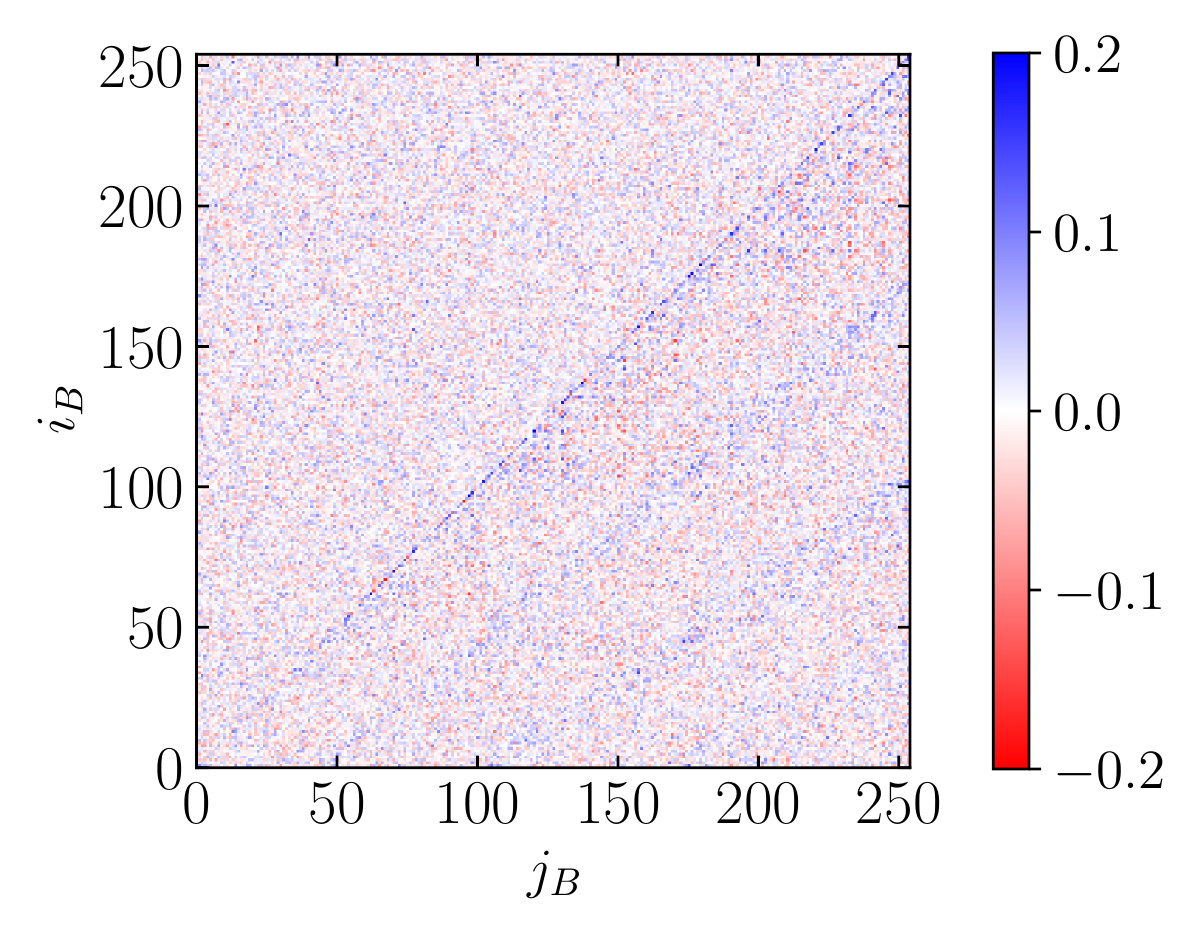}
\caption{\textbf{Half-inverse test for the \qtwo set}, considering different numbers of realizations. Upper left: $100$ realizations. upper right: $250$ realizations. Bottom left: $500$ realizations. Bottom right: $1000$ realizations.}\label{fig:halfi}
\end{figure}

As a further check of the inverse, we can compute the relation  $\mathcal H = \mathbf{C}_{th}^{-1/2}\,\mathbf{C}_{N-body}\,\mathbf{C}_{th}^{-1/2} - \mathbf{I}$ , where $\mathbf{C}_{th}$ is our model covariance and $\mathbf{C}_{N-body}$ is the covariance computed from the simulations \cite{Slepian:2015hca}. If the model and covariance computed from the simulations were identical, this relation should be zero. However, we expect that our model is accurate within $\approx 20\%$ of the covariance made out of a very large number of realizations. Following \cite{Hou:2022wfj}, we can check whether the noise scales as $1/\sqrt{N_{sims}}$. Since $\mathcal{H}$ is symmetric by construction, we can plot only the lower half-triangle of $\mathcal{H}$ and compare it to a noise scaling exactly as $1/\sqrt{N_{ sims}}$, which we put on the upper half-triangle of the same matrix. We show this check for the \qtwo set of simulations for $N_{sims}=100$, $250$, $500$ and $1000$ realizations in Figure \ref{fig:halfi}\footnote{Note that, for ease of visualization, in this case the triangle index is growing from bottom to top in the y-axis, differently than previous plots.}. It shows that increasing the number of simulations, (and therefore lowering the noise which scales as $1/\sqrt{N_{\rm sims}}$), an additional structure emerges, which is particularly evident on the diagonal (lower right panel). There is also further structure on off-diagonal cells parallel to the diagonal. We reserve a better study of these features to future projects.

\bibliographystyle{utphys}
\bibliography{references.bib}

\providecommand{\href}[2]{#2}\begingroup\raggedright\begin{thebibliography}{100}

\bibitem{EUCLID:2011zbd}
{\bfseries EUCLID} Collaboration, R.~Laureijs {\em et~al.}, ``{Euclid
  Definition Study Report},'' \href{http://arxiv.org/abs/1110.3193}{{\ttfamily
  arXiv:1110.3193 [astro-ph.CO]}}.

\bibitem{DESI:2016fyo}
{\bfseries DESI} Collaboration, A.~Aghamousa {\em et~al.}, ``{The DESI
  Experiment Part I: Science,Targeting, and Survey Design},''
  \href{http://arxiv.org/abs/1611.00036}{{\ttfamily arXiv:1611.00036
  [astro-ph.IM]}}.

\bibitem{Abate:2012za}
{\bfseries LSST Dark Energy Science} Collaboration, A.~Abate {\em et~al.},
  ``{Large Synoptic Survey Telescope: Dark Energy Science Collaboration},''
\href{http://arxiv.org/abs/1211.0310}{{\ttfamily arXiv:1211.0310
  [astro-ph.CO]}}.

\bibitem{Dore:2014cca}
O.~Dor\'e {\em et~al.}, ``{Cosmology with the SPHEREX All-Sky Spectral
  Survey},''
\href{http://arxiv.org/abs/1412.4872}{{\ttfamily arXiv:1412.4872
  [astro-ph.CO]}}.

\bibitem{PhysRevLett.73.215}
J.~N. Fry, ``Gravity, bias, and the galaxy three-point correlation function,''
  \href{http://dx.doi.org/10.1103/PhysRevLett.73.215}{{\em Phys. Rev. Lett.}
  {\bfseries 73} (Jul, 1994) 215--219}.
  \url{https://link.aps.org/doi/10.1103/PhysRevLett.73.215}.

\bibitem{Matarrese:1997sk}
S.~Matarrese, L.~Verde, and A.~F. Heavens, ``{Large scale bias in the universe:
  Bispectrum method},'' \href{http://dx.doi.org/10.1093/mnras/290.4.651}{{\em
  Mon. Not. Roy. Astron. Soc.} {\bfseries 290} (1997) 651--662},
  \href{http://arxiv.org/abs/astro-ph/9706059}{{\ttfamily
  arXiv:astro-ph/9706059}}.

\bibitem{Scoccimarro:2000sp}
R.~Scoccimarro, H.~A. Feldman, J.~N. Fry, and J.~A. Frieman, ``{The Bispectrum
  of IRAS redshift catalogs},'' \href{http://dx.doi.org/10.1086/318284}{{\em
  Astrophys. J.} {\bfseries 546} (2001) 652},
  \href{http://arxiv.org/abs/astro-ph/0004087}{{\ttfamily
  arXiv:astro-ph/0004087}}.

\bibitem{Scoccimarro:2000ee}
R.~Scoccimarro and H.~M.~P. Couchman, ``{A fitting formula for the nonlinear
  evolution of the bispectrum},''
  \href{http://dx.doi.org/10.1046/j.1365-8711.2001.04281.x}{{\em Mon. Not. Roy.
  Astron. Soc.} {\bfseries 325} (2001) 1312},
\href{http://arxiv.org/abs/astro-ph/0009427}{{\ttfamily arXiv:astro-ph/0009427
  [astro-ph]}}.

\bibitem{Takada:2002qq}
M.~Takada and B.~Jain, ``{The Three - point correlation function in
  cosmology},'' \href{http://dx.doi.org/10.1046/j.1365-8711.2003.06321.x}{{\em
  Mon. Not. Roy. Astron. Soc.} {\bfseries 340} (2003) 580--608},
  \href{http://arxiv.org/abs/astro-ph/0209167}{{\ttfamily
  arXiv:astro-ph/0209167}}.

\bibitem{Szapudi:2004gg}
I.~Szapudi, ``{Three - point statistics from a new perspective},''
  \href{http://dx.doi.org/10.1086/420894}{{\em Astrophys. J. Lett.} {\bfseries
  605} (2004) L89}, \href{http://arxiv.org/abs/astro-ph/0404476}{{\ttfamily
  arXiv:astro-ph/0404476}}.

\bibitem{Gaztanaga:2005an}
E.~Gaztanaga, P.~Norberg, C.~M. Baugh, and D.~J. Croton, ``{Statistical
  analysis of galaxy surveys. 2. The 3-point galaxy correlation function
  measured from the 2dFGRS},''
  \href{http://dx.doi.org/10.1111/j.1365-2966.2005.09583.x}{{\em Mon. Not. Roy.
  Astron. Soc.} {\bfseries 364} (2005) 620--634},
  \href{http://arxiv.org/abs/astro-ph/0506249}{{\ttfamily
  arXiv:astro-ph/0506249}}.

\bibitem{Nichol:2006mg}
R.~C. Nichol {\em et~al.}, ``{The effect of large-scale structure on the sdss
  galaxy three-point correlation function},''
  \href{http://dx.doi.org/10.1111/j.1365-2966.2006.10239.x}{{\em Mon. Not. Roy.
  Astron. Soc.} {\bfseries 368} (2006) 1507--1514},
  \href{http://arxiv.org/abs/astro-ph/0602548}{{\ttfamily
  arXiv:astro-ph/0602548}}.

\bibitem{Sefusatti:2006pa}
E.~Sefusatti, M.~Crocce, S.~Pueblas, and R.~Scoccimarro, ``{Cosmology and the
  Bispectrum},'' \href{http://dx.doi.org/10.1103/PhysRevD.74.023522}{{\em Phys.
  Rev. D} {\bfseries 74} (2006) 023522},
  \href{http://arxiv.org/abs/astro-ph/0604505}{{\ttfamily
  arXiv:astro-ph/0604505}}.

\bibitem{Sefusatti:2007ih}
E.~Sefusatti and E.~Komatsu, ``{The bispectrum of galaxies from high-redshift
  galaxy surveys: Primordial non-Gaussianity and non-linear galaxy bias},''
  \href{http://dx.doi.org/10.1103/PhysRevD.76.083004}{{\em Phys. Rev.}
  {\bfseries D76} (2007) 083004},
\href{http://arxiv.org/abs/0705.0343}{{\ttfamily arXiv:0705.0343 [astro-ph]}}.

\bibitem{Smith:2007sb}
R.~E. Smith, R.~K. Sheth, and R.~Scoccimarro, ``{An analytic model for the
  bispectrum of galaxies in redshift space},''
  \href{http://dx.doi.org/10.1103/PhysRevD.78.023523}{{\em Phys. Rev. D}
  {\bfseries 78} (2008) 023523},
  \href{http://arxiv.org/abs/0712.0017}{{\ttfamily arXiv:0712.0017
  [astro-ph]}}.

\bibitem{Liguori:2010hx}
M.~Liguori, E.~Sefusatti, J.~R. Fergusson, and E.~P.~S. Shellard, ``{Primordial
  non-Gaussianity and Bispectrum Measurements in the Cosmic Microwave
  Background and Large-Scale Structure},''
  \href{http://dx.doi.org/10.1155/2010/980523}{{\em Adv. Astron.} {\bfseries
  2010} (2010) 980523},
\href{http://arxiv.org/abs/1001.4707}{{\ttfamily arXiv:1001.4707
  [astro-ph.CO]}}.

\bibitem{Pollack:2011xp}
J.~E. Pollack, R.~E. Smith, and C.~Porciani, ``{Modelling large-scale halo bias
  using the bispectrum},''
  \href{http://dx.doi.org/10.1111/j.1365-2966.2011.20279.x}{{\em Mon. Not. Roy.
  Astron. Soc.} {\bfseries 420} (2012) 3469},
\href{http://arxiv.org/abs/1109.3458}{{\ttfamily arXiv:1109.3458
  [astro-ph.CO]}}.

\bibitem{Assassi:2014fva}
V.~Assassi, D.~Baumann, D.~Green, and M.~Zaldarriaga, ``{Renormalized Halo
  Bias},'' \href{http://dx.doi.org/10.1088/1475-7516/2014/08/056}{{\em JCAP}
  {\bfseries 1408} (2014) 056},
\href{http://arxiv.org/abs/1402.5916}{{\ttfamily arXiv:1402.5916
  [astro-ph.CO]}}.

\bibitem{Baldauf:2014qfa}
T.~Baldauf, L.~Mercolli, M.~Mirbabayi, and E.~Pajer, ``{The Bispectrum in the
  Effective Field Theory of Large Scale Structure},''
  \href{http://dx.doi.org/10.1088/1475-7516/2015/05/007}{{\em JCAP} {\bfseries
  05} (2015) 007}, \href{http://arxiv.org/abs/1406.4135}{{\ttfamily
  arXiv:1406.4135 [astro-ph.CO]}}.

\bibitem{Angulo:2014tfa}
R.~E. Angulo, S.~Foreman, M.~Schmittfull, and L.~Senatore, ``{The One-Loop
  Matter Bispectrum in the Effective Field Theory of Large Scale Structures},''
  \href{http://dx.doi.org/10.1088/1475-7516/2015/10/039}{{\em JCAP} {\bfseries
  10} (2015) 039}, \href{http://arxiv.org/abs/1406.4143}{{\ttfamily
  arXiv:1406.4143 [astro-ph.CO]}}.

\bibitem{McCullagh:2015oga}
N.~McCullagh, D.~Jeong, and A.~S. Szalay, ``{Toward accurate modelling of the
  non-linear matter bispectrum: standard perturbation theory and transients
  from initial conditions},''
  \href{http://dx.doi.org/10.1093/mnras/stv2525}{{\em Mon. Not. Roy. Astron.
  Soc.} {\bfseries 455} no.~3, (2016) 2945--2958},
  \href{http://arxiv.org/abs/1507.07824}{{\ttfamily arXiv:1507.07824
  [astro-ph.CO]}}.

\bibitem{Lazanu:2015rta}
A.~Lazanu, T.~Giannantonio, M.~Schmittfull, and E.~P.~S. Shellard, ``{The
  matter bispectrum of large-scale structure: three-dimensional comparison
  between theoretical models and numerical simulations},''
\href{http://arxiv.org/abs/1510.04075}{{\ttfamily arXiv:1510.04075
  [astro-ph.CO]}}.

\bibitem{deBelsunce:2018xtd}
R.~de~Belsunce and L.~Senatore, ``{Tree-Level Bispectrum in the Effective Field
  Theory of Large-Scale Structure extended to Massive Neutrinos},''
  \href{http://dx.doi.org/10.1088/1475-7516/2019/02/038}{{\em JCAP} {\bfseries
  02} (2019) 038}, \href{http://arxiv.org/abs/1804.06849}{{\ttfamily
  arXiv:1804.06849 [astro-ph.CO]}}.

\bibitem{Sugiyama:2018yzo}
N.~S. Sugiyama, S.~Saito, F.~Beutler, and H.-J. Seo, ``{A complete FFT-based
  decomposition formalism for the redshift-space bispectrum},''
  \href{http://dx.doi.org/10.1093/mnras/sty3249}{{\em Mon. Not. Roy. Astron.
  Soc.} {\bfseries 484} no.~1, (2019) 364--384},
  \href{http://arxiv.org/abs/1803.02132}{{\ttfamily arXiv:1803.02132
  [astro-ph.CO]}}.

\bibitem{Eggemeier:2018qae}
A.~Eggemeier, R.~Scoccimarro, and R.~E. Smith, ``{Bias Loop Corrections to the
  Galaxy Bispectrum},''
  \href{http://dx.doi.org/10.1103/PhysRevD.99.123514}{{\em Phys. Rev. D}
  {\bfseries 99} no.~12, (2019) 123514},
  \href{http://arxiv.org/abs/1812.03208}{{\ttfamily arXiv:1812.03208
  [astro-ph.CO]}}.

\bibitem{Floerchinger:2019eoj}
S.~Floerchinger, M.~Garny, A.~Katsis, N.~Tetradis, and U.~A. Wiedemann, ``{The
  dark matter bispectrum from effective viscosity and one-particle irreducible
  vertices},'' \href{http://dx.doi.org/10.1088/1475-7516/2019/09/047}{{\em
  JCAP} {\bfseries 09} (2019) 047},
  \href{http://arxiv.org/abs/1907.10729}{{\ttfamily arXiv:1907.10729
  [astro-ph.CO]}}.

\bibitem{Steele:2020tak}
T.~Steele and T.~Baldauf, ``{Precise Calibration of the One-Loop Bispectrum in
  the Effective Field Theory of Large Scale Structure},''
  \href{http://dx.doi.org/10.1103/PhysRevD.103.023520}{{\em Phys. Rev. D}
  {\bfseries 103} no.~2, (2021) 023520},
  \href{http://arxiv.org/abs/2009.01200}{{\ttfamily arXiv:2009.01200
  [astro-ph.CO]}}.

\bibitem{Alkhanishvili:2021pvy}
D.~Alkhanishvili, C.~Porciani, E.~Sefusatti, M.~Biagetti, A.~Lazanu, A.~Oddo,
  and V.~Yankelevich, ``{The reach of next-to-leading-order perturbation theory
  for the matter bispectrum},''
  \href{http://arxiv.org/abs/2107.08054}{{\ttfamily arXiv:2107.08054
  [astro-ph.CO]}}.

\bibitem{Schmittfull:2014tca}
M.~Schmittfull, T.~Baldauf, and U.~Seljak, ``{Near optimal bispectrum
  estimators for large-scale structure},''
  \href{http://dx.doi.org/10.1103/PhysRevD.91.043530}{{\em Phys. Rev. D}
  {\bfseries 91} no.~4, (2015) 043530},
  \href{http://arxiv.org/abs/1411.6595}{{\ttfamily arXiv:1411.6595
  [astro-ph.CO]}}.

\bibitem{Karagiannis:2018jdt}
D.~Karagiannis, A.~Lazanu, M.~Liguori, A.~Raccanelli, N.~Bartolo, and L.~Verde,
  ``{Constraining primordial non-Gaussianity with bispectrum and power spectrum
  from upcoming optical and radio surveys},''
  \href{http://dx.doi.org/10.1093/mnras/sty1029}{{\em Mon. Not. Roy. Astron.
  Soc.} {\bfseries 478} no.~1, (2018) 1341--1376},
  \href{http://arxiv.org/abs/1801.09280}{{\ttfamily arXiv:1801.09280
  [astro-ph.CO]}}.

\bibitem{Yankelevich:2018uaz}
V.~Yankelevich and C.~Porciani, ``{Cosmological information in the
  redshift-space bispectrum},''
  \href{http://dx.doi.org/10.1093/mnras/sty3143}{{\em Mon. Not. Roy. Astron.
  Soc.} {\bfseries 483} no.~2, (2019) 2078--2099},
  \href{http://arxiv.org/abs/1807.07076}{{\ttfamily arXiv:1807.07076
  [astro-ph.CO]}}.

\bibitem{Gualdi:2021yvq}
D.~Gualdi, H.~Gil-Marin, and L.~Verde, ``{Joint analysis of anisotropic power
  spectrum, bispectrum and trispectrum: application to N-body simulations},''
  \href{http://arxiv.org/abs/2104.03976}{{\ttfamily arXiv:2104.03976
  [astro-ph.CO]}}.

\bibitem{Verde:2001sf}
L.~Verde {\em et~al.}, ``{The 2dF Galaxy Redshift Survey: The Bias of galaxies
  and the density of the Universe},''
  \href{http://dx.doi.org/10.1046/j.1365-8711.2002.05620.x}{{\em Mon. Not. Roy.
  Astron. Soc.} {\bfseries 335} (2002) 432},
  \href{http://arxiv.org/abs/astro-ph/0112161}{{\ttfamily
  arXiv:astro-ph/0112161}}.

\bibitem{Gil-Marin:2014sta}
H.~Gil-Mar\'in, J.~Nore\~na, L.~Verde, W.~J. Percival, C.~Wagner, M.~Manera,
  and D.~P. Schneider, ``{The power spectrum and bispectrum of SDSS DR11 BOSS
  galaxies – I. Bias and gravity},''
  \href{http://dx.doi.org/10.1093/mnras/stv961}{{\em Mon. Not. Roy. Astron.
  Soc.} {\bfseries 451} no.~1, (2015) 539--580},
\href{http://arxiv.org/abs/1407.5668}{{\ttfamily arXiv:1407.5668
  [astro-ph.CO]}}.

\bibitem{Gil-Marin:2014baa}
H.~Gil-Mar\'\i{}n, L.~Verde, J.~Nore\~na, A.~J. Cuesta, L.~Samushia, W.~J.
  Percival, C.~Wagner, M.~Manera, and D.~P. Schneider, ``{The power spectrum
  and bispectrum of SDSS DR11 BOSS galaxies \textendash{} II. Cosmological
  interpretation},'' \href{http://dx.doi.org/10.1093/mnras/stv1359}{{\em Mon.
  Not. Roy. Astron. Soc.} {\bfseries 452} no.~2, (2015) 1914--1921},
  \href{http://arxiv.org/abs/1408.0027}{{\ttfamily arXiv:1408.0027
  [astro-ph.CO]}}.

\bibitem{Slepian:2015hca}
Z.~Slepian {\em et~al.}, ``{The large-scale 3-point correlation function of the
  SDSS BOSS DR12 CMASS galaxies},''
\href{http://arxiv.org/abs/1512.02231}{{\ttfamily arXiv:1512.02231
  [astro-ph.CO]}}.

\bibitem{Gil-Marin:2016wya}
H.~Gil-Mar\'\i{}n, W.~J. Percival, L.~Verde, J.~R. Brownstein, C.-H. Chuang,
  F.-S. Kitaura, S.~A. Rodr\'\i{}guez-Torres, and M.~D. Olmstead, ``{The
  clustering of galaxies in the SDSS-III Baryon Oscillation Spectroscopic
  Survey: RSD measurement from the power spectrum and bispectrum of the DR12
  BOSS galaxies},'' \href{http://dx.doi.org/10.1093/mnras/stw2679}{{\em Mon.
  Not. Roy. Astron. Soc.} {\bfseries 465} no.~2, (2017) 1757--1788},
  \href{http://arxiv.org/abs/1606.00439}{{\ttfamily arXiv:1606.00439
  [astro-ph.CO]}}.

\bibitem{Slepian:2016kfz}
Z.~Slepian {\em et~al.}, ``{Detection of baryon acoustic oscillation features
  in the large-scale three-point correlation function of SDSS BOSS DR12 CMASS
  galaxies},'' \href{http://dx.doi.org/10.1093/mnras/stx488}{{\em Mon. Not.
  Roy. Astron. Soc.} {\bfseries 469} no.~2, (2017) 1738--1751},
  \href{http://arxiv.org/abs/1607.06097}{{\ttfamily arXiv:1607.06097
  [astro-ph.CO]}}.

\bibitem{Monaco:2016pys}
P.~Monaco, ``{Approximate methods for the generation of dark matter halo
  catalogs in the age of precision cosmology},''
  \href{http://dx.doi.org/10.3390/galaxies4040053}{{\em Galaxies} {\bfseries 4}
  no.~4, (2016) 53}, \href{http://arxiv.org/abs/1605.07752}{{\ttfamily
  arXiv:1605.07752 [astro-ph.CO]}}.

\bibitem{Colavincenzo:2018cgf}
M.~Colavincenzo {\em et~al.}, ``{Comparing approximate methods for mock
  catalogues and covariance matrices \textendash{} III: bispectrum},''
  \href{http://dx.doi.org/10.1093/mnras/sty2964}{{\em Mon. Not. Roy. Astron.
  Soc.} {\bfseries 482} no.~4, (2019) 4883--4905},
  \href{http://arxiv.org/abs/1806.09499}{{\ttfamily arXiv:1806.09499
  [astro-ph.CO]}}.

\bibitem{Scoccimarro:2000sn}
R.~Scoccimarro, ``{The bispectrum: from theory to observations},''
  \href{http://dx.doi.org/10.1086/317248}{{\em Astrophys. J.} {\bfseries 544}
  (2000) 597}, \href{http://arxiv.org/abs/astro-ph/0004086}{{\ttfamily
  arXiv:astro-ph/0004086}}.

\bibitem{Sefusatti:2004xz}
E.~Sefusatti and R.~Scoccimarro, ``{Galaxy bias and halo-occupation numbers
  from large-scale clustering},''
  \href{http://dx.doi.org/10.1103/PhysRevD.71.063001}{{\em Phys. Rev. D}
  {\bfseries 71} (2005) 063001},
  \href{http://arxiv.org/abs/astro-ph/0412626}{{\ttfamily
  arXiv:astro-ph/0412626}}.

\bibitem{Chan:2016ehg}
K.~C. Chan and L.~Blot, ``{Assessment of the Information Content of the Power
  Spectrum and Bispectrum},''
  \href{http://dx.doi.org/10.1103/PhysRevD.96.023528}{{\em Phys. Rev. D}
  {\bfseries 96} no.~2, (2017) 023528},
  \href{http://arxiv.org/abs/1610.06585}{{\ttfamily arXiv:1610.06585
  [astro-ph.CO]}}.

\bibitem{Byun:2017fkz}
J.~Byun, A.~Eggemeier, D.~Regan, D.~Seery, and R.~E. Smith, ``{Towards optimal
  cosmological parameter recovery from compressed bispectrum statistics},''
  \href{http://dx.doi.org/10.1093/mnras/stx1681}{{\em Mon. Not. Roy. Astron.
  Soc.} {\bfseries 471} no.~2, (2017) 1581--1618},
  \href{http://arxiv.org/abs/1705.04392}{{\ttfamily arXiv:1705.04392
  [astro-ph.CO]}}.

\bibitem{Chan:2017fiv}
K.~C. Chan, A.~Moradinezhad~Dizgah, and J.~Nore\~na, ``{Bispectrum Supersample
  Covariance},'' \href{http://dx.doi.org/10.1103/PhysRevD.97.043532}{{\em Phys.
  Rev. D} {\bfseries 97} no.~4, (2018) 043532},
  \href{http://arxiv.org/abs/1709.02473}{{\ttfamily arXiv:1709.02473
  [astro-ph.CO]}}.

\bibitem{Hahn:2019zob}
C.~Hahn, F.~Villaescusa-Navarro, E.~Castorina, and R.~Scoccimarro,
  ``{Constraining $M_\nu$ with the bispectrum. Part I. Breaking parameter
  degeneracies},'' \href{http://dx.doi.org/10.1088/1475-7516/2020/03/040}{{\em
  JCAP} {\bfseries 03} (2020) 040},
  \href{http://arxiv.org/abs/1909.11107}{{\ttfamily arXiv:1909.11107
  [astro-ph.CO]}}.

\bibitem{Oddo:2019run}
A.~Oddo, E.~Sefusatti, C.~Porciani, P.~Monaco, and A.~G. Sánchez, ``{Toward a
  robust inference method for the galaxy bispectrum: likelihood function and
  model selection},''
  \href{http://dx.doi.org/10.1088/1475-7516/2020/03/056}{{\em JCAP} {\bfseries
  03} (2020) 056}, \href{http://arxiv.org/abs/1908.01774}{{\ttfamily
  arXiv:1908.01774 [astro-ph.CO]}}.

\bibitem{Gualdi:2020ymf}
D.~Gualdi and L.~Verde, ``{Galaxy redshift-space bispectrum: the Importance of
  Being Anisotropic},''
  \href{http://dx.doi.org/10.1088/1475-7516/2020/06/041}{{\em JCAP} {\bfseries
  06} (2020) 041}, \href{http://arxiv.org/abs/2003.12075}{{\ttfamily
  arXiv:2003.12075 [astro-ph.CO]}}.

\bibitem{Sugiyama:2019ike}
N.~S. Sugiyama, S.~Saito, F.~Beutler, and H.-J. Seo, ``{Perturbation theory
  approach to predict the covariance matrices of the galaxy power spectrum and
  bispectrum in redshift space},''
  \href{http://dx.doi.org/10.1093/mnras/staa1940}{{\em Mon. Not. Roy. Astron.
  Soc.} {\bfseries 497} no.~2, (2020) 1684--1711},
  \href{http://arxiv.org/abs/1908.06234}{{\ttfamily arXiv:1908.06234
  [astro-ph.CO]}}.

\bibitem{Byun:2020rgl}
J.~Byun, A.~Oddo, C.~Porciani, and E.~Sefusatti, ``{Towards cosmological
  constraints from the compressed modal bispectrum: a robust comparison of
  real-space bispectrum estimators},''
  \href{http://dx.doi.org/10.1088/1475-7516/2021/03/105}{{\em JCAP} {\bfseries
  03} (2021) 105}, \href{http://arxiv.org/abs/2010.09579}{{\ttfamily
  arXiv:2010.09579 [astro-ph.CO]}}.

\bibitem{Hahn:2020lou}
C.~Hahn and F.~Villaescusa-Navarro, ``{Constraining $M_\nu$ with the
  bispectrum. Part II. The information content of the galaxy bispectrum
  monopole},'' \href{http://dx.doi.org/10.1088/1475-7516/2021/04/029}{{\em
  JCAP} {\bfseries 04} (2021) 029},
  \href{http://arxiv.org/abs/2012.02200}{{\ttfamily arXiv:2012.02200
  [astro-ph.CO]}}.

\bibitem{Oddo:2021iwq}
A.~Oddo, F.~Rizzo, E.~Sefusatti, C.~Porciani, and P.~Monaco, ``{Cosmological
  parameters from the likelihood analysis of the galaxy power spectrum and
  bispectrum in real space},''
  \href{http://arxiv.org/abs/2108.03204}{{\ttfamily arXiv:2108.03204
  [astro-ph.CO]}}.

\bibitem{Kitaura:2015uqa}
F.-S. Kitaura {\em et~al.}, ``{The clustering of galaxies in the SDSS-III
  Baryon Oscillation Spectroscopic Survey: mock galaxy catalogues for the BOSS
  Final Data Release},'' \href{http://dx.doi.org/10.1093/mnras/stv2826}{{\em
  Mon. Not. Roy. Astron. Soc.} {\bfseries 456} no.~4, (2016) 4156--4173},
  \href{http://arxiv.org/abs/1509.06400}{{\ttfamily arXiv:1509.06400
  [astro-ph.CO]}}.

\bibitem{Eggemeier:2021cam}
A.~Eggemeier, R.~Scoccimarro, R.~E. Smith, M.~Crocce, A.~Pezzotta, and A.~G.
  S\'anchez, ``{Testing one-loop galaxy bias: Joint analysis of power spectrum
  and bispectrum},'' \href{http://dx.doi.org/10.1103/PhysRevD.103.123550}{{\em
  Phys. Rev. D} {\bfseries 103} no.~12, (2021) 123550},
  \href{http://arxiv.org/abs/2102.06902}{{\ttfamily arXiv:2102.06902
  [astro-ph.CO]}}.

\bibitem{MoradinezhadDizgah:2020whw}
A.~Moradinezhad~Dizgah, M.~Biagetti, E.~Sefusatti, V.~Desjacques, and
  J.~Nore\~na, ``{Primordial Non-Gaussianity from Biased Tracers: Likelihood
  Analysis of Real-Space Power Spectrum and Bispectrum},''
  \href{http://dx.doi.org/10.1088/1475-7516/2021/05/015}{{\em JCAP} {\bfseries
  05} (2021) 015}, \href{http://arxiv.org/abs/2010.14523}{{\ttfamily
  arXiv:2010.14523 [astro-ph.CO]}}.

\bibitem{Ivanov:2021kcd}
M.~M. Ivanov, O.~H.~E. Philcox, T.~Nishimichi, M.~Simonovi\'c, M.~Takada, and
  M.~Zaldarriaga, ``{Precision analysis of the redshift-space galaxy
  bispectrum},'' \href{http://arxiv.org/abs/2110.10161}{{\ttfamily
  arXiv:2110.10161 [astro-ph.CO]}}.

\bibitem{Gualdi:2017iey}
D.~Gualdi, M.~Manera, B.~Joachimi, and O.~Lahav, ``{Maximal compression of the
  redshift space galaxy power spectrum and bispectrum},''
  \href{http://dx.doi.org/10.1093/mnras/sty261}{{\em Mon. Not. Roy. Astron.
  Soc.} {\bfseries 476} no.~3, (2018) 4045--4070},
  \href{http://arxiv.org/abs/1709.03600}{{\ttfamily arXiv:1709.03600
  [astro-ph.CO]}}.

\bibitem{Child:2018kec}
H.~L. Child, Z.~Slepian, and M.~Takada, ``{A Physical Picture of Bispectrum
  Baryon Acoustic Oscillations in the Interferometric Basis},''
  \href{http://arxiv.org/abs/1811.12396}{{\ttfamily arXiv:1811.12396
  [astro-ph.CO]}}.

\bibitem{Child:2018klv}
H.~L. Child, M.~Takada, T.~Nishimichi, T.~Sunayama, Z.~Slepian, S.~Habib, and
  K.~Heitmann, ``{Bispectrum as Baryon Acoustic Oscillation Interferometer},''
  \href{http://dx.doi.org/10.1103/PhysRevD.98.123521}{{\em Phys. Rev. D}
  {\bfseries 98} no.~12, (2018) 123521},
  \href{http://arxiv.org/abs/1806.11147}{{\ttfamily arXiv:1806.11147
  [astro-ph.CO]}}.

\bibitem{Gualdi:2018pyw}
D.~Gualdi, H.~Gil-Mar\'\i{}n, R.~L. Schuhmann, M.~Manera, B.~Joachimi, and
  O.~Lahav, ``{Enhancing BOSS bispectrum cosmological constraints with maximal
  compression},'' \href{http://dx.doi.org/10.1093/mnras/stz051}{{\em Mon. Not.
  Roy. Astron. Soc.} {\bfseries 484} no.~3, (2019) 3713--3730},
  \href{http://arxiv.org/abs/1806.02853}{{\ttfamily arXiv:1806.02853
  [astro-ph.CO]}}.

\bibitem{Gualdi:2019sfc}
D.~Gualdi, H.~Gil-Mar\'\i{}n, M.~Manera, B.~Joachimi, and O.~Lahav, ``{GEOMAX:
  beyond linear compression for three-point galaxy clustering statistics},''
  \href{http://dx.doi.org/10.1093/mnras/staa1941}{{\em Mon. Not. Roy. Astron.
  Soc.} {\bfseries 497} no.~1, (2020) 776--792},
  \href{http://arxiv.org/abs/1912.01011}{{\ttfamily arXiv:1912.01011
  [astro-ph.CO]}}.

\bibitem{Gualdi:2019ybt}
D.~Gualdi, H.~Gil-Mar\'\i{}n, M.~Manera, B.~Joachimi, and O.~Lahav,
  ``{Geometrical compression: a new method to enhance the BOSS galaxy
  bispectrum monopole constraints},''
  \href{http://dx.doi.org/10.1093/mnrasl/sly242}{{\em Mon. Not. Roy. Astron.
  Soc.} {\bfseries 484} no.~1, (2019) L29--L34},
  \href{http://arxiv.org/abs/1901.00987}{{\ttfamily arXiv:1901.00987
  [astro-ph.CO]}}.

\bibitem{Barreira:2019icq}
A.~Barreira, ``{The squeezed matter bispectrum covariance with responses},''
  \href{http://dx.doi.org/10.1088/1475-7516/2019/03/008}{{\em JCAP} {\bfseries
  03} (2019) 008}, \href{http://arxiv.org/abs/1901.01243}{{\ttfamily
  arXiv:1901.01243 [astro-ph.CO]}}.

\bibitem{Shirasaki:2020vkk}
M.~Shirasaki, N.~S. Sugiyama, R.~Takahashi, and F.-S. Kitaura, ``{Constraining
  primordial non-Gaussianity with postreconstructed galaxy bispectrum in
  redshift space},'' \href{http://dx.doi.org/10.1103/PhysRevD.103.023506}{{\em
  Phys. Rev. D} {\bfseries 103} no.~2, (2021) 023506},
  \href{http://arxiv.org/abs/2010.04567}{{\ttfamily arXiv:2010.04567
  [astro-ph.CO]}}.

\bibitem{Barreira:2021ueb}
A.~Barreira, ``{Predictions for local PNG bias in the galaxy power spectrum and
  bispectrum and the consequences for $f_{\rm NL}$ constraints},''
  \href{http://arxiv.org/abs/2107.06887}{{\ttfamily arXiv:2107.06887
  [astro-ph.CO]}}.

\bibitem{Wadekar:2019rdu}
D.~Wadekar and R.~Scoccimarro, ``{Galaxy power spectrum multipoles covariance
  in perturbation theory},''
  \href{http://dx.doi.org/10.1103/PhysRevD.102.123517}{{\em Phys. Rev. D}
  {\bfseries 102} no.~12, (2020) 123517},
  \href{http://arxiv.org/abs/1910.02914}{{\ttfamily arXiv:1910.02914
  [astro-ph.CO]}}.

\bibitem{Sefusatti:2006eu}
E.~Sefusatti, C.~Vale, K.~Kadota, and J.~Frieman, ``{Primordial non-Gaussianity
  and Dark Energy constraints from Cluster Surveys},''
  \href{http://dx.doi.org/10.1086/511331}{{\em Astrophys. J.} {\bfseries 658}
  (2007) 669--679},
\href{http://arxiv.org/abs/astro-ph/0609124}{{\ttfamily arXiv:astro-ph/0609124
  [astro-ph]}}.

\bibitem{Kayo:2012nm}
I.~Kayo, M.~Takada, and B.~Jain, ``{Information content of weak lensing power
  spectrum and bispectrum: including the non-Gaussian error covariance
  matrix},'' \href{http://dx.doi.org/10.1093/mnras/sts340}{{\em Mon. Not. Roy.
  Astron. Soc.} {\bfseries 429} (2013) 344--371},
  \href{http://arxiv.org/abs/1207.6322}{{\ttfamily arXiv:1207.6322
  [astro-ph.CO]}}.

\bibitem{Kayo:2013aha}
I.~Kayo and M.~Takada, ``{Cosmological parameters from weak lensing power
  spectrum and bispectrum tomography: including the non-Gaussian errors},''
  \href{http://arxiv.org/abs/1306.4684}{{\ttfamily arXiv:1306.4684
  [astro-ph.CO]}}.

\bibitem{Takada:2013wfa}
M.~Takada and W.~Hu, ``{Power Spectrum Super-Sample Covariance},''
  \href{http://dx.doi.org/10.1103/PhysRevD.87.123504}{{\em Phys. Rev. D}
  {\bfseries 87} no.~12, (2013) 123504},
  \href{http://arxiv.org/abs/1302.6994}{{\ttfamily arXiv:1302.6994
  [astro-ph.CO]}}.

\bibitem{Barreira:2017sqa}
A.~Barreira and F.~Schmidt, ``{Responses in Large-Scale Structure},''
  \href{http://dx.doi.org/10.1088/1475-7516/2017/06/053}{{\em JCAP} {\bfseries
  06} (2017) 053}, \href{http://arxiv.org/abs/1703.09212}{{\ttfamily
  arXiv:1703.09212 [astro-ph.CO]}}.

\bibitem{dePutter:2018jqk}
R.~de~Putter, ``{Primordial physics from large-scale structure beyond the power
  spectrum},'' \href{http://arxiv.org/abs/1802.06762}{{\ttfamily
  arXiv:1802.06762 [astro-ph.CO]}}.

\bibitem{Villaescusa-Navarro:2019bje}
F.~Villaescusa-Navarro {\em et~al.}, ``{The Quijote simulations},''
  \href{http://dx.doi.org/10.3847/1538-4365/ab9d82}{{\em Astrophys. J. Suppl.}
  {\bfseries 250} no.~1, (2020) 2},
  \href{http://arxiv.org/abs/1909.05273}{{\ttfamily arXiv:1909.05273
  [astro-ph.CO]}}.

\bibitem{Salopek:1990jq}
D.~S. Salopek and J.~R. Bond, ``{Nonlinear evolution of long wavelength metric
  fluctuations in inflationary models},''
\href{http://dx.doi.org/10.1103/PhysRevD.42.3936}{{\em Phys. Rev.} {\bfseries
  D42} (1990) 3936--3962}.

\bibitem{Bartolo:2001cw}
N.~Bartolo, S.~Matarrese, and A.~Riotto, ``{Nongaussianity from inflation},''
  \href{http://dx.doi.org/10.1103/PhysRevD.65.103505}{{\em Phys. Rev.}
  {\bfseries D65} (2002) 103505},
\href{http://arxiv.org/abs/hep-ph/0112261}{{\ttfamily arXiv:hep-ph/0112261
  [hep-ph]}}.

\bibitem{Bernardeau:2002jf}
F.~Bernardeau and J.-P. Uzan, ``{Inflationary models inducing non-Gaussian
  metric fluctuations},''
  \href{http://dx.doi.org/10.1103/PhysRevD.67.121301}{{\em Phys. Rev.}
  {\bfseries D67} (2003) 121301},
\href{http://arxiv.org/abs/astro-ph/0209330}{{\ttfamily arXiv:astro-ph/0209330
  [astro-ph]}}.

\bibitem{Bernardeau:2002jy}
F.~Bernardeau and J.-P. Uzan, ``{NonGaussianity in multifield inflation},''
  \href{http://dx.doi.org/10.1103/PhysRevD.66.103506}{{\em Phys. Rev.}
  {\bfseries D66} (2002) 103506},
\href{http://arxiv.org/abs/hep-ph/0207295}{{\ttfamily arXiv:hep-ph/0207295
  [hep-ph]}}.

\bibitem{Creminelli:2004yq}
P.~Creminelli and M.~Zaldarriaga, ``{Single field consistency relation for the
  3-point function},''
  \href{http://dx.doi.org/10.1088/1475-7516/2004/10/006}{{\em JCAP} {\bfseries
  0410} (2004) 006},
\href{http://arxiv.org/abs/astro-ph/0407059}{{\ttfamily arXiv:astro-ph/0407059
  [astro-ph]}}.

\bibitem{Komatsu:2001rj}
E.~Komatsu and D.~N. Spergel, ``{Acoustic signatures in the primary microwave
  background bispectrum},''
  \href{http://dx.doi.org/10.1103/PhysRevD.63.063002}{{\em Phys. Rev. D}
  {\bfseries 63} (2001) 063002},
  \href{http://arxiv.org/abs/astro-ph/0005036}{{\ttfamily
  arXiv:astro-ph/0005036}}.

\bibitem{Planck:2019kim}
{\bfseries Planck} Collaboration, Y.~Akrami {\em et~al.}, ``{Planck 2018
  results. IX. Constraints on primordial non-Gaussianity},''
  \href{http://dx.doi.org/10.1051/0004-6361/201935891}{{\em Astron. Astrophys.}
  {\bfseries 641} (2020) A9}, \href{http://arxiv.org/abs/1905.05697}{{\ttfamily
  arXiv:1905.05697 [astro-ph.CO]}}.

\bibitem{Castorina:2019wmr}
E.~Castorina {\em et~al.}, ``{Redshift-weighted constraints on primordial
  non-Gaussianity from the clustering of the eBOSS DR14 quasars in Fourier
  space},'' \href{http://dx.doi.org/10.1088/1475-7516/2019/09/010}{{\em JCAP}
  {\bfseries 09} (2019) 010}, \href{http://arxiv.org/abs/1904.08859}{{\ttfamily
  arXiv:1904.08859 [astro-ph.CO]}}.

\bibitem{Esposito:2019jkb}
A.~Esposito, L.~Hui, and R.~Scoccimarro, ``{Nonperturbative test of consistency
  relations and their violation},''
  \href{http://dx.doi.org/10.1103/PhysRevD.100.043536}{{\em Phys. Rev. D}
  {\bfseries 100} no.~4, (2019) 043536},
  \href{http://arxiv.org/abs/1905.11423}{{\ttfamily arXiv:1905.11423
  [astro-ph.CO]}}.

\bibitem{Marinucci:2019wdb}
M.~Marinucci, T.~Nishimichi, and M.~Pietroni, ``{Measuring Bias via the
  Consistency Relations of the Large Scale Structure},''
  \href{http://dx.doi.org/10.1103/PhysRevD.100.123537}{{\em Phys. Rev. D}
  {\bfseries 100} no.~12, (2019) 123537},
  \href{http://arxiv.org/abs/1907.09866}{{\ttfamily arXiv:1907.09866
  [astro-ph.CO]}}.

\bibitem{Marinucci:2020weg}
M.~Marinucci, T.~Nishimichi, and M.~Pietroni, ``{Model independent measurement
  of the growth rate from the consistency relations of the LSS},''
  \href{http://dx.doi.org/10.1088/1475-7516/2020/07/054}{{\em JCAP} {\bfseries
  07} (2020) 054}, \href{http://arxiv.org/abs/2005.09574}{{\ttfamily
  arXiv:2005.09574 [astro-ph.CO]}}.

\bibitem{Scoccimarro:1997st}
R.~Scoccimarro, S.~Colombi, J.~N. Fry, J.~A. Frieman, E.~Hivon, and A.~Melott,
  ``{Nonlinear evolution of the bispectrum of cosmological perturbations},''
  \href{http://dx.doi.org/10.1086/305399}{{\em Astrophys. J.} {\bfseries 496}
  (1998) 586}, \href{http://arxiv.org/abs/astro-ph/9704075}{{\ttfamily
  arXiv:astro-ph/9704075}}.

\bibitem{Scoccimarro:1999kp}
R.~Scoccimarro, M.~Zaldarriaga, and L.~Hui, ``{Power spectrum correlations
  induced by nonlinear clustering},''
  \href{http://dx.doi.org/10.1086/308059}{{\em Astrophys. J.} {\bfseries 527}
  (1999) 1}, \href{http://arxiv.org/abs/astro-ph/9901099}{{\ttfamily
  arXiv:astro-ph/9901099}}.

\bibitem{Scoccimarro:2003wn}
R.~Scoccimarro, E.~Sefusatti, and M.~Zaldarriaga, ``{Probing primordial
  non-Gaussianity with large - scale structure},''
  \href{http://dx.doi.org/10.1103/PhysRevD.69.103513}{{\em Phys. Rev.}
  {\bfseries D69} (2004) 103513},
\href{http://arxiv.org/abs/astro-ph/0312286}{{\ttfamily arXiv:astro-ph/0312286
  [astro-ph]}}.

\bibitem{Chiang:2015pwa}
C.-T. Chiang, {\em {Position-dependent power spectrum: a new observable in the
  large-scale structure}}.
\newblock PhD thesis, Munich U., 2015.
\newblock \href{http://arxiv.org/abs/1508.03256}{{\ttfamily arXiv:1508.03256
  [astro-ph.CO]}}.
\newblock
\url{http://inspirehep.net/record/1387739/files/arXiv:1508.03256.pdf}.
\newblock

\bibitem{Barreira:2017fjz}
A.~Barreira, E.~Krause, and F.~Schmidt, ``{Complete super-sample lensing
  covariance in the response approach},''
  \href{http://dx.doi.org/10.1088/1475-7516/2018/06/015}{{\em JCAP} {\bfseries
  06} (2018) 015}, \href{http://arxiv.org/abs/1711.07467}{{\ttfamily
  arXiv:1711.07467 [astro-ph.CO]}}.

\bibitem{Gil-Marin:2011jtv}
H.~Gil-Marin, C.~Wagner, F.~Fragkoudi, R.~Jimenez, and L.~Verde, ``{An improved
  fitting formula for the dark matter bispectrum},''
  \href{http://dx.doi.org/10.1088/1475-7516/2012/02/047}{{\em JCAP} {\bfseries
  02} (2012) 047}, \href{http://arxiv.org/abs/1111.4477}{{\ttfamily
  arXiv:1111.4477 [astro-ph.CO]}}.

\bibitem{Takahashi:2019hth}
R.~Takahashi, T.~Nishimichi, T.~Namikawa, A.~Taruya, I.~Kayo, K.~Osato,
  Y.~Kobayashi, and M.~Shirasaki, ``{Fitting the nonlinear matter bispectrum by
  the Halofit approach},''
  \href{http://dx.doi.org/10.3847/1538-4357/ab908d}{{\em Astrophys. J.}
  {\bfseries 895} no.~2, (2020) 113},
  \href{http://arxiv.org/abs/1911.07886}{{\ttfamily arXiv:1911.07886
  [astro-ph.CO]}}.

\bibitem{Bertolini:2016bmt}
D.~Bertolini, K.~Schutz, M.~P. Solon, and K.~M. Zurek, ``{The Trispectrum in
  the Effective Field Theory of Large Scale Structure},''
  \href{http://dx.doi.org/10.1088/1475-7516/2016/06/052}{{\em JCAP} {\bfseries
  06} (2016) 052}, \href{http://arxiv.org/abs/1604.01770}{{\ttfamily
  arXiv:1604.01770 [astro-ph.CO]}}.

\bibitem{Gualdi:2020eag}
D.~Gualdi, S.~Novell, H.~Gil-Mar\'\i{}n, and L.~Verde, ``{Matter trispectrum:
  theoretical modelling and comparison to N-body simulations},''
  \href{http://dx.doi.org/10.1088/1475-7516/2021/01/015}{{\em JCAP} {\bfseries
  01} (2021) 015}, \href{http://arxiv.org/abs/2009.02290}{{\ttfamily
  arXiv:2009.02290 [astro-ph.CO]}}.

\bibitem{Steele:2021lnz}
T.~Steele and T.~Baldauf, ``{Precise Calibration of the One-Loop Trispectrum in
  the Effective Field Theory of Large Scale Structure},''
  \href{http://dx.doi.org/10.1103/PhysRevD.103.103518}{{\em Phys. Rev. D}
  {\bfseries 103} no.~10, (2021) 103518},
  \href{http://arxiv.org/abs/2101.10289}{{\ttfamily arXiv:2101.10289
  [astro-ph.CO]}}.

\bibitem{Creminelli:2013mca}
P.~Creminelli, J.~Noreña, M.~Simonović, and F.~Vernizzi, ``{Single-Field
  Consistency Relations of Large Scale Structure},''
  \href{http://dx.doi.org/10.1088/1475-7516/2013/12/025}{{\em JCAP} {\bfseries
  1312} (2013) 025},
\href{http://arxiv.org/abs/1309.3557}{{\ttfamily arXiv:1309.3557
  [astro-ph.CO]}}.

\bibitem{10.1214/aoms/1177729893}
J.~Sherman and W.~J. Morrison, ``{Adjustment of an Inverse Matrix Corresponding
  to a Change in One Element of a Given Matrix},''
  \href{http://dx.doi.org/10.1214/aoms/1177729893}{{\em The Annals of
  Mathematical Statistics} {\bfseries 21} no.~1, (1950) 124 -- 127}.
  \url{https://doi.org/10.1214/aoms/1177729893}.

\bibitem{10.1214/aoms/1177729698}
M.~S. Bartlett, ``{An Inverse Matrix Adjustment Arising in Discriminant
  Analysis},'' \href{http://dx.doi.org/10.1214/aoms/1177729698}{{\em The Annals
  of Mathematical Statistics} {\bfseries 22} no.~1, (1951) 107 -- 111}.
  \url{https://doi.org/10.1214/aoms/1177729698}.

\bibitem{Springel:2005mi}
V.~Springel, ``{The Cosmological simulation code GADGET-2},''
  \href{http://dx.doi.org/10.1111/j.1365-2966.2005.09655.x}{{\em Mon. Not. Roy.
  Astron. Soc.} {\bfseries 364} (2005) 1105--1134},
\href{http://arxiv.org/abs/astro-ph/0505010}{{\ttfamily arXiv:astro-ph/0505010
  [astro-ph]}}.

\bibitem{Crocce:2006ve}
M.~Crocce, S.~Pueblas, and R.~Scoccimarro, ``{Transients from Initial
  Conditions in Cosmological Simulations},''
  \href{http://dx.doi.org/10.1111/j.1365-2966.2006.11040.x}{{\em Mon. Not. Roy.
  Astron. Soc.} {\bfseries 373} (2006) 369--381},
\href{http://arxiv.org/abs/astro-ph/0606505}{{\ttfamily arXiv:astro-ph/0606505
  [astro-ph]}}.

\bibitem{Lewis:1999bs}
A.~Lewis, A.~Challinor, and A.~Lasenby, ``{Efficient computation of CMB
  anisotropies in closed FRW models},''
  \href{http://dx.doi.org/10.1086/309179}{{\em Astrophys. J.} {\bfseries 538}
  (2000) 473--476},
\href{http://arxiv.org/abs/astro-ph/9911177}{{\ttfamily arXiv:astro-ph/9911177
  [astro-ph]}}.

\bibitem{Sefusatti:2015aex}
E.~Sefusatti, M.~Crocce, R.~Scoccimarro, and H.~Couchman, ``{Accurate
  Estimators of Correlation Functions in Fourier Space},''
  \href{http://dx.doi.org/10.1093/mnras/stw1229}{{\em Mon. Not. Roy. Astron.
  Soc.} {\bfseries 460} no.~4, (2016) 3624--3636},
  \href{http://arxiv.org/abs/1512.07295}{{\ttfamily arXiv:1512.07295
  [astro-ph.CO]}}.

\bibitem{Smith:2008ut}
R.~E. Smith, ``{Covariance of cross-correlations: towards efficient measures
  for large-scale structure},''
  \href{http://dx.doi.org/10.1111/j.1365-2966.2009.15490.x}{{\em Mon. Not. Roy.
  Astron. Soc.} {\bfseries 400} (2009) 851},
  \href{http://arxiv.org/abs/0810.1960}{{\ttfamily arXiv:0810.1960
  [astro-ph]}}.

\bibitem{Meerburg:2019qqi}
P.~D. Meerburg {\em et~al.}, ``{Primordial Non-Gaussianity},''
\href{http://arxiv.org/abs/1903.04409}{{\ttfamily arXiv:1903.04409
  [astro-ph.CO]}}.

\bibitem{Dalal:2007cu}
N.~Dalal, O.~Dor\'e, D.~Huterer, and A.~Shirokov, ``{The imprints of primordial
  non-gaussianities on large-scale structure: scale dependent bias and
  abundance of virialized objects},''
  \href{http://dx.doi.org/10.1103/PhysRevD.77.123514}{{\em Phys. Rev.}
  {\bfseries D77} (2008) 123514},
\href{http://arxiv.org/abs/0710.4560}{{\ttfamily arXiv:0710.4560 [astro-ph]}}.

\bibitem{Matarrese:2008nc}
S.~Matarrese and L.~Verde, ``{The effect of primordial non-Gaussianity on halo
  bias},'' \href{http://dx.doi.org/10.1086/587840}{{\em Astrophys. J.}
  {\bfseries 677} (2008) L77--L80},
\href{http://arxiv.org/abs/0801.4826}{{\ttfamily arXiv:0801.4826 [astro-ph]}}.

\bibitem{Slosar:2008hx}
A.~Slosar, C.~Hirata, U.~Seljak, S.~Ho, and N.~Padmanabhan, ``{Constraints on
  local primordial non-Gaussianity from large scale structure},''
  \href{http://dx.doi.org/10.1088/1475-7516/2008/08/031}{{\em JCAP} {\bfseries
  0808} (2008) 031},
\href{http://arxiv.org/abs/0805.3580}{{\ttfamily arXiv:0805.3580 [astro-ph]}}.

\bibitem{Biagetti:2019bnp}
M.~Biagetti, ``{The Hunt for Primordial Interactions in the Large Scale
  Structures of the Universe},''
  \href{http://dx.doi.org/10.3390/galaxies7030071}{{\em Galaxies} {\bfseries 7}
  no.~3, (2019) 71},
\href{http://arxiv.org/abs/1906.12244}{{\ttfamily arXiv:1906.12244
  [astro-ph.CO]}}.

\bibitem{Baldauf:2010vn}
T.~Baldauf, U.~Seljak, and L.~Senatore, ``{Primordial non-Gaussianity in the
  Bispectrum of the Halo Density Field},''
  \href{http://dx.doi.org/10.1088/1475-7516/2011/04/006}{{\em JCAP} {\bfseries
  1104} (2011) 006},
\href{http://arxiv.org/abs/1011.1513}{{\ttfamily arXiv:1011.1513
  [astro-ph.CO]}}.

\bibitem{Sefusatti:2010ee}
E.~Sefusatti, M.~Crocce, and V.~Desjacques, ``{The Matter Bispectrum in N-body
  Simulations with non-Gaussian Initial Conditions},''
  \href{http://dx.doi.org/10.1111/j.1365-2966.2010.16723.x}{{\em Mon. Not. Roy.
  Astron. Soc.} {\bfseries 406} (2010) 1014--1028},
\href{http://arxiv.org/abs/1003.0007}{{\ttfamily arXiv:1003.0007
  [astro-ph.CO]}}.

\bibitem{Sefusatti:2011gt}
E.~Sefusatti, M.~Crocce, and V.~Desjacques, ``{The Halo Bispectrum in N-body
  Simulations with non-Gaussian Initial Conditions},''
  \href{http://dx.doi.org/10.1111/j.1365-2966.2012.21271.x}{{\em Mon. Not. Roy.
  Astron. Soc.} {\bfseries 425} (2012) 2903},
\href{http://arxiv.org/abs/1111.6966}{{\ttfamily arXiv:1111.6966
  [astro-ph.CO]}}.

\bibitem{Yokoyama:2013mta}
S.~Yokoyama, T.~Matsubara, and A.~Taruya, ``{Halo/galaxy bispectrum with
  primordial non-Gaussianity from integrated perturbation theory},''
  \href{http://dx.doi.org/10.1103/PhysRevD.89.043524}{{\em Phys. Rev.}
  {\bfseries D89} no.~4, (2014) 043524},
\href{http://arxiv.org/abs/1310.4925}{{\ttfamily arXiv:1310.4925
  [astro-ph.CO]}}.

\bibitem{Tasinato:2013vna}
G.~Tasinato, M.~Tellarini, A.~J. Ross, and D.~Wands, ``{Primordial
  non-Gaussianity in the bispectra of large-scale structure},''
  \href{http://dx.doi.org/10.1088/1475-7516/2014/03/032}{{\em JCAP} {\bfseries
  1403} (2014) 032},
\href{http://arxiv.org/abs/1310.7482}{{\ttfamily arXiv:1310.7482
  [astro-ph.CO]}}.

\bibitem{Dizgah:2015kqi}
A.~M. Dizgah, K.~C. Chan, J.~Nore\~na, M.~Biagetti, and V.~Desjacques,
  ``{Squeezing the halo bispectrum: a test of bias models},''
\href{http://arxiv.org/abs/1512.06084}{{\ttfamily arXiv:1512.06084
  [astro-ph.CO]}}.

\bibitem{Hashimoto:2015tnv}
I.~Hashimoto, A.~Taruya, T.~Matsubara, T.~Namikawa, and S.~Yokoyama,
  ``{Constraining higher-order parameters for primordial non-Gaussianities from
  power spectra and bispectra of imaging surveys},''
  \href{http://dx.doi.org/10.1103/PhysRevD.93.103537}{{\em Phys. Rev. D}
  {\bfseries 93} no.~10, (2016) 103537},
  \href{http://arxiv.org/abs/1512.08352}{{\ttfamily arXiv:1512.08352
  [astro-ph.CO]}}.

\bibitem{Tellarini:2016sgp}
M.~Tellarini, A.~J. Ross, G.~Tasinato, and D.~Wands, ``{Galaxy bispectrum,
  primordial non-Gaussianity and redshift space distortions},''
  \href{http://dx.doi.org/10.1088/1475-7516/2016/06/014}{{\em JCAP} {\bfseries
  06} (2016) 014}, \href{http://arxiv.org/abs/1603.06814}{{\ttfamily
  arXiv:1603.06814 [astro-ph.CO]}}.

\bibitem{Hashimoto:2016lmh}
I.~Hashimoto, S.~Mizuno, and S.~Yokoyama, ``{Constraining equilateral-type
  primordial non-Gaussianities from imaging surveys},''
  \href{http://dx.doi.org/10.1103/PhysRevD.94.043532}{{\em Phys. Rev. D}
  {\bfseries 94} no.~4, (2016) 043532},
  \href{http://arxiv.org/abs/1605.07348}{{\ttfamily arXiv:1605.07348
  [astro-ph.CO]}}.

\bibitem{Yamauchi:2016wuc}
D.~Yamauchi, S.~Yokoyama, and K.~Takahashi, ``{Multitracer technique for galaxy
  bispectrum: An application to constraints on nonlocal primordial
  non-Gaussianities},''
  \href{http://dx.doi.org/10.1103/PhysRevD.95.063530}{{\em Phys. Rev. D}
  {\bfseries 95} no.~6, (2017) 063530},
  \href{http://arxiv.org/abs/1611.03590}{{\ttfamily arXiv:1611.03590
  [astro-ph.CO]}}.

\bibitem{DiDio:2016gpd}
E.~Di~Dio, H.~Perrier, R.~Durrer, G.~Marozzi, A.~Moradinezhad~Dizgah,
  J.~Nore\~na, and A.~Riotto, ``{Non-Gaussianities due to Relativistic
  Corrections to the Observed Galaxy Bispectrum},''
  \href{http://dx.doi.org/10.1088/1475-7516/2017/03/006}{{\em JCAP} {\bfseries
  03} (2017) 006}, \href{http://arxiv.org/abs/1611.03720}{{\ttfamily
  arXiv:1611.03720 [astro-ph.CO]}}.

\bibitem{Chiang:2017vsq}
C.-T. Chiang, A.~M. Cieplak, F.~Schmidt, and A.~Slosar, ``{Response approach to
  the squeezed-limit bispectrum: application to the correlation of quasar and
  Lyman-$\alpha$ forest power spectrum},''
  \href{http://dx.doi.org/10.1088/1475-7516/2017/06/022}{{\em JCAP} {\bfseries
  06} (2017) 022}, \href{http://arxiv.org/abs/1701.03375}{{\ttfamily
  arXiv:1701.03375 [astro-ph.CO]}}.

\bibitem{An:2017rwo}
H.~An, M.~McAneny, A.~K. Ridgway, and M.~B. Wise, ``{Non-Gaussian Enhancements
  of Galactic Halo Correlations in Quasi-Single Field Inflation},''
  \href{http://dx.doi.org/10.1103/PhysRevD.97.123528}{{\em Phys. Rev.}
  {\bfseries D97} no.~12, (2018) 123528},
\href{http://arxiv.org/abs/1711.02667}{{\ttfamily arXiv:1711.02667 [hep-ph]}}.

\bibitem{MoradinezhadDizgah:2018ssw}
A.~Moradinezhad~Dizgah, H.~Lee, J.~B. Mu\~noz, and C.~Dvorkin, ``{Galaxy
  Bispectrum from Massive Spinning Particles},''
  \href{http://dx.doi.org/10.1088/1475-7516/2018/05/013}{{\em JCAP} {\bfseries
  05} (2018) 013}, \href{http://arxiv.org/abs/1801.07265}{{\ttfamily
  arXiv:1801.07265 [astro-ph.CO]}}.

\bibitem{MoradinezhadDizgah:2018pfo}
A.~Moradinezhad~Dizgah, G.~Franciolini, A.~Kehagias, and A.~Riotto,
  ``{Constraints on long-lived, higher-spin particles from galaxy
  bispectrum},'' \href{http://dx.doi.org/10.1103/PhysRevD.98.063520}{{\em Phys.
  Rev. D} {\bfseries 98} no.~6, (2018) 063520},
  \href{http://arxiv.org/abs/1805.10247}{{\ttfamily arXiv:1805.10247
  [astro-ph.CO]}}.

\bibitem{Barreira:2020ekm}
A.~Barreira, ``{On the impact of galaxy bias uncertainties on primordial
  non-Gaussianity constraints},''
  \href{http://dx.doi.org/10.1088/1475-7516/2020/12/031}{{\em JCAP} {\bfseries
  12} (2020) 031}, \href{http://arxiv.org/abs/2009.06622}{{\ttfamily
  arXiv:2009.06622 [astro-ph.CO]}}.

\bibitem{Biagetti:2020skr}
M.~Biagetti, A.~Cole, and G.~Shiu, ``{The Persistence of Large Scale Structures
  I: Primordial non-Gaussianity},''
  \href{http://dx.doi.org/10.1088/1475-7516/2021/04/061}{{\em JCAP} {\bfseries
  04} (2021) 061}, \href{http://arxiv.org/abs/2009.04819}{{\ttfamily
  arXiv:2009.04819 [astro-ph.CO]}}.

\bibitem{Scoccimarro:2011pz}
R.~Scoccimarro, L.~Hui, M.~Manera, and K.~C. Chan, ``{Large-scale Bias and
  Efficient Generation of Initial Conditions for Non-Local Primordial
  Non-Gaussianity},'' \href{http://dx.doi.org/10.1103/PhysRevD.85.083002}{{\em
  Phys. Rev.} {\bfseries D85} (2012) 083002},
\href{http://arxiv.org/abs/1108.5512}{{\ttfamily arXiv:1108.5512
  [astro-ph.CO]}}.

\bibitem{Blas:2011rf}
D.~Blas, J.~Lesgourgues, and T.~Tram, ``{The Cosmic Linear Anisotropy Solving
  System (CLASS) II: Approximation schemes},''
  \href{http://dx.doi.org/10.1088/1475-7516/2011/07/034}{{\em JCAP} {\bfseries
  1107} (2011) 034},
\href{http://arxiv.org/abs/1104.2933}{{\ttfamily arXiv:1104.2933
  [astro-ph.CO]}}.

\bibitem{Behroozi:2011ju}
P.~S. Behroozi, R.~H. Wechsler, and H.-Y. Wu, ``{The Rockstar Phase-Space
  Temporal Halo Finder and the Velocity Offsets of Cluster Cores},''
  \href{http://dx.doi.org/10.1088/0004-637X/762/2/109}{{\em Astrophys. J.}
  {\bfseries 762} (2013) 109},
\href{http://arxiv.org/abs/1110.4372}{{\ttfamily arXiv:1110.4372
  [astro-ph.CO]}}.

\bibitem{Davis:1985rj}
M.~Davis, G.~Efstathiou, C.~S. Frenk, and S.~D.~M. White, ``{The Evolution of
  Large Scale Structure in a Universe Dominated by Cold Dark Matter},''
  \href{http://dx.doi.org/10.1086/163168}{{\em Astrophys. J.} {\bfseries 292}
  (1985) 371--394}.

\bibitem{Hou:2022wfj}
J.~Hou, Z.~Slepian, and R.~N. Cahn, ``{Measurement of Parity-Odd Modes in the
  Large-Scale 4-Point Correlation Function of SDSS BOSS DR12 CMASS and LOWZ
  Galaxies},'' \href{http://arxiv.org/abs/2206.03625}{{\ttfamily
  arXiv:2206.03625 [astro-ph.CO]}}.

\end{thebibliography}\endgroup
\end{document}